\documentclass[12pt,reqno,a4paper]{article}
\usepackage[OT1]{fontenc}
\usepackage{authblk,amsthm,amsmath, amsfonts, caption, multirow, float, graphicx, tikz, pdfpages, setspace, enumitem, fancyhdr, longtable} 
\usepackage[misc]{ifsym}
\usepackage[compact]{titlesec}
\usepackage[round]{natbib}
\usepackage[colorlinks,citecolor=blue,urlcolor=blue]{hyperref}
\usepackage[margin=1in,headsep=10mm,footskip=15mm]{geometry}
\titlespacing*{\section}
{0pt}{2.5ex plus 1ex minus .2ex}{1.8ex plus .2ex}
\titlespacing*{\subsection}
{0pt}{1.5ex plus 1ex minus .2ex}{0.8ex plus .2ex}
\titlespacing*{\subsubsection}
{0pt}{1.5ex plus 1ex minus .2ex}{0.8ex plus .2ex}

\numberwithin{equation}{section}
\theoremstyle{plain}

\newtheorem{thm}{Theorem}[section]

\newtheorem*{result*}{Result}

\theoremstyle{definition}

\newtheorem{example}[thm]{Example}
\newtheorem*{example*}{Example}

\title{\vspace{-6em}{\small Running headline: Bayesian model selection for SECR models}\\[2em]Bayesian Model Selection for a Class of Spatially-Explicit Capture Recapture Models} 
\author[1]{Soumen Dey} 
\author[1]{Mohan Delampady} 
\author[1]{Arjun M. Gopalaswamy \thanks{\emph{arjungswamy@gmail.com} ({\scriptsize \Letter})}}
\affil[1]{\footnotesize Statistics and Mathematics Unit, Indian Statistical Institute, Bangalore 560059, India}

\newcommand{\w}{\omega}

\newcommand{\V}{\ensuremath{\mathcal{V}}}
\newcommand{\E}{\ensuremath{\mathbb{E}}}
\newcommand{\Var}{\ensuremath{\mathbb{V}}\mbox{ar}}

\newcommand{\dic}{\scriptstyle \text{DIC}}
\newcommand{\waic}{\scriptstyle \text{WAIC}}

\begin{document}

\setlength{\abovedisplayskip}{5pt}
\setlength{\belowdisplayskip}{5pt}
\date{}
\maketitle
\vspace{-10mm}
\begin{center}	{\bf Abstract } \end{center}
{\footnotesize\begin{enumerate}
		
\item  A vast amount of ecological knowledge generated over the past two decades has hinged upon the ability of model selection methods to discriminate among various ecological hypotheses. The last decade has seen the rise of Bayesian hierarchical models in ecology. Consequently, commonly used tools, such as the AIC, become largely inapplicable and there appears to be no consensus about a particular model selection tool that can be universally applied. 

\item We focus on a specific class of competing Bayesian spatially explicit capture recapture (SECR) models and first apply some of the recommended Bayesian model selection tools: (1) Bayes Factor - using (a) Gelfand-Dey (b) harmonic mean methods, (2) Deviance Information Criterion (DIC), (3) Watanabe-Akaike's Information Criteria (WAIC) and (4) the posterior predictive loss function. In all, we evaluate 25 variants of model selection tools in our study. 

\item We evaluate these model selection tools from the standpoint of model selection and parameter estimation by contrasting the choice recommended by a tool with a `true' model. In all, we generate 120 simulated data sets using the true model and assess the frequency with which the true model is selected and how well the tool estimates $N$ (population size).

\item We find that when information content is low in the data, no particular model selection tool can be recommended to help realise, simultaneously, both the goals of model selection and parameter estimation. In such scenarios, we recommend that practitioners utilise our application of Bayes Factor (Gelfand-Dey with MAP approximation or the harmonic mean) for parameter estimation and recommend the posterior predictive loss approach for model selection when information content is low. Generally, when we consider both the objectives together, we recommend the use of our application of Bayes Factor for Bayesian SECR models.

\item Our study highlights the aspect that although new model selection tools are emerging (eg: WAIC) in the applied statistics literature, an uncritical absorption of these new tools (i.e. without assessing their efficacies for the problem at hand) into ecological practice may provide misleading inferences. 


	\end{enumerate} }

\section{Introduction}\label{intro.BMSE}



Following the highly influential text \cite{burnham2002model} on model selection, ecologists and conservation biologists have drastically shifted their inferential practice from the `hypothesis testing' approach to the more appropriate `hypothesis discrimination' approach \citep{ellison2004bayesian, diaz2005statistical}. Till date, \cite{burnham2002model} has been cited 43047 times (as on September 22, 2018 - Google Scholar), demonstrating the impact of this text. One may argue, that this contribution has helped increase the pace of growth in ecological knowledge because it has paved the way for researchers to draw inferences, more authentically, because of the ability to now assess the influence of various competing a priori hypotheses (models) without altering the study question to suit the restrictive hypothesis testing paradigm \citep{bolker2008ecological}. 

Since a large amount of ecological data are based on observations it permitted for ecologists to take the approach of `detectives'  rather than `hypothetico-deductive' scientists by formulating models using likelihood functions to confront various a priori hypotheses using observational data  \citep{hilborn1997ecological}. And by maximizing the likelihood and using a model selection tool, such as the Akaike's Information Criterion \citep{burnham2002model}, researchers found a way to place increased faith on models favoured by such criteria. Thus, a vast amount of ecological knowledge generated, has relied on the robustness of such model selection tools in accurately discriminating hypotheses. 

Recently, there has been an increased use of hierarchical models in ecology since they appear to address two important issues: (1) ecological scales are naturally hierarchical in structure and (2) hierarchical models form a natural way of incorporating the observation process \citep{royledorazio2008hierarchical}. And with powerful tools such as MCMC, it is now possible to confront complex ecological models with data in a Bayesian inferential framework \citep{bolker2008ecological}. However, it remains unclear as to how to discriminate among competing hypotheses (models) because popular model selection criteria (such as AIC, BIC or DIC) \citep{burnham2002model} aren't easy to apply or work poorly for complex hierarchical models \citep{millar2009comparison}. 

It is well known that, asymptotically, Bayes factor is the preferred model selection tool due to its consistency property \citep{GDS2006Bayesian, MR2723361, bernardo2009bayesian}. However, this property holds only under certain regularity conditions that are often difficult to verify for complex models \citep{berger2003approximations, dass2004note, ghosh2001model}. But more prominently, there is vast literature expressing the difficulties in computing the marginal likelihood in applied problems \citep{weinberg2012computing, chan2015marginal, wang2018new}. Hence, it becomes necessary to also consider alternatives to Bayes factor or to find novel ways of applying them in practice. 

Recently, \cite{hooten2015guide} summarised a wide array of Bayesian model selection methods that are available to ecologists. However, the generality of these recommendations remain unknown.
Given such innate difficulties involved in discovering the `ideal' model selection tool both from the standpoint of theory \emph{and} its application to a broad class of models, it appears to be prudent to explore the model selection issue by conditioning, at least, on a particular \emph{class} of models.   

Here, we evaluate various Bayesian model selection tools on a class of Bayesian spatially explicit capture recapture (SECR) models that are now used frequently for animal density estimation \citep{royle2013spatial}. Although, previously, \cite{goldberg2015examining} has attempted to apply Bayes Factor (Gelfand-Dey estimator) in an abundance estimation problem for leopards (\textit Panthera pardus) for model selection, their approach of computing the ratio term in the estimator seems inaccurate in the context of how the denominator has to be computed according to \cite{gelfand1994bayesian}. 

Thus, we evaluate various Bayesian model selection tools by:  
\begin{enumerate}
\item Defining a class of competing models (in our case these include the model developed in \cite{dey2017spatial} along with simplified alternatives) that vary both in terms of structural and model complexity.
\item Simulating data sets from a `true' model.
\item Practically implementing a variety of Bayesian model selection tools, and in specific cases, also proposing alternatives. 
\item Assessing the efficacy of these implementations from the standpoint of model selection and parameter estimation. 
\item Providing recommendations to practitioners based on our results.  
\end{enumerate}

\section{Methods}\label{methods.BMSE}


We describe here the true (data-generating) model which will be used for simulation of the data sets. We then fit candidate models to these data sets and apply and assess efficiencies of various model selection tools.





\subsection{The candidate model set}\label{candmodelset.BMSE}


\cite{dey2017spatial} develops a Bayesian SECR model for partially identified individuals as a Bayesian hierarchical model  \citep{royledorazio2008hierarchical}. We summarise the description of the model here below. 

\subsubsection{Sampling situation}{\ }\\
The notations used in this article are described in Tables~\ref{par.definitons1} and \ref{par.definitons2}. However, we describe a few variables and parameters for ease in the model description below.  
 
Consider a capture-recapture survey of a species with naturally marked individuals in which two detectors are collocated at $J$ trap stations (within a bounded geographic region $\mathcal{V} \, \subset \mathbb{R}^2$) and kept active for $K$ sampling occasions. An individual can be completely identified if both the detectors record the individual simultaneously at least once during the course of study \citep{royle2015spatial}. We assume that each detector captures some mutually exclusive attributes of an individual. These capture outcomes are recorded as binary observations $y_{ijk}^{(1)}$ and $y_{ijk}^{(2)}$ for an individual $i$ at trap station $\mathbf{x}_j$ on sampling occasion $k$ corresponding to detectors 1 and 2 respectively. The paired Bernoulli outcomes $y_{ijk} = (y_{ijk}^{(1)}, y_{ijk}^{(2)})$ give rise to bilateral spatial capture-recapture data for each individual $i$ at location $\mathbf{x}_j$ on occasion $k$. The array of a bilateral capture history for an individual $i$ is denoted by  $\mathbf{Y_{\emph{i}, obs}} = (\mathbf{Y^{(1)}_{\emph{i}, obs}}, \mathbf{Y^{(2)}_{\emph{i}, obs}}) = ((y_{ijk}^{(1)}, y_{ijk}^{(2)}))_ {j,k}$, which is of dimension $2\times J \times K$. 
Below we provide an example of a sample data set coming out of a spatial capture-recapture survey with two detectors deployed at each station.
\begin{example} \label{example:sampledata}
	Suppose a capture-recapture survey is conducted where a pair of detectors (1 and 2) are deployed at each of the 3 ($ = J$) trap stations and kept active for 4 ($ = K$) sampling occasions. Two individuals get fully identified based on their obtained capture histories (captured in both cameras at least once during the survey). The capture history for each of these two fully-identified individuals is of dimension $2 \times 3 \times 4$. The detection histories are tabulated in Table~\ref{sampledata}. Here, individual 1 is fully-identified owing to the capture event at trap 2 on occasion 4. Individual 1 is also fully-identified as it is captured at trap 2 on occasion 4. Due to absence of simultaneous capture events in the detection histories of the partially-identified individuals, we are uncertain about whether these histories correspond to two different individuals or to the same individual.
\end{example}
\subsubsection{Model likelihoods}{\ }\\
When detection rates in recorded samples are low due to failure or malfunction of detectors, capture-recapture data may comprise of individuals with uncertain identities or `partially identified individuals'. \cite{dey2017spatial} separately accounts for the process of animal arrival within detection region of a detector and detection process by conditioning on animal arrival - thus modelling the underlying mechanism by which we obtain different events leading to partial identification.

The probability of animal arrival $\eta_{j}(\mathbf{s}_i)$ (termed as `trap entry probability') is modelled as a decreasing function of Euclidean distance $d(\mathbf{s}_i, \mathbf{x}_j) = \Arrowvert \mathbf{s}_i - \mathbf{x}_j \Arrowvert$ between individual activity centre $\mathbf{s}_i$ and trap station $\mathbf{x}_j$ : $\eta_{j}(\mathbf{s}_i) = \w_0\, \exp(-d(\mathbf{s}_i, \mathbf{x}_j)^2/(2\sigma^2))$. Here, $\w_0$ is regarded as the ``baseline trap entry probability'' and $\sigma$ quantifies the rate of decline in trap entry probability as $d(\mathbf{s}_i, \mathbf{x}_j)$ increases. The observation process is parameterised in terms of detection probability $\phi$ which denotes the probability that any arbitrary individual $i$ is detected by a detector on some occasion $k$ given its arrival at that trap.

The obtained capture history observations $\mathbf{Y_{obs}^{(1)}}=((y_{ijk}^{(1)}))_{i,j,k}$ and $\mathbf{Y_{obs}^{(2)}}=((y_{ijk}^{(2)}))_{i,j,k}$ from the two detectors 1 and 2 during a spatial capture-recapture survey may not be synchronised as detectors often perform imperfectly. These two data arrays are then augmented with zero capture histories so that each of them is of dimension $M \times J \times K$, $M$ being an upper bound of the population size. This also makes the dimension of the likelihood fixed in each iteration of Markov chain Monte Carlo algorithm which in turn eases computation. A vector of $M$ latent binary variables $z = (z_1, . . . , z_M)'$ is introduced where $z_i = 1$ implies that individual $i$ is a member of the population. We assume that each $z_i$ is a Bernoulli random variable with parameter $\psi$ and is independent of other $z_j$'s.  Here $\psi$ is the proportion of individuals that are real and present within $\V$. Thus, the true population size $N$ follows the Binomial distribution with parameters $M$ and $\psi$. The individuals from the two lists obtained from detector 1 and detector 2 are linked probabilistically, by introducing a latent identity variable $\mathbf{L}=(\text{L}_1, \text{L}_2, \dots, \text{L}_M)'$ which is a one-to-one mapping from an index set of individuals capture by detector 2 to $\{1, 2, \dots,M\}$ giving the true index of each of detector 2 individuals. Without loss of generality, the true identity of each individual in the population is defined to be in the row-order of capture histories of detector 1. Then the rows of detector 2 data set $\mathbf{Y^{(2)}}$ are reordered as indicated by $\mathbf{L}$ to synchronise with the individuals of the detector 1 data set $\mathbf{Y^{(1)}}$. We denote this newly ordered detector 2 data set as $\mathbf{Y^{(2*)}}$. The joint density of $\mathbf{Y^*} := ( \mathbf{Y^{(1)}}, \mathbf{Y^{(2*)}} ) = ((y_{ijk}^{(1)}, y_{ijk}^{(2*)}))$  is as given below: 
{\small \begin{align} \label{intlik.lr.dey1}
& f(\mathbf{Y^*} \, | \, \phi, \w_0, \sigma, \mathbf{z}, \mathbf{S}, \mathbf{L})  
  =\prod_{i=1}^{M}\,  \Big{\{} 
\phi^{y_{i\cdot \cdot}} (1-\phi)^{2n_{i\cdot}-y_{i\cdot \cdot}} \,
\prod_{j=1}^{J} \,  \eta_j(\mathbf{s}_i)^{n_{ij}} \{(1-\eta_j(\mathbf{s}_i)) + \eta_j(\mathbf{s}_i) (1-\phi)^2 \}^{K-n_{ij}}\Big{\}}^{z_i},
\end{align}}where $y_{i\cdot \cdot} = y_{i\cdot \cdot}^{(1)} + y_{i \cdot \cdot}^{(2*)}$, $n_{ij}=\sum_{k=1}^K I(y_{ijk}^{(1)} + y_{ijk}^{(2*)} >0)$ is the number of times individual $i$ got detected on at least one the detectors over $K$ occasions and $n_{i\cdot} = \sum_{j=1}^J n_{ij}$. 

Prior to \cite{dey2017spatial}, \cite{royle2015spatial} proposed an SECR model for partially identified individuals coming from spatial capture-recapture surveys. The joint density of $\mathbf{Y^*}$ under \cite{royle2015spatial} is the following:
{\small \begin{align} \label{intlik.lr.royle1}
& f_\text{R}(\mathbf{Y^*} \, | \,  p_0, \sigma, \mathbf{z}, \mathbf{S}, \mathbf{L}) 
=\prod_{i=1}^{M}\, \prod_{j=1}^{J} \, \big{\{} p_{j}(\mathbf{s}_i) ^{y_{ij \cdot}} (1-p_{j}(\mathbf{s}_i))^{2K - y_{ij \cdot}} \big{\}}^{z_i},
\end{align}}where $y_{ij \cdot} = y_{ij \cdot}^{(1)} + y_{ij \cdot}^{(2*)}$ and $p_{j}(\mathbf{s}_i) = p_0\, \exp(-d(\mathbf{s}_i, \mathbf{x}_j)^2/(2\sigma^2))$ denotes the probability that an individual $i$ is detected at a trap station $\mathbf{x}_j$ on some occasion $k$. 

Note that, unlike model (\ref{intlik.lr.dey1}), here movement through detection region is considered inherently as a part of the observation process and $p_0$ is regarded as ``baseline detection probability'' and $\sigma$, although related to animal movement, is regarded as the \emph{rate of decline in detection probability}. Qualitatively, the absence of $\phi$ in (\ref{intlik.lr.royle1}), distinguishes this model from (\ref{intlik.lr.dey1}), and can be regarded as a less general model. 

Both the models, (\ref{intlik.lr.dey1}) and (\ref{intlik.lr.royle1}), can be extended by introducing a binary covariate on sex category $u$ on spatial animal movement, $\sigma$, as in \cite{sollmann2011improving}. 
We define $\sigma$ as a function of the latent structural vector $\mathbf{u} = (u_1, u_2, \dots, u_M)'$:
$\sigma (u_i) = \sigma_m$,  if $u_i=1$, i.e., individual $i$ is a male; $\sigma (u_i) = \sigma_f$,  if $u_i=0$, i.e., individual $i$ is a female. $u_i$'s are independently and identically distributed Bernoulli random variables with parameter $\theta$, $\theta$ being the probability that an arbitrary individual in the population is male. 
Let $\mathbf{u_{obs}}\, (\subset \mathbf{u})$ be a vector of binary observations on sex category of the captured individuals. The vector of latent missing observations in $\mathbf{u}$ is denoted by
 $\mathbf{u_0}$. Assuming that covariate information on individual sex category is available,
the joint density of $\mathbf{Y^*}$ and $\mathbf{u}$ under \cite{dey2017spatial} and \cite{royle2015spatial} are, respectively, the following: 
{\small \begin{align} 
	& f(\mathbf{Y^*}, \mathbf{u_{obs}}\, | \, \theta, \phi, \w_0, \sigma_m, \sigma_f, \mathbf{u_0}, \mathbf{z}, \mathbf{S}, \mathbf{L}) \nonumber \\
	& =\prod_{i=1}^{M}\,  \Big{[} \Big{\{}  \theta^{u_i} (1-\theta)^{1-u_i} \phi^{y_{i\cdot \cdot}} (1-\phi)^{2n_{i\cdot}-y_{i\cdot \cdot}} \, \prod_{j=1}^{J} \,  \eta_j(\mathbf{s}_i, u_i)^{n_{ij}} \{(1-\eta_j(\mathbf{s}_i, u_i)) + \eta_j(\mathbf{s}_i, u_i) (1-\phi)^2 \}^{K-n_{ij}}\Big{\}}^{z_i} \Big{]},
	\label{intlik.lr.dey2} \\
	& f_\text{R}(\mathbf{Y^*}, \mathbf{u_{obs}} \, | \, \theta, p_0, \sigma_m, \sigma_f, \mathbf{u_0}, \mathbf{z}, \mathbf{S}, \mathbf{L}) 
	=\prod_{i=1}^{M}\, \Big{[} \big{\{} \theta^{u_i} (1-\theta)^{1-u_i} \, \prod_{j=1}^{J} \, p_{j}(\mathbf{s}_i, u_i) ^{y_{ij \cdot}} (1-p_{j}(\mathbf{s}_i, u_i))^{2K - y_{ij \cdot}} \big{\}}^{z_i} \Big{]},
	\label{intlik.lr.royle2}
	\end{align}where  $\eta_j(\mathbf{s}_i, u_i) = \w_0\, \exp(-d(\mathbf{s}_i, \mathbf{x}_j)^2/(2\sigma(u_i)^2))$ denotes the probability that an individual $i$ passes through a trap station $\mathbf{x}_j$ on some occasion $k$, $p_{j}(\mathbf{s}_i, u_i) = p_0\, \exp(-d(\mathbf{s}_i, \mathbf{x}_j)^2/(2\sigma(u_i)^2))$ denotes the probability that an individual $i$ is detected at $\mathbf{x}_j$ on occasion $k$. 
	
The prior distributions for the model parameters $\phi$, $\w_0$, $p_0$, $\psi$, $\theta$, $\sigma$, $\sigma_m$ are assumed to be independent and are provided in Table~\ref{table.prior}. 
To ensure that the marginal distribution of the data is well defined, we have assumed proper priors for each of these parameters \citep{gopalaswamy2016examining}. We assume a uniform prior over the entire state space $\V$ for each location of activity centre $\mathbf{s}_i$ and that these $\mathbf{s}_i$'s are independently distributed. $\mathbf{L}$ is assumed to have a uniform prior distribution over the permutation space of $\{1,\dots,M\}$.
The prior specifications remain the same for all the model fits. The MCMC algorithm used to sample from the respective posterior density under each model is detailed in Appendix~D. 

Thus, we have the four models, \\ 
$M_1$ : Model with density (\ref{intlik.lr.dey2}),
$M_2$ : Model with density (\ref{intlik.lr.royle2}),
$M_3$ : Model with density (\ref{intlik.lr.dey1}) and
$M_4$ : Model with density (\ref{intlik.lr.royle1}).

Among these models, $M_1$ can be regarded as the most complex, general, model and $M_4$ can be regarded as the simplest. Therefore, in our study, we simulate data from $M_1$ as the \emph{true model}.

\subsection{Candidate model selection tools}\label{msm.BMSE}
We have considered four different Bayesian model selection methods for application and evaluation: \textit{Bayes factors}, \textit{Deviance Information Criterion} (DIC), \textit{Watanabe-Akaike information criterion} (WAIC) and \textit{posterior predictive loss}. 
Two popular model selection tools (AIC and BIC; \cite{burnham2002model}) are not used here because because they impose restrictive assumptions on the parameter space as the sample size increases - situations often encountered in many hierarchical models \citep{royledorazio2008hierarchical}.
 For example, in the SECR models we study here, the concept of `number of parameters' is unclear and we therefore cannot apply criteria such as AIC and BIC directly.
 


\subsubsection{Bayes factors}\label{bfapp}

Model comparison using Bayes factors requires the computation of the marginal likelihood, which involves the integration $m(\mathbf{Y}\, | \, M_i) = \int f(\mathbf{Y} \, | \, \boldsymbol{\mu}, M_i) \, \pi(\boldsymbol{\mu}) \, d\boldsymbol{\mu}$ where $f(\mathbf{Y} \, | \, \boldsymbol{\mu}, M_i)$  denotes the model density and $\pi(\boldsymbol{\mu})$ denotes the prior density of the parameters $\boldsymbol{\mu}$ under $M_i$.
This integration is difficult to compute in practice unless the models are very simple in structure, which is often not the case in ecology. Therefore, computation of the marginal likelihood of data using MCMC methods remains to be a holy grail and is an active area of research \citep{wang2016warp, wang2018new}.



\subsubsection*{Estimation of marginal likelihood of data}\label{computemarglik}

Under our model settings, $\mathbf{Y} = (\mathbf{Y^*}, \mathbf{u_{obs}})$ for models $M_1$, $M_2$ and $\mathbf{Y} = \mathbf{Y^*}$ for the other models $M_3$, $M_4$. $\boldsymbol{\mu}$ denotes the collection of all parameters and latent variables for each model as a generic notation. Specifically, let $\boldsymbol{\mu} = (\boldsymbol{\mu_p}, \boldsymbol{\mu_s})$, where $\boldsymbol{\mu_p}$ is the collection of scalar parameters and $\boldsymbol{\mu_s}$ is the collection of all latent variables.
The Gelfand-Dey estimator of marginal likelihood of data $m(\mathbf{Y})$ is expressed as:
\begin{align}\label{GDest1}
\hat{m}_{\text{GD}}(\mathbf{Y}) = \bigg{[}\, \frac{1}{N_{iter}} \sum_{d=1}^{N_{iter}} \frac{g(\boldsymbol{\mu}^{(d)})}{f(\mathbf{Y} \, | \, \boldsymbol{\mu}^{(d)})\, \pi(\boldsymbol{\mu}^{(d)})} \, \bigg{]}^{-1}, 
\end{align}
where $\{\boldsymbol{\mu}^{(d)} : d = 1,\dots,N_{iter}\}$ is a set of MCMC draws from the posterior $\pi(\boldsymbol{\mu} \, | \, \mathbf{Y})$ and $g(\boldsymbol{\mu})$ is a tuning density. 
It is to be noted that by specifying $g(\boldsymbol{\mu}) = \pi(\boldsymbol{\mu})$ in (\ref{GDest1}), we obtain the harmonic mean estimator of the marginal likelihood 
\begin{align}\label{HMest1}
\hat{m}_{\text{HM}}(\mathbf{Y}) = \bigg{[}\, \frac{1}{N_{iter}} \sum_{d=1}^{N_{iter}} \frac{1}{f(\mathbf{Y} \, | \, \boldsymbol{\mu}^{(d)})} \, \bigg{]}^{-1}. 
\end{align}
Details on these estimators and their properties can be found in \cite{gelfand1994bayesian}, \cite{kass1995bayes}. 
For our problem at hand, the computation of (\ref{GDest1}) requires us to obtain the integrated likelihoods (marginals) under the different models that we consider. This becomes particularly tricky in the presence of high-dimensional latent variables such as $\mathbf{u_0}$, $\mathbf{z}$, $\mathbf{S}$, $\mathbf{L}$ which are elements of $\boldsymbol{\mu_s}$.


We have developed two \textit{approximate} approaches to compute the Gelfand-Dey estimator: the maximum a posteriori (MAP) approximation approach and the integrated likelihood (IL) approach. 

\noindent\textbf{Approach 1: MAP approximation}

\noindent In this approach, we fix the high-dimensional variables at their MAP estimates $\boldsymbol{\hat{\mu}_s}$, assuming that their posterior distributions are well summarised by these estimates which are derived from the MCMC draws. 
The Gelfand-Dey estimator is then computed using the formula,
\begin{align}\label{GDest2}
	\hat{m}_{\text{GD}}(\mathbf{Y}) = \bigg{[}\, \frac{1}{N_{iter}} \sum_{d=1}^{N_{iter}} \frac{g(\boldsymbol{\mu_p}^{(d)})}{f(\mathbf{Y} \, | \, \boldsymbol{\mu_p}^{(d)}, \boldsymbol{\hat{\mu}_s})\, \pi(\boldsymbol{\mu_p}^{(d)})} \, \bigg{]}^{-1}. 
	\end{align}
where $\{(\boldsymbol{\mu_p}^{(d)},\boldsymbol{\mu_s}^{(d)}) \, : \, d = 1,\dots,N_{iter}\}$ is a set of MCMC draws from the posterior $\pi(\boldsymbol{\mu_p},\boldsymbol{\mu_s} \, | \, \mathbf{Y} )$.
We begin with $(\boldsymbol{\mu_p}^{(d_0)},\boldsymbol{\mu_s}^{(d_0)})$ as an initial estimate of $(\boldsymbol{\mu_p}, \boldsymbol{\mu_s})$ where
$$f(\mathbf{Y}\, | \, \boldsymbol{\mu_p}^{(d_0)},\boldsymbol{\mu_s}^{(d_0)}) \, \pi(\boldsymbol{\mu_p}^{(d_0)},\boldsymbol{\mu_s}^{(d_0)}) =\displaystyle{\max_d} \{f(\mathbf{Y}\, | \, \boldsymbol{\mu_p}^{(d)},\boldsymbol{\mu_s}^{(d)}) \, \pi(\boldsymbol{\mu_p}^{(d)},\boldsymbol{\mu_s}^{(d)})\}.$$
This estimate of posterior mode of $(\boldsymbol{\mu_p}, \boldsymbol{\mu_s})$ may not be optimal since in our high dimensional parameter setting, an MCMC sample of a practical size may not be enough to extensively explore the posterior surface. We, therefore, fix one of the parameters $\boldsymbol{\mu_s} =  \boldsymbol{\mu_s}^{(d_0)}$ and explore the posterior surface to find $d_1$ such that 
$f( \mathbf{Y} \, | \, \boldsymbol{\mu_p}^{(d_1)},\boldsymbol{\mu_s}^{(d_0)}) \, \pi(\boldsymbol{\mu_p}^{(d_1)},\boldsymbol{\mu_s}^{(d_0)}) =$ $\displaystyle{\max_d} \{f( \mathbf{Y} \, | \, \boldsymbol{\mu_p}^{(d)},\boldsymbol{\mu_s}^{(d_0)}) \, \pi(\boldsymbol{\mu_p}^{(d)},\boldsymbol{\mu_s}^{(d_0)})\}.$
In this way we obtain an improved MAP estimate of $(\boldsymbol{\mu_p}, \boldsymbol{\mu_s})$,  $(\boldsymbol{\mu_p}^{(d_1)},\boldsymbol{\mu_s}^{(d_0)})$, if
$$f( \mathbf{Y} \, | \, \boldsymbol{\mu_p}^{(d_1)},\boldsymbol{\mu_s}^{(d_0)}) \, \pi(\boldsymbol{\mu_p}^{(d_1)},\boldsymbol{\mu_s}^{(d_0)}) > f( \mathbf{Y} \, | \, \boldsymbol{\mu_p}^{(d_0)},\boldsymbol{\mu_s}^{(d_0)}) \, \pi(\boldsymbol{\mu_p}^{(d_0)},\boldsymbol{\mu_s}^{(d_0)}).$$
Similarly, we then fix $\boldsymbol{\mu_p} =  \boldsymbol{\mu_p}^{(d_1)}$ and find $\boldsymbol{\mu_s}^{(d_2)}$. This procedure is continued iteratively to eventually give us the best MAP estimate of the posterior mode ($\boldsymbol{\hat{\mu}_p},\boldsymbol{\hat{\mu}_s}$). 
Suitable transformations of the parameters ensure that all the points in $\{(\boldsymbol{\mu_p}^{(a)}$, $\boldsymbol{\mu_s}^{(b)}) \, : \, a,b = 1,\dots, N_{iter}; a \neq b\}$ belong to the posterior support. The above mentioned resampling procedure is detailed in Appendix~A.


\noindent\textbf{Approach 2: Integrated likelihood (IL) approximation}

\noindent Ideally, we would like to compute the marginal likelihood $m(\mathbf{Y})$ by integrating out all the latent variables with respect to their corresponding prior distributions from the model likelihoods. However, in the case of model likelihoods (\ref{intlik.lr.dey1}-\ref{intlik.lr.royle2}) this integration is not possible for the permutation vector $\mathbf{L}$.
The integration over the variables $\mathbf{u_0}$ and $\mathbf{z}$ can be performed analytically. The integration over $\mathbf{S}$ is evaluated numerically by partitioning the region $\V$ into a sufficiently fine grid and then evaluating a Riemann sum (as direct integration can not be expressed in a closed form). This integrated likelihood can then be used in (\ref{GDest1}) for estimating the marginal likelihood
\begin{align}\label{GDIL1}
\hat{m}_{\text{GD}}(\mathbf{Y}) = \bigg{[}\, \frac{1}{N_{iter}} \sum_{d=1}^{N_{iter}} \frac{g(\boldsymbol{\mu_p}^{(d)}, \mathbf{L}^{(d)})}{f(\mathbf{Y} \, | \, \boldsymbol{\mu_p}^{(d)}, \mathbf{L}^{(d)})\, \pi(\boldsymbol{\mu_p}^{(d)}, \mathbf{L}^{(d)})} \, \bigg{]}^{-1}.
\end{align}
One downside of IL approximation approach is the lack of clarity about the interdependencies between the latent variables $\mathbf{u_0}$, $\mathbf{z}$, $\mathbf{S}$ and $\mathbf{L}$ after carrying out the integrations. The derivations of the integrated likelihoods for each of the four models $M_1 - M_4$ are given in Appendix~B by ignoring any possible interdependencies.

We have also assessed the robustness of the Gelfand-Dey estimator by computing the marginal likelihood estimates using different tuning densities, e.g., multivariate normal density, multivariate-t density with varying degrees of freedom and the truncated normal density following the suggestion of \cite{geweke1999using}. These technical details are described in Appendix~C. 

\subsubsection{Deviance information criterion (DIC)} \label{dicapp}

Deviance is defined as $D(\boldsymbol{\mu}) = -2\log f(\mathbf{Y} \, | \, \boldsymbol{\mu})$.
Deviance information criterion (DIC) is then defined as $\text{DIC} = D(\hat{\boldsymbol{\mu}}) + 2 \, p_{\dic}$
where $\hat{\boldsymbol{\mu}}$ is an estimate of $\boldsymbol{\mu}$. A model with smaller DIC value is preferred. 
Model comparison using DIC is not invariant to parameterisation and depends on the components of the model density to be considered as the likelihood. \cite{spiegelhalter2002DIC} suggests practitioners to carefully decide on the parameters of interest so that they can avoid this potential pitfall. This piece of advice is often not straightforward to implement in practice, especially when there exists inherent ambiguity in the interpretation of latent parameters. However \cite{celeux2006deviance} suggests several forms for DIC that can be used for different hierarchical models but does not recommend any particular form as the best. 

For the computation of deviance, we have used the MAP estimate of $\boldsymbol{\mu}$ to obtain $\hat{\boldsymbol{\mu}}$ instead of the posterior mean, due to the presence of binary latent variables and unknown permutation vectors in the likelihood. We have then computed two versions of $p_{\dic}$ \citep{hooten2015guide, gelman2014bayesian} using MCMC draws $\{\boldsymbol{\mu}^{(d)} : d = 1,\dots,N_{iter}\}$ from $\pi(\mu \, | \,\mathbf{Y})$ as follows:
\begin{align}\label{diceq}
	&\hat{p}_{\dic1} = 2\log f(\mathbf{Y} \, | \, \hat{\boldsymbol{\mu}}) -  \frac{2}{N_{iter}} \sum_{d=1}^{N_{iter}}  \log f(\mathbf{Y} \, | \, \boldsymbol{\mu}^{(d)}),\nonumber \\
	&\hat{p}_{\dic2} = 2 \Big{[}\, \frac{1}{N_{iter}} \sum_{d=1}^{N_{iter}} \Big{(}\log f(\mathbf{Y} \, | \, \boldsymbol{\mu}^{(d)}) - \frac{1}{N_{iter}} \sum_{d=1}^{N_{iter}}  \log f(\mathbf{Y} \, | \, \boldsymbol{\mu}^{(d)})\Big{)}^2\Big{]}.
	\end{align}


\subsubsection{Watanabe-Akaike information criterion (WAIC)} \label{waicapp}

The Watanabe-Akaike information criterion (WAIC) is a Bayesian version of AIC as it uses the posterior predictive distribution of the data to estimate the out-of-sample predictive accuracy of the model. \cite{watanabe2010asymptotic} introduced the WAIC criterion based on the assumption of independence between data points and has shown its asymptotic equivalence with cross-validation. In our model formulations, we have assumed that the different data points correspond to capture-recapture data set for each of the $M$ individuals. WAIC is then defined as $\text{WAIC} = -2\sum_{i=1}^{M} \log  \E_{\boldsymbol{\mu}\, | \,\mathbf{Y} } \big{(}f(\mathbf{Y}_i \, | \, \boldsymbol{\mu})\big{)} + 2 \, p_{\waic}.$ A model with smaller WAIC value is preferred. In computing WAIC we partition data $Y$ in terms of individuals $(Y_1,Y_2,\dots,Y_M )$. We compute the two commonly used versions of $p_{\waic}$ \citep{hooten2015guide} using MCMC draws $\{\boldsymbol{\mu}^{(d)} : d = 1,\dots,N_{iter}\}$ from $\pi(\mu \, | \,\mathbf{Y})$ as follows:
{\small \begin{align}\label{waiceq}
	&\hat{p}_{\waic1} = 2\sum_{i=1}^{M} \Big{\{}\log \Big{(}\frac{1}{N_{iter}} \sum_{d=1}^{N_{iter}} f (\mathbf{Y}_i \, | \, \boldsymbol{\mu}^{(d)})\Big{)} -  \frac{1}{N_{iter}} \sum_{d=1}^{N_{iter}} \log f(\mathbf{Y}_i \, | \, \boldsymbol{\mu}^{(d)})\Big{\}},\nonumber\\
	&\hat{p}_{\waic2} = \sum_{i=1}^{M} \Big{\{}\frac{1}{N_{iter}} \sum_{d=1}^{N_{iter}}\Big{(}\log f(\mathbf{Y}_i \, | \, \boldsymbol{\mu}^{(d)}) - \frac{1}{N_{iter}} \sum_{d=1}^{N_{iter}} \log f(\mathbf{Y}_i \, | \, \boldsymbol{\mu}^{(d)}) \Big{)}^2\Big{\}}. 
	\end{align}}
We propose another version for $p_{\waic}$ based on absolute error loss:
{\small \begin{align}\label{waiceq2}
 \hat{p}_{\waic3} = 2\sum_{i=1}^{M}\Big{\{}\frac{1}{N_{iter}} \sum_{d=1}^{N_{iter}}\Big{|}\log f(\mathbf{Y}_i \, | \, \boldsymbol{\mu}^{(d)}) - \frac{1}{N_{iter}} \sum_{d=1}^{N_{iter}} \log f(\mathbf{Y}_i \, | \, \boldsymbol{\mu}^{(d)}) \Big{|}\Big{\}}.
\end{align}}

\subsubsection{Posterior predictive loss} \label{pplapp}

\cite{gelfand1998model} derived a model selection criterion, popularly known as the posterior predictive loss criterion, by adopting a decision theoretic approach for measuring predictive accuracy of a model.  The posterior predictive loss $D_\infty$ criterion (based on a square error loss function) under our model setting is defined as follows:
\begin{align}\label{pplinf}
D_{\infty} = \sum_{i=1}^{2MJK} (y_{i, \text{vec}} - \E(y_{i,\text{rep}}\, | \, \mathbf{Y_{vec}}))^2 + \sum_{i=1}^{2MJK}  \Var(y_{i,\text{rep}}\, | \, \mathbf{Y_{vec}}),
\end{align}
where $\mathbf{Y_{vec}} = (y_{1, \text{vec}}, y_{2, \text{vec}}, \dots, y_{2MJK, \text{vec}})'$ is a vector of length $2MJK$ obtained by vectorizing observed data array $\mathbf{Y}$ and $\mathbf{Y_{rep}} = (y_{1, \text{rep}}, y_{2, \text{rep}}, \dots, y_{2MJK, \text{rep}})'$ denotes replicate of the observed data. The first term in the $D_\infty$ criterion  (see (\ref{pplinf})) is the goodness-of-fit term while the second term can be interpreted as a penalty term for model complexity. The model with the smallest $D_\infty$ is to be preferred. In our analysis, the data set is obtained by vectorising the two binary data arrays and placing one after the other.

We compute the above expectation $\E(y_{i,\text{rep}}\, | \, \mathbf{Y_{\text{vec}}})$ and variance $\Var(y_{i,\text{rep}}\, | \, \mathbf{Y_{\text{vec}}})$ using MCMC draws. Given an MCMC sample  $\{\boldsymbol{\mu}^{(d)} : d = 1,\dots,N_{iter}\}$ from $\pi(\mu \, | \,\mathbf{Y})$, we simulate $\mathbf{Y}_{\mathbf{rep}}^{(d)}$ from $f(\mathbf{Y} \, | \, \boldsymbol{\mu}^{(d)})$ for each $d = 1,\dots,N_{iter}$.
For instance, in model $M_1$, $\mu$ denotes the collection of the parameters $\psi$, $\theta$, $\phi$, $\w_0$, $\sigma_m$, $\sigma_f$, $\mathbf{u_0}$, $\mathbf{z}$, $\mathbf{S}$, $\mathbf{L}$. 
Then $\E(y_{i,\text{rep}}\, | \, \mathbf{Y_{\text{vec}}}) \approx N_{iter}^{-1} \sum_{d=1}^{N_{iter}} y^{(d)}_{i,\text{rep}}$ and
$\Var(y_{i,\text{rep}}\, | \, \mathbf{Y_{\text{vec}}}) \approx N_{iter}^{-1} \sum_{i=1}^{N_{iter}} \Big{(}y^{(d)}_{i,\text{rep}} - N_{iter}^{-1} \sum_{d=1}^{N_{iter}} y^{(d)}_{i,\text{rep}} \Big{)}^2$.

We summarise the various model selection methods and their variants in Table~\ref{msc}. Considering all these model selection tools and their variants (from approximation approaches to setting tuning densities), our evaluation is carried out on 25 unique tools.

\section{Evaluation of the Performance of Model Selection Methods}\label{mstperformance}
\subsection{Simulation design and simulation scenarios}\label{simdesign.BMSE}

We have conducted simulations for 12 scenarios (provided in Table~ \ref{t.simscenarios.BMSE}) grouped into 2 equal sized sets, to assess the performance of the models proposed here. We set $\sigma_m$ = 0.3 and $\sigma_f$ = 0.15 for the first set of 6 scenarios, $\sigma_m$ = 0.4 and $\sigma_f$ = 0.2 for the second set of 6 scenarios. We set $(\w_0, \phi)$  = \{(0.01, 0.3), (0.05, 0.3), (0.05, 0.5), (0.03, 0.8), (0.01, 0.9), (0.05, 0.9)\}, which gives us 6 different scenarios for each of the two sets corresponding to the values taken by $\w_0$ and $\phi$. We assume that a total of 100 individuals are residing inside the state space of which 40 are male. Each of the simulation experiments is conducted within a rectangular state space of dimension 5 unit $\times$ 7 unit (Figure~\ref{statespace.BMSE}), after setting a buffer of 1 unit in both horizontal and vertical directions, a $10 \times 16$ trapping array of total $J=160$ trap stations has been set (trap spacing is 0.3 unit on $X$ axis and 0.3125 unit on $Y$ axis). This meets the requirement suggested in \cite{karanth2017methods}.  Each of the traps remains active for $K=50$ sampling occasions simultaneously. For parameter estimation, we set the maximum possible number of individuals present in the population ($M$) at 400 for all the scenarios. The experiment is repeated $n_{sim} = 10$ times. The MCMC samples for each of the parameters are obtained (each of length 30000) and the estimates are computed using those chains with a burn-in of 10000.

Capture-recapture data sets are simulated independently under each of the 12 simulation scenarios (Table~\ref{t.simscenarios.BMSE}) under model $M_1$. 
Recall that, model $M_1$ corresponds to the statistical model in (\ref{intlik.lr.dey2}) with $\sigma$ parameter modelled in terms of individual covariate on sex category $\mathbf{u}$ (see Section~\ref{candmodelset.BMSE}). Then each simulated data set is fitted with all the four competing model $M_1$, $M_2$, $M_3$ and $M_4$. 
\subsection{Defining performance measures}

\paragraph{Probability of selecting the true model}
In our study, since all the data sets are simulated from model $M_1$, it is considered as the true model. We have computed the proportion of times a model selection method chooses $M_1$ as the best. This proportion will serve as an estimate for the probability of selecting the true model. Additionally we have also computed the proportion of times a model selection method chooses $M_2$, $M_3$ and $M_4$ respectively. 
The computed proportions of selecting these models from the simulation study will indicate the efficacy of the model selection methods.

\paragraph{Assessing the quality of parameter estimation.}
The abundance parameter $N$ carries a lot of significance in ecology and conservation. Due to its importance, ecologists place their interests in the robustness and accuracy of its estimate, and will therefore rely on a model selection method that will achieve this. The precision and accuracy of the parameter estimates indicates the quality of the model fit and we assess this by computing the \textit{average root mean square error} (average RMSE). 

Suppose $\{\mu^{(td)} \, : \, d = 1, \dots, N_{iter}\}$ denotes a set of MCMC draws from the posterior distribution of an arbitrary parameter $\mu$ for the $t$-th simulated data set, $t = 1,\dots,n_{sim}$. \textit{Mean square error} (MSE) of $\mu$ for $t$-th simulated data set is estimated as $MSE(\mu,\, t) = N_{iter}^{-1} \sum_{d=1}^{N_{iter}} (\mu^{(td)} - \mu)^2.$
Average RMSE is calculated by first averaging the estimated MSEs of different simulations and then taking the square root of the average: $Average \, RMSE(\mu) = \sqrt{n_{sim}^{-1} \sum_{t=1}^{n_{sim}} MSE(\mu,\, t)}.$

The quantities related to the various Bayesian model selection methods (Section~\ref{msm.BMSE}) are computed for each of the four models for the simulated data and analysis sets. First, we analyse the quality of fit of the competing models to the simulated data sets. We compute the proportion of times they favour any of these four models out of the $n_{sim}$ simulations and study the consistency of these different model selection methods. These proportions are then compared with the computed average RMSEs of the parameters to study the behaviour of the model selection methods with respect to varying information content. We also generate pairwise correlation plots from the MCMC draws to study the extent of identifiability issues between various pairs of parameters as a consequence of lack of information content in the data. 

\section{Results of the Simulation Study}\label{simresults.BMSE}


Our results suggest that the choice of the tuning density had no impact on model selection (Figures~\ref{fig.modelsel1.BMSE} and \ref{fig.modelsel2.BMSE}). Thus, we focus only on plot (a) in Figure~\ref{fig.modelsel1.BMSE} for our inferences on model selection by Bayes factor (GD-MAP approximation). Here, we observe that Bayes factor GD-MAP approximation favours $M_4$ in $70\%$ of the times under scenario 1 and favours $M_3$ in at least $70\%$ of the times under scenarios 2, 3, 4 and 7. Bayes factor is in favour of $M_1$ more than $80\%$ of the times under all the remaining scenarios 5-6 and 8-12. 

The plots corresponding to the Bayes factors (GD-IL approximation) are shown in Figures~\ref{fig.modelsel3.BMSE} and \ref{fig.modelsel4.BMSE}. As seen here, GD-IL approximation performs slightly worse than the GD-MAP approximation. However,  we observe that the model choices by the harmonic mean estimator of Bayes Factor performs well and favours the true model $M_1$ in majority of the scenarios (Figure~\ref{fig.modelsel_hm.BMSE}). 

We have considered three forms of WAIC and two forms of DIC. The corresponding plots are shown in Figure~\ref{fig.modelsel5.BMSE} (a)-(e). WAIC1, WAIC3, DIC1 and DIC2 exhibit very similar tendencies in their model choices by favouring $M_3$ under all the scenarios except scenario 1 and 3. These four methods favour $M_4$ under scenario 1. Under scenario 3, WAIC1 and DIC1 favour $M_3$ in majority of the times whereas WAIC3 and DIC2 favour $M_4$. Like Bayes factor, these model selection methods (DICs and WAICs) also tend to favour simpler models under scenarios 1 and 3. WAIC2 largely agrees with the other WAICs (and DICs) but more often selects the true model when data sets are more informative (Figure~\ref{fig.modelsel5.BMSE} (a)-(c)). In general, DICs and WAICs seem to discourage the presence of high dimensional latent variables. 

The plot showing the proportion of different model choices by posterior predictive loss is given in Figure~\ref{fig.modelsel5.BMSE}(f). The posterior predictive loss criterion $D_\infty$ favours models with individual sex-covariates ($M_1$ and $M_2$). Posterior predictive loss also appears to select the true model some of the times, even when there isn't sufficient information in the data (scenarios 1 and 3). 
Figures~\ref{fig.rmse1.BMSE} and \ref{fig.rmse2.BMSE} show the plots for the average root mean square error (average RMSE) for the parameters $N$, $\psi$, $N_{Male}$, $\theta$, $\phi$, $\w_0$, $p_0$, $\sigma_m$, $\sigma_f$ and $\sigma$. We focus our discussion based on the efficiency and accuracy in estimating $N$. In Figure~\ref{fig.rmse1.BMSE}(a) we see that the average RMSE of $N$ is substantially higher in scenarios 1 - 3 relative to scenarios 4 - 12.  Figures~\ref{fig.scatplotsc1m1.BMSE} - \ref{fig.scatplotsc12m1.BMSE} are a representative set of scatter plots (scenarios 1, 2, 3, 7, 9, 12) meant to highlight the range of the correlation coefficient as the information content in the data sets increases.
 We observe that, the scatter plots of ($N$, $\theta$) and ($N$, $\sigma_f$) show substantial correlation under scenarios 1 - 3 (see Figures~~\ref{fig.scatplotsc1m1.BMSE} - \ref{fig.scatplotsc3m1.BMSE}). This high correlation may indicate that these parameter estimates are of poor accuracy due to inadequate sample size. However these correlations decrease for the other scenarios (e.g., 9 and 12, see Figures~\ref{fig.scatplotsc9m1.BMSE}, \ref{fig.scatplotsc12m1.BMSE}), likely due to increased information content in the data. 

\section{Conclusions and Discussion}\label{conclusions.BMSE}

Contemporary practice of ecology and conservation biology relies largely on the use of model selection for hypotheses discrimination \citep{ellison2004bayesian, diaz2005statistical}. Simultaneously, there has been major growth in the use of hierarchical models in ecology, especially within the realm of Bayesian inference \citep{royledorazio2008hierarchical, kery2010introduction}. These models have now enabled statistical ecologists to fairly easily formulate complex ecological models and elegantly deal with the sampling process and also fit these complex models using powerful tools such as MCMC \citep[e.g.,][]{kery2010introduction}. However, the lack of availability of readymade model selection tools when practicing Bayesian inference has sometimes motivated ecologists to continue using likelihood-based inferences, merely because one can use well known model selection tools such as the AIC \citep{burnham2002model} for inference. 

To provide a context for this argument, in the spatial capture-recapture literature, we have essentially seen the development of three important likelihood functions: (1) \cite{borchers2008spatially}, (2) \cite{royle2009bayesian} and (3) \cite{royle2015likelihood}. Inferences, for the models (1) and (3) are by maximizing the likelihood, while the inference for (2) is Bayesian. We note with interest, that one of the reasons for the development of (3), was motivated on the pretext that model selection is much easier (using known tools such as the AIC) for practitioners using the likelihood approach, in spite of the problem having been already solved in the Bayesian context \citep{royle2013spatial}. It is specially of concern in the context of the models we study here in that investigators may be forced to integrate out $\mathbf{s}_i$'s (activity centres of individuals) in order to construct tractable likelihoods and thus oversimplifying ecological reality.

In this study, we have tried to implement some selected Bayesian model selection methods on a specific class of advanced Bayesian SECR models (\citealp{royle2015spatial} - with and without sex covariates; \citealp{dey2017spatial} - with and without sex covariates) dealing with partially identified individuals. We have found our Bayes factor implementation using the Gelfand-Dey estimator (using the MAP approximation approach) to be the preferred choice as a model selection method over a wide range of simulation scenarios. This approach appears to work particularly well when information content in the data is moderate to high. The IL approximation approach also worked well, but not as well as the MAP approximation approach perhaps because, there exists interdependency between some of the latent variables.

However, implementing Bayes factors for model selection using the Gelfand-Dey estimator (with MAP approximation) can be computationally intensive and complicated while setting up. Interestingly, our study demonstrates that obtaining Bayes factors using harmonic mean approach for marginal likelihood computation is less demanding but yet serves as a very good model selection method. This finding deviates from the popular view among applied scientists that it is futile to estimate the marginal likelihood using the harmomic mean approach \citep{lartillot2006computing, xie2011marginal}. We surmise that this finding may be attributed to the fact that when we bind the priors (as we have done, but for other reasons) and not permit extremely low probabilities to occur at the tails, many of the criticisms \citep{lartillot2006computing, xie2011marginal} may become irrelevant in practice, but this would require further enquiry.  




As our simulation study shows, the two goals of model selection and parameter estimation cannot be simultaneously achieved under certain circumstances (scenarios 1 - 3), especially when information content is low (indicated by high RMSE values and correlation coefficients).
Hence, researchers have to clearly prioritize their objectives prior to data analysis. If the goal, for example, is to find a model that best estimates population size $N$, then we recommend the use of Bayes factor (Gelfand-Dey with MAP approximation) or the Bayes factor (harmonic mean, due to its simplicity) because these appear to provide the most reliable estimates of $N$ over all the simulation scenarios. However, if researchers are only interested to select the true model, especially when data are less informative (scenarios 1 - 3), we recommend the posterior predictive loss approach since they favour the true model nearly a 1/3rd of the times in situations with such low information content. Of course, the dual objectives of model selection and parameter estimation are met when information content is moderate or high (scenarios 4 - 12) and as stated previously, we recommend either Bayes factor (Gelfand-Dey with MAP approximation) or Bayes factor (harmonic mean) in such cases. However, the posterior predictive loss (with the squared error loss function) as used here, does not select the true model when information content is moderate to high.    

We also do not recommend the use of DICs or WAICs, since they do not appear to outcompete other model selection tools (either from the standpoint of model selection or parameter estimation) in any of the simulation scenarios. This is an interesting finding, because tools such as WAIC are much newer tools developed by applied Bayesians to provide answers to a wide range of contemporary model selection problems involving hierarchical models \citep{hooten2015guide, gelman2014bayesian}. Thus, our study brings back focus on the need to assess the strength of inference from a model selection method by \textit{conditioning} on a true model and consequently evaluating a \textit{competing} set of model selection methods prior to selecting the most appropriate one for the problem on hand.  

Our approach does not, strictly speaking, permit us to draw conclusions and make inferences on the most suitable model selection tools beyond the restrictive set of competing models and the settings we have used in this study. However, at the risk of making a claim beyond our case study, we would recommend the use of the Bayes Factor (Gelfand-Dey with MAP approximation or the harmonic mean estimator) for most Bayesian SECR models and perhaps to  much larger class of hierarchical models in ecology \citep{royle2013spatial}. We note with interest that it is also unclear whether the routinely used AIC works as an appropriate model selection tool for MLE-based SECR models as discussed in \cite{efford2014compensatory}.

\section{Acknowledgements}
We thank the Indian Statistical Institute for financial and administrative support. AMG thanks the Wildlife Conservation Society, New York for partial funding support.
\section{Authors contribution}
SD, MD and AMG conceived the ideas and designed methodology; SD, MD and AMG analysed the simulated data; SD, MD and AMG did the writing of the manuscript. All
authors contributed critically to the drafts and gave final approval for publication.



\pagebreak
\begin{table}[!htb]
	\centering 
	\caption{An example of detection histories for two fully identified individuals and partially identified individuals is presented. The circled 1's indicate the simultaneous captures of an individual by the detectors 1 and 2.}
	{\footnotesize \begin{tabular}{l @{\extracolsep{15pt}} c @{\extracolsep{15pt}} c @{\extracolsep{15pt}} c @{\extracolsep{15pt}} c @{\extracolsep{15pt}} c @{\extracolsep{30pt}} c @{\extracolsep{15pt}} c @{\extracolsep{15pt}} c @{\extracolsep{15pt}} c @{\extracolsep{15pt}} c} 
			\\[-1.8ex]\hline 
			\hline \\[-1.8ex] 
			& & \multicolumn{4}{c}{Detector 1} & & \multicolumn{4}{c}{Detector 2}\\
			& Occasion & 1 & 2 & 3 & 4 & Occasion & 1 & 2 & 3 & 4 \\
			& Trap & & & & & Trap & & & &   \\ 
			\hline \\[-1.8ex] 
			\multirow{3}{*}{Fully-identified individual 1} 
			& 1 & 0 & 1 & 0 & 1 & 1 & 0 & 0 & 1 & 0  \\
			& 2 & 1 & 0 & 0 & \raisebox{0.5pt}{\textcircled{\raisebox{-0.9pt} {1}}} & 2 & 0 & 0 & 0 & \raisebox{0.5pt}{\textcircled{\raisebox{-0.9pt} {1}}}  \\ 
			& 3 & 0 & 0 & 1 & 1 & 3 & 1 & 0 & 0 & 0  \\  
			\hline \\[-1.8ex] 
			\multirow{3}{*}{Fully-identified individual 2} 
			& 1 & \raisebox{0.5pt}{\textcircled{\raisebox{-0.9pt} {1}}} & 0 & 0 & 0 & 1 &  \raisebox{0.5pt}{\textcircled{\raisebox{-0.9pt} {1}}} & 1 & 0 & 0 \\ 
			& 2 & 0 & 0 & 0 & 1 & 2 & 0 & 0 & 0 & 0  \\
			& 3 & 1 & 1 & 0 & 0 & 3 & 0 & 0 & 1 & 0  \\
			\hline \\[-1.8ex]
			\multirow{3}{*}{Partially-identified individual} 
			& 1 & 1 & 0 & 0 & 1 & 1 & - & - & - & -  \\ 
			& 2 & 0 & 0 & 1 & 0 & 2 & - & - & - & -  \\
			& 3 & 0 & 0 & 0 & 0 & 3 & - & - & - & - \\  
			\hline \\[-1.8ex] 
			\multirow{3}{*}{Partially-identified individual} 
			& 1 &- & - & - & - & 1 & 0 & 0 & 1 & 0  \\ 
			& 2 & - & - & - & - & 2 & 1 & 0 & 0 & 0  \\
			& 3 & - & - & - & - & 3 & 0 & 0 & 1 & 0  \\  
			\hline \\[-1.8ex] 
	\end{tabular}}
	\label{sampledata} 
\end{table}
\pagebreak

\begin{table}[H] 
	\centering
	\caption{Notations of variables and parameters used in this article. Bold symbols represent collections (vectors).}
	\footnotesize{
		\begin{tabular}{ l @{\extracolsep{30pt}} l } 
			\\[-1.8ex]\hline 
			\hline \\[-1.8ex]
			\textbf{Variables and parameters} & \textbf{Definition}\\ 
			\hline \\[-1.8ex]
			$\mathcal{V}$ & A bounded geographic region of scientific or operational relevance\\
			& where a population of individuals of certain species reside. \\ [+1.5ex]
			$N \sim \mathrm{Binomial}(M,\psi)$ & Population size of the superpopulation, i.e., the number of\\  &  individuals  within $\mathcal{V}$.  \\ [+1.5ex]
			$M$ & Maximum number of individuals within the state space $\mathcal{V}$. \\ & This is a fixed quantity defined by the investigator. \\ [+1.5ex]
			$\psi$ & Proportion of individuals that are real and present within $\mathcal{V}$.\\ [+1.5ex]
			$\theta$ & Probability that an individual is male.\\  [+1.5ex]
			$J$ & Number of trap stations in $\mathcal{V}$.\\ [+1.5ex]
			$K$ & Number of sampling occasions.\\ [+1.5ex]
			$R$ & Maximum permissible value of movement range for each\\
			&  individual during the survey.\\[+1.5ex]
			$\w_0$ & Baseline trap entry probability in the models $M_1$ and $M_2$, \\
			& i.e., probability  that an individual passes through a trap station \\
			& assuming its centre of activity is also located at that trap station.\\ [+1.5ex]
			$p_0$ & Baseline detection probability in the models $M_2$ and $M_4$,  \\
			&i.e., probability  that an individual is detected by a detector \\
			&assuming   its centre of activity is also located at that trap station.\\ [+1.5ex]
			$\sigma$ & $\sigma$ measures the spatial extent of movement around individual \\  
			& activity centre. $\sigma = \sigma_m$ for male individuals,  $\sigma = \sigma_f$ for\\
			& female individuals.\\ [+1.5ex]
			$d_{ij} = d(\mathbf{s}_i, \mathbf{x}_j) =  \Arrowvert \mathbf{s}_i - \mathbf{x}_j \Arrowvert$ & Euclidean distance between points $\mathbf{s}_i$ and $\mathbf{x}_j$.\\ [+1.5ex]
			$\eta_j(\mathbf{s}_i, u_i) = \w_0\, \exp(-\frac{d(\mathbf{s}_i, \mathbf{x}_j)^2}{2\sigma(u_i)^2})$ & Probability that an individual $i$ passes through a trap station \\
			& $\mathbf{x}_j$ on some occasion $k$ and $\sigma$ is modelled as a function \\
			& of individual covariate on sex category $u_i$.  \\ [+1.5ex]
			$\eta_j(\mathbf{s}_i) = \w_0\, \exp(-\frac{d(\mathbf{s}_i, \mathbf{x}_j)^2}{2\sigma^2})$ & Probability that an individual $i$ passes through a trap station \\
			&$\mathbf{x}_j$ on  some occasion $k$.  \\ [+1.5ex]
			$\phi$ & Probability that an individual $i$ is detected by a detector on \\
			& some occasion $k$ given that it is present at that trap.\\ [+1.5ex]
				\\[-1.8ex]\hline 
						\hline \\[-1.8ex]
						\textbf{Notations pertaining to} & \textbf{Definition}\\ 
							\textbf{model selection tools} & \\ 
						\hline \\[-1.8ex]
$\hat{m}_{\text{GD}}(\mathbf{Y})$ & Gelfand-Dey estimator of the marginal likelihood of data $m(\mathbf{Y})$.\\ [+1.5ex]
$\hat{m}_{\text{HM}}(\mathbf{Y})$ & Harmonic mean estimator of the marginal likelihood of data $m(\mathbf{Y})$.\\ [+1.5ex]
$p_{\dic}$ & Correction term for bias due to overfitting in DIC criterion. \\ [+1.5ex]
$p_{\waic}$ & Correction term for bias due to overfitting in WAIC criterion. \\ [+1.5ex]
$D_\infty$ & Posterior predictive loss criterion. \\[+1.5ex]
			\hline \\[-1.8ex] 
	\end{tabular}}
	\label{par.definitons1}
\end{table}
\pagebreak


\begin{table}[H] 
	\centering
	\caption{Notations of latent variables and data used in this article. Bold symbols represent collections (vectors).}
	\resizebox{\textwidth}{0.47\textheight}{
		\begin{tabular}{ l @{\extracolsep{100pt}} l } 
			\\[-1.8ex]\hline 
			\hline \\[-1.8ex]
			\textbf{Latent variables} & \textbf{Definition}\\ 
			\hline \\[-1.8ex]
			$\mathbf{S}$ & Locations of the activity centres of $N$ animals within $\mathcal{V}$.\\ [+1.5ex]
			$\mathbf{s}_i = (s_{i1}, s_{i2})^\prime$ & Location of individual $i$'s activity centre. \\ [+1.5ex]
			$\mathbf{z}=(z_1, z_2, \dots, z_M)^\prime$ & A vector of Bernoulli variables, $z_i=1$ if individual $i$ is present. \\ [+1.5ex]
			$\mathbf{u}=(u_1,  \dots, u_M)^\prime$ & A vector of Bernoulli variables, $u_i=1$ if individual $i$ is male \\ 
			& in the population and $u_i=0$ if it is a female.\\ [+1.5ex]
			$\mathbf{u_0}\ (\subset \mathbf{u})$ & Vector of ``missing'' binary observations on sexes of the\\
			& list of $M$ individuals.\\ [+1.5ex]
			$\mathbf{L}=(\text{L}_1, \text{L}_2, \dots, \text{L}_M)^\prime$ & A one to one mapping from index set of individuals captured \\
			& by  detector 2 to $\{1,2,\dots,M\}$ providing the true index of each\\
			&  detector 2 individuals.\\ [+1.5ex]
			\\[-1.8ex]\hline 
			\hline \\[-1.8ex]
			\textbf{Data } & \textbf{Definition}\\ 
			\hline \\[-1.8ex]
			$\mathbf{x}_j= (x_{j1}, x_{j2})^\prime$ & Location of $j^{\,\text{th}}$ trap station for detectors.\\  [+1.5ex]
			$\mathbf{u_{obs}} \ (\subset \mathbf{u})$ & Vector of ``recorded'' binary observations on sexes of the\\
			& captured individuals.\\ [+1.5ex]
			$y_{ijk}^{(1)}$ & $y_{ijk}^{(1)}=1$ if individual $i$ is detected in detector 1 at trap station \\
			& $\mathbf{x}_j$ on occasion $k$, $y_{ijk}^{(1)}=0$ if not detected in detector 1.\\
			$y_{i\cdot \cdot}^{(1)}=\displaystyle{\sum_{j=1}^J \sum_{k=1}^K y_{ijk}^{(1)}} $ & Number of times individual $i$ got detected\\ [-1.8ex]
			&  in detector 1 over $J$ trap stations and $K$ occasions.\\ [+1.5ex]
			$y_{ijk}^{(2)}$ & $y_{ijk}^{(2)}=1$ if individual $i$ is detected in detector 2 at trap station  \\
			& $\mathbf{x}_j$ on occasion $k$, $y_{ijk}^{(2)}=0$ if not detected in detector 2.\\
			$y_{i\cdot \cdot}^{(2)}=\displaystyle{\sum_{j=1}^J \sum_{k=1}^K y_{ijk}^{(2)}} $ & Number of times individual $i$ got detected\\ [-1.8ex]
			&  in detector 2 over $J$ trap stations and $K$ occasions.\\ [+1.5ex]
			$n$  & Number of fully identified individuals, each of them is captured\\
			& by both the detectors on at least one occasion.\\ [+1.5ex]
			$\mathbf{Y_{obs}^{(1)}} = ((y_{ijk}^{(1)}))$ & Array of individual specific capture histories  obtained by\\
			& detector 1 (dimension $n \times J \times K$).\\ [+1.5ex]
			$\mathbf{Y_{obs}^{(2)}} = ((y_{ijk}^{(2)}))$ & Array of individual specific capture histories  obtained by\\
			& detector 2 (dimension $n \times J \times K$).\\ [+1.5ex] 
			$\mathbf{Y^{(1)}}$ & Zero augmented array of individual specific capture histories \\
			& corresponding to detector 1 (dimension $M \times J \times K$).\\ [+1.5ex]
			$\mathbf{Y^{(2)}}$ & Zero augmented array of individual specific capture histories \\
			& corresponding to detector 2 (dimension $M \times J \times K$).\\ [+1.5ex]
			$\mathbf{Y^{(2*)}}$ & Reordered $\mathbf{Y^{(2)}}$ according to $\mathbf{L}$ (dimension $M \times J \times K$).\\ [+1.5ex]
			$n_{ij}=\displaystyle{\sum_{k=1}^K} \,I(y_{ijk}^{(1)} + y_{ijk}^{(2)} >0)$ & Number of times individual $i$ got detected at trap $j$ \\[-1.8ex]
			& on at least one of its sides over $K$ occasions. \\ 
			$n_{i\cdot}=\displaystyle{\sum_{j=1}^J} n_{ij}$ & Number of times individual $i$ got detected on at least\\[-1.8ex]
			& one of its sides over $J$ traps and $K$ occasions. \\ [+1.5ex]
			\hline \\[-1.8ex] 
	\end{tabular}}
	\label{par.definitons2}
\end{table}
\pagebreak
\begin{table}[!htbp]
	\centering
	\caption{Prior distributions of model parameters. $R$ is high enough to expect that it would be impossible for animals to exhibit movement as widely as this scale during sampling.}
	\footnotesize{\begin{tabular}{ l @{\extracolsep{15pt}} l  @{\extracolsep{50pt}} l   @{\extracolsep{15pt}} l } 
			\\[-1.8ex]\hline 
			\hline \\[-1.8ex]
			\textbf{Parameters} & \textbf{Prior} & \textbf{Parameters} & \textbf{Prior}\\ 
			\hline \\[-1.8ex]
			$\phi$ & Uniform$(0,\, 1)$ & 	$\sigma_m$ & Uniform$(0,\, R)$  \\[1.5ex]
			$\w_0$ & Uniform$(0,\, 1)$ & $\sigma_f$ & Uniform$(0,\, R)$ \\ [+1.5ex]
			$p_0$ & Uniform$(0,\, 1)$  &  $\theta$ & Uniform$(0,\, 1)$ \\ [+1.5ex]
			$\sigma$ & Uniform$(0,\, R)$ &  $\psi$ & Uniform$(0,\, 1)$ \\ [+0.5ex] 
	\hline \\[-1.8ex] 
	\end{tabular}}
	\label{table.prior}
\end{table}
\pagebreak
\begin{table}[!htb]
	\centering 
	\caption{Bayesian model selection methods used in this study.} 
		\resizebox{\textwidth}{0.17\textheight}{
	 \begin{tabular}{c @{\extracolsep{5pt}} c @{\extracolsep{5pt}} c @{\extracolsep{5pt}} c @{\extracolsep{5pt}} c @{\extracolsep{5pt}} c} 
			\\[-1.8ex]\hline 
			\hline \\[-1.8ex] 
			Sl. no. & Model selection & Variant & Approximation & Choices of & Eq. No. \\ 
			& method && method & tuning density & \\ 
			\hline \\[-1.8ex] 
			1. & Bayes factor & Gelfand-Dey  & MAP & Multivariate normal density, multivar-  & (\ref{GDest2})\\
& & estimator & & iate-$t$ density with degrees of  freedom& \\
& & & &  10, 100, 500, 1000, 10000  and truncated& \\
& & & &   multivariate normal density with confid-& \\
& & & &   ence coefficients 0.90, 0,95, 0.99.  & \\
	2 & Bayes factor & Gelfand-Dey  & IL & -Do-  & (\ref{GDIL1})\\
	& & estimator & &  & \\
3. & Bayes factor & Harmonic mean & - & - & (\ref{HMest1}) \\
	& & estimator & &  & \\
4. & DIC & $p_{DIC1}$ & MAP  & - & (\ref{diceq}) \\
5. & DIC & $p_{DIC2}$ &  MAP & - & (\ref{diceq}) \\
6. & WAIC & $p_{WAIC1}$ & -  & - & (\ref{waiceq}) \\
7. & WAIC& $p_{WAIC2}$ &  - & - & (\ref{waiceq}) \\
8. & WAIC & $p_{WAIC3}$ & - & - & (\ref{waiceq2}) \\
9. &  Posterior & - &  -  & -  & (\ref{pplinf}) \\
	& predictive loss & &  & & \\
			\hline \\[-1.8ex] 
	\end{tabular}}
	\label{msc} 
\end{table}
\pagebreak
\begin{table}[!htb] 
	\centering 
	\caption{Parameter specifications corresponding to different simulation scenarios.} 
	\label{simsetting.bmse} 
	\footnotesize{\begin{tabular}{@{\extracolsep{5pt}} cccccccc} 
			\\[-1.8ex]\hline 
			\hline \\[-1.8ex] 
			Scenario & $M$ & $N$ & $N_{Male}$ & $\w_0$ & $\phi$ & $\sigma_m$ & $\sigma_f$\\ 
			\hline \\[-1.8ex]
			1 & 400 & 100 & 40 & 0.01 & 0.3 & 0.3 & 0.15   \\ 
			2 & 400 & 100 & 40 & 0.01 & 0.9 & 0.3 & 0.15  \\ 
			3 & 400 & 100 & 40 & 0.01 & 0.3 & 0.4 & 0.20  \\ 
			4 & 400 & 100 & 40 & 0.01 & 0.9 & 0.4 & 0.20  \\ 
			5 & 400 & 100 & 40 & 0.03 & 0.8 & 0.3 & 0.15  \\ 
			6 & 400 & 100 & 40 & 0.03 & 0.8 & 0.4 & 0.20  \\ 
			7 & 400 & 100 & 40 & 0.05 & 0.3 & 0.3 & 0.15   \\ 
			8 & 400 & 100 & 40 & 0.05 & 0.5 & 0.3 & 0.15  \\ 
			9 & 400 & 100 & 40 & 0.05 & 0.9 & 0.3 & 0.15  \\ 
			10 & 400 & 100 & 40 & 0.05 & 0.3 & 0.4 & 0.20 \\ 
			11 & 400 & 100 & 40 & 0.05 & 0.5 & 0.4 & 0.20  \\ 
			12 & 400 & 100 & 40 & 0.05 & 0.9 & 0.4 & 0.20  \\ 
			\hline \\[-1.8ex] 
	\end{tabular} }
	\label{t.simscenarios.BMSE}
\end{table}  
\pagebreak

\begin{figure}[!htb] 
	\centering
	\includegraphics[width=270pt,height=270pt]{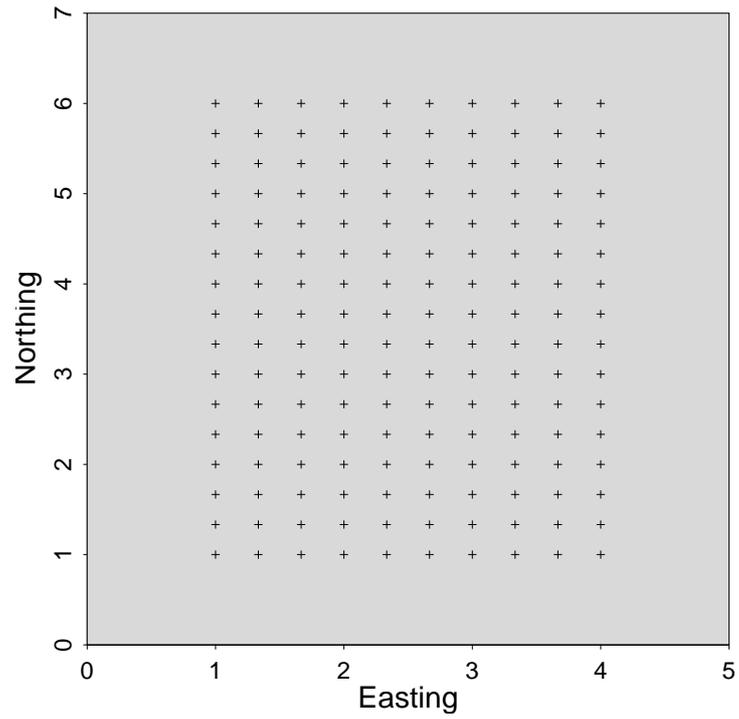}
	\caption{Array of trap locations (denoted by `+') within the state space $(0,5) \times (0,7)$.}
	\label{statespace.BMSE}
\end{figure}	


\begin{figure}[H] 
	\centering
	\begin{tabular}{l @{\extracolsep{0pt}} l }
		{\hspace{100pt} (a)} &  {\hspace{100pt} (b)} \\ [-0.5em]
		\includegraphics[width=220pt,height=180pt]{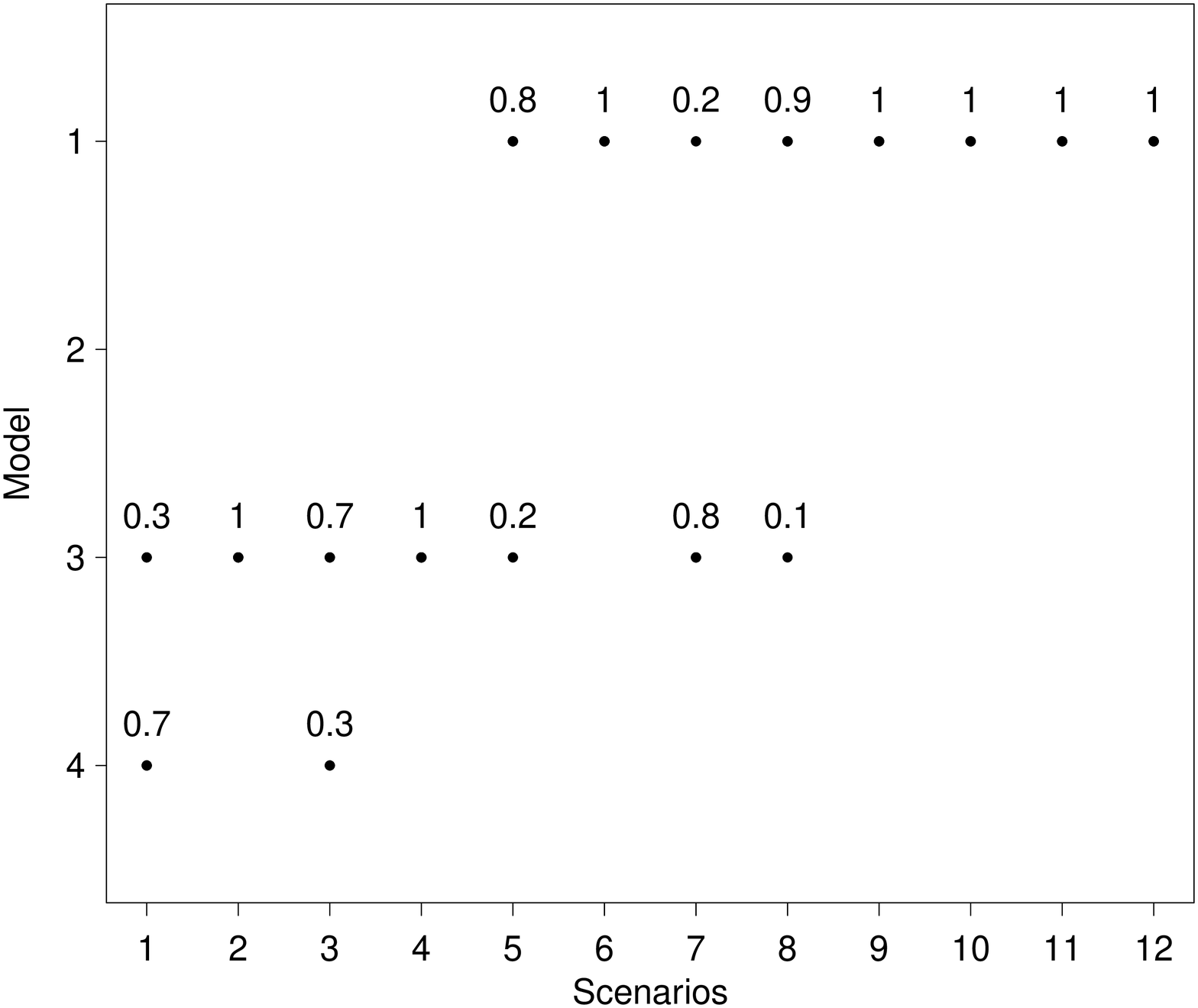} &  \includegraphics[width=220pt,height=180pt]{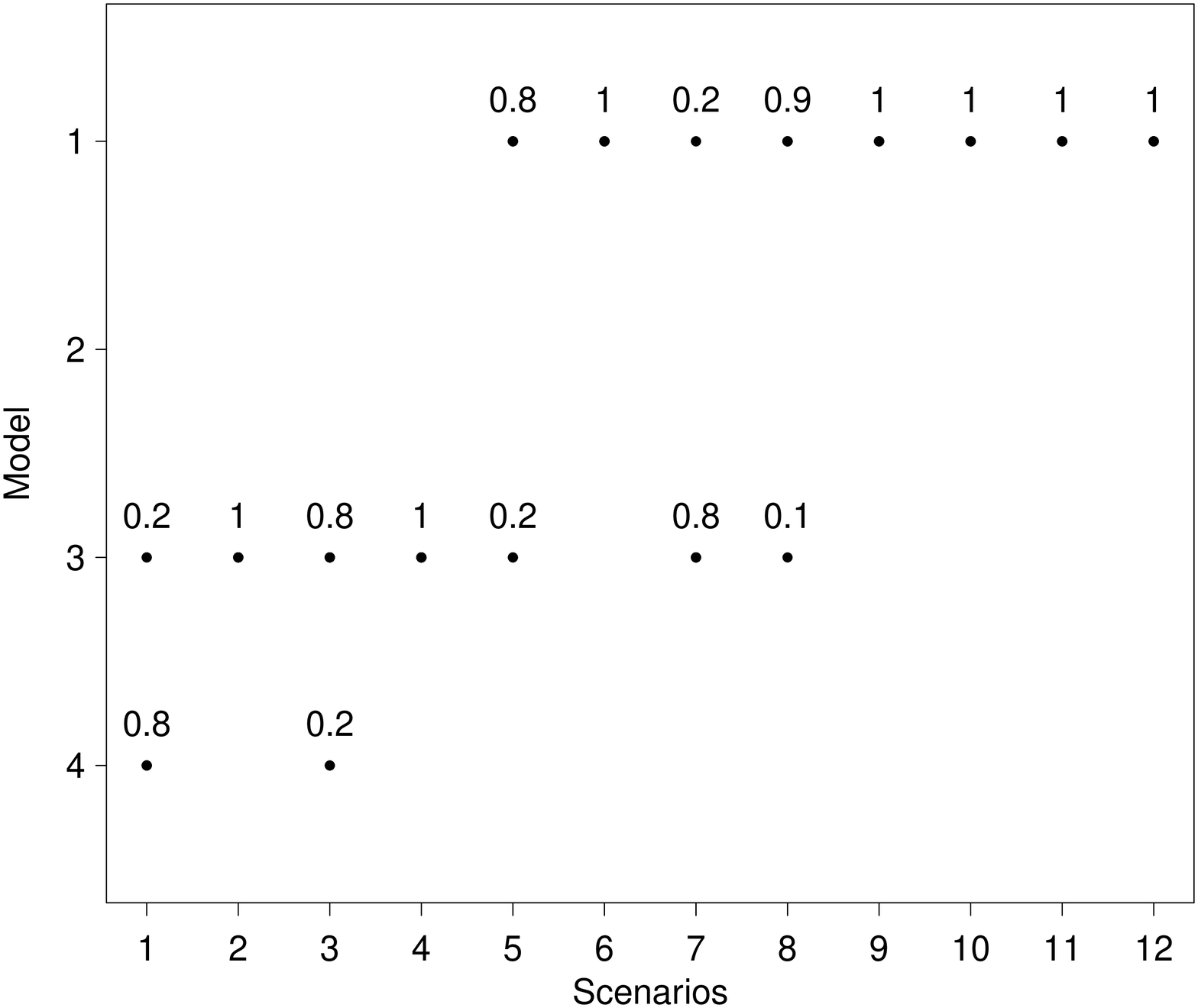} \\ [0.6em]
		{\hspace{100pt} (c)} &  {\hspace{100pt} (d)} \\ [-0.5em]
		\includegraphics[width=220pt,height=180pt]{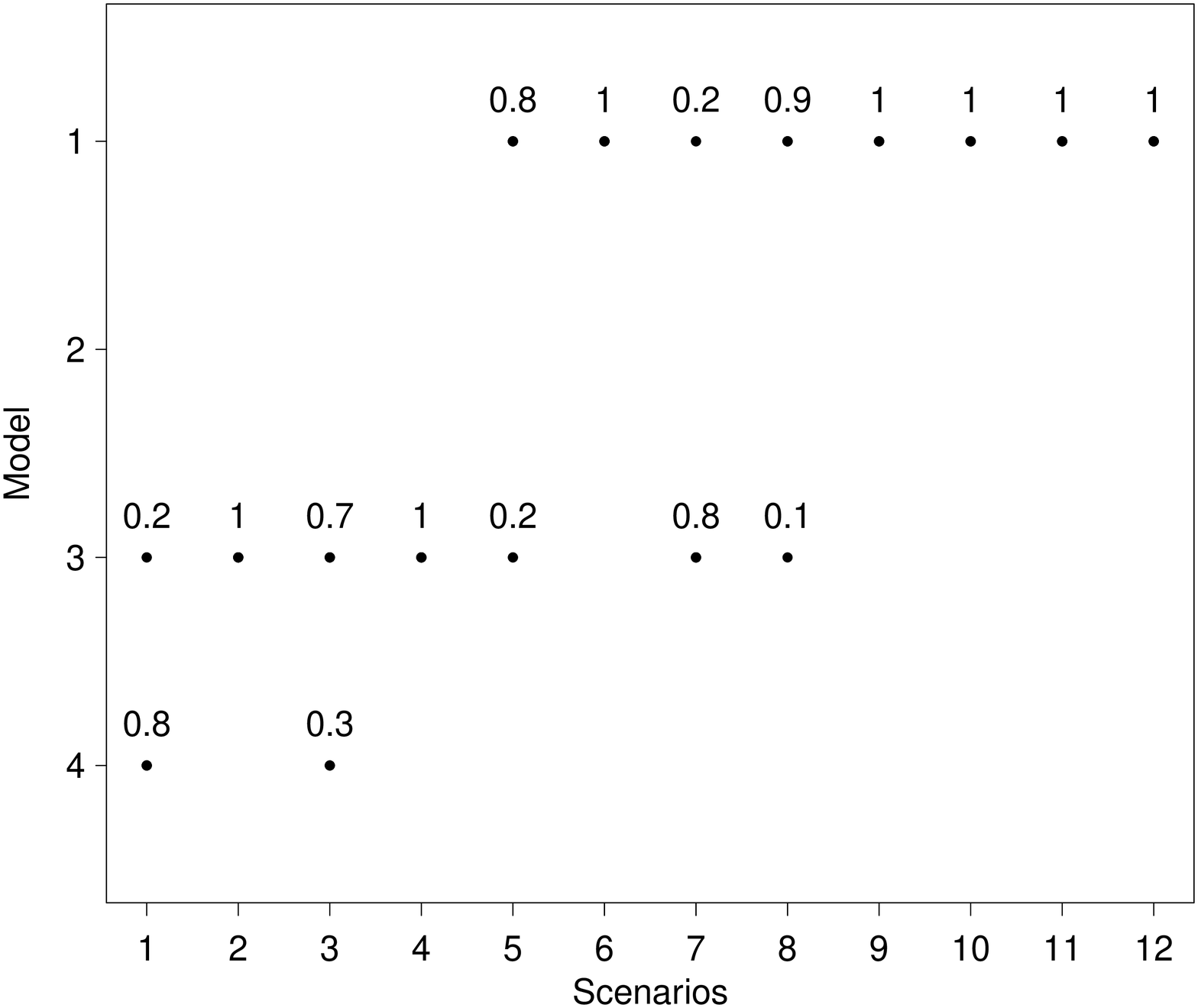} & \includegraphics[width=220pt,height=180pt]{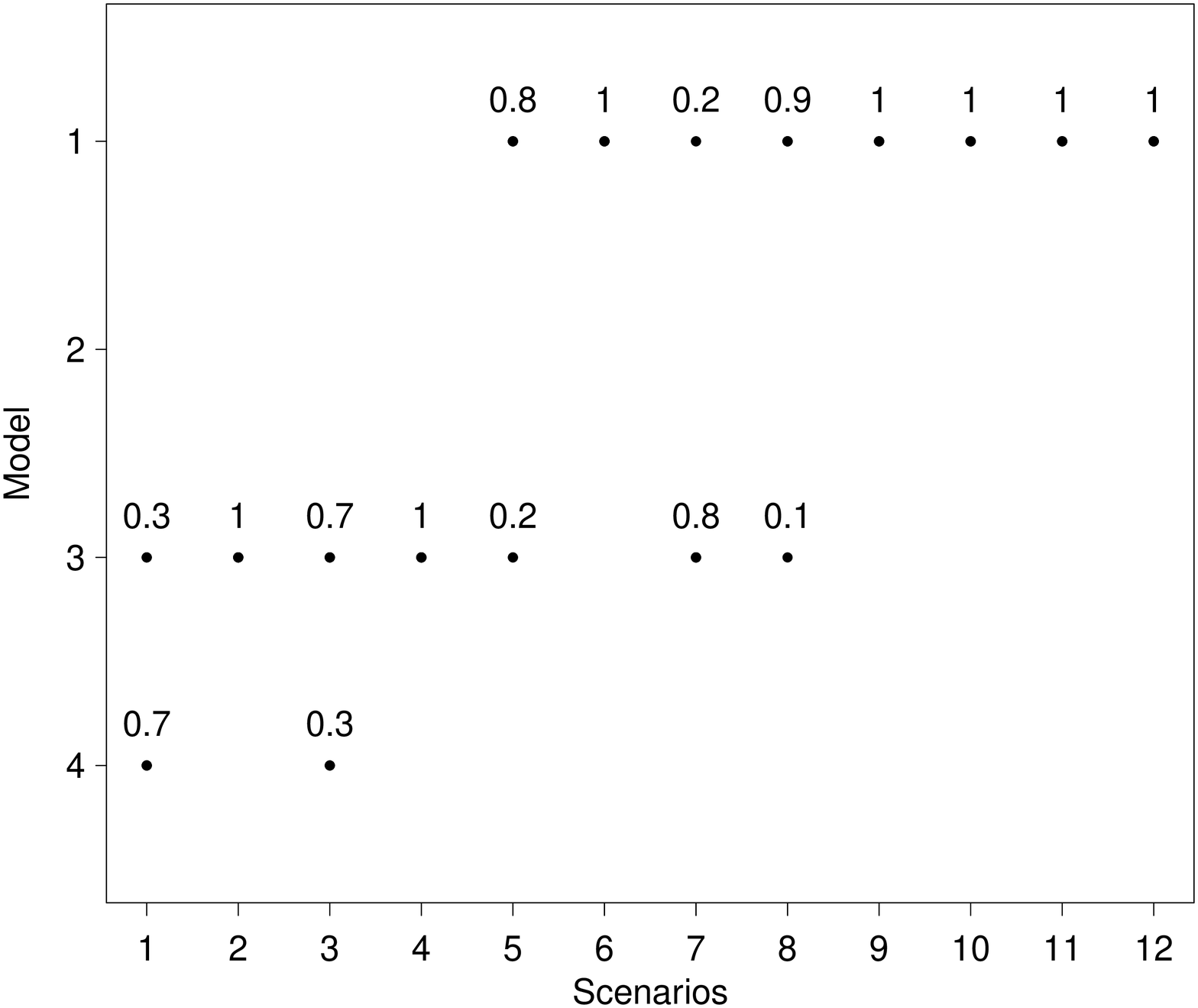}  \\ [0.6em]  
		{\hspace{100pt} (e)} &  {\hspace{100pt} (f)} \\ [-0.5em]
		\includegraphics[width=220pt,height=180pt]{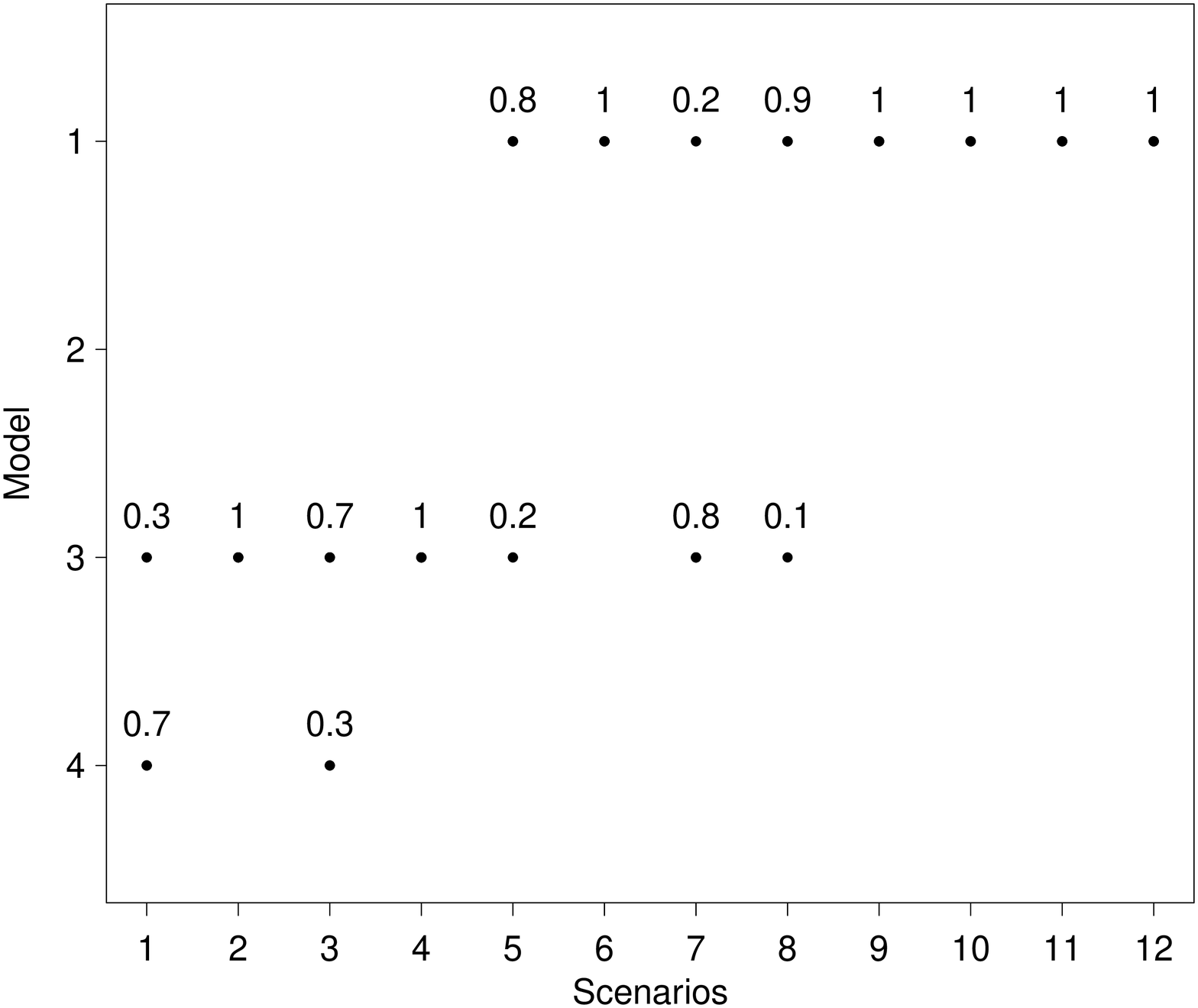} & \includegraphics[width=220pt,height=180pt]{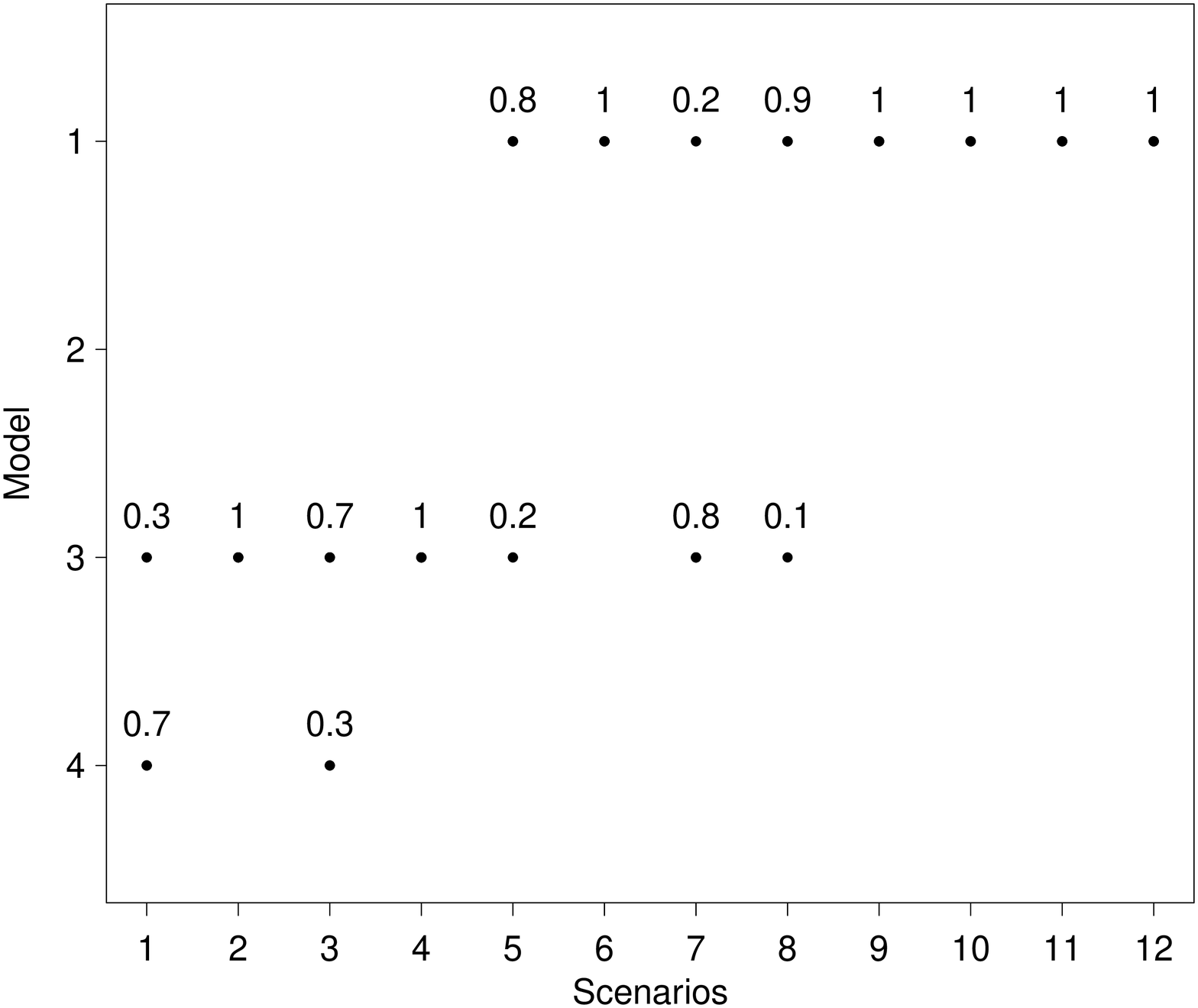}  \\ [0.6em]          
	\end{tabular}
	\caption{\small{The proportion of times Bayes factor favours any particular model using the MAP approximation approach. Plot (a): Gelfand-Dey estimator with a multivariate normal density for $g$. Plots (b)-(f) : Gelfand-Dey estimator with five different choices for $g$, viz., densities of multivariate-$t$ distribution with degrees of freedom 10, 100, 500, 1000, 10000 respectively.}}
	\label{fig.modelsel1.BMSE}
\end{figure}
\newpage

\begin{figure}[H] 
	\centering
	\begin{tabular}{l @{\extracolsep{0pt}} l }
		{\hspace{100pt} (a)} &  {\hspace{100pt} (b)} \\ [-0.5em]
		\includegraphics[width=220pt,height=200pt]{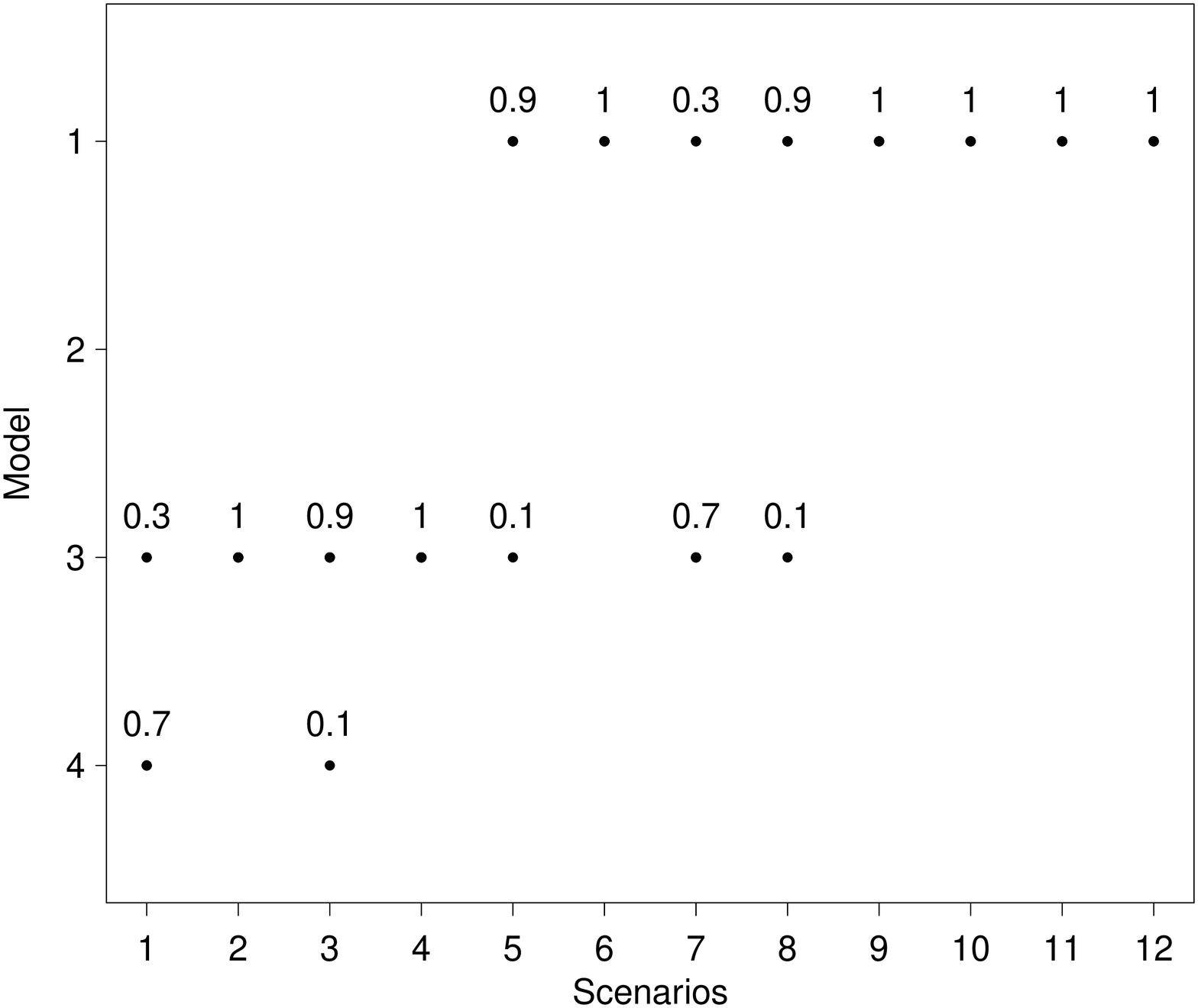} &
		\includegraphics[width=220pt,height=200pt]{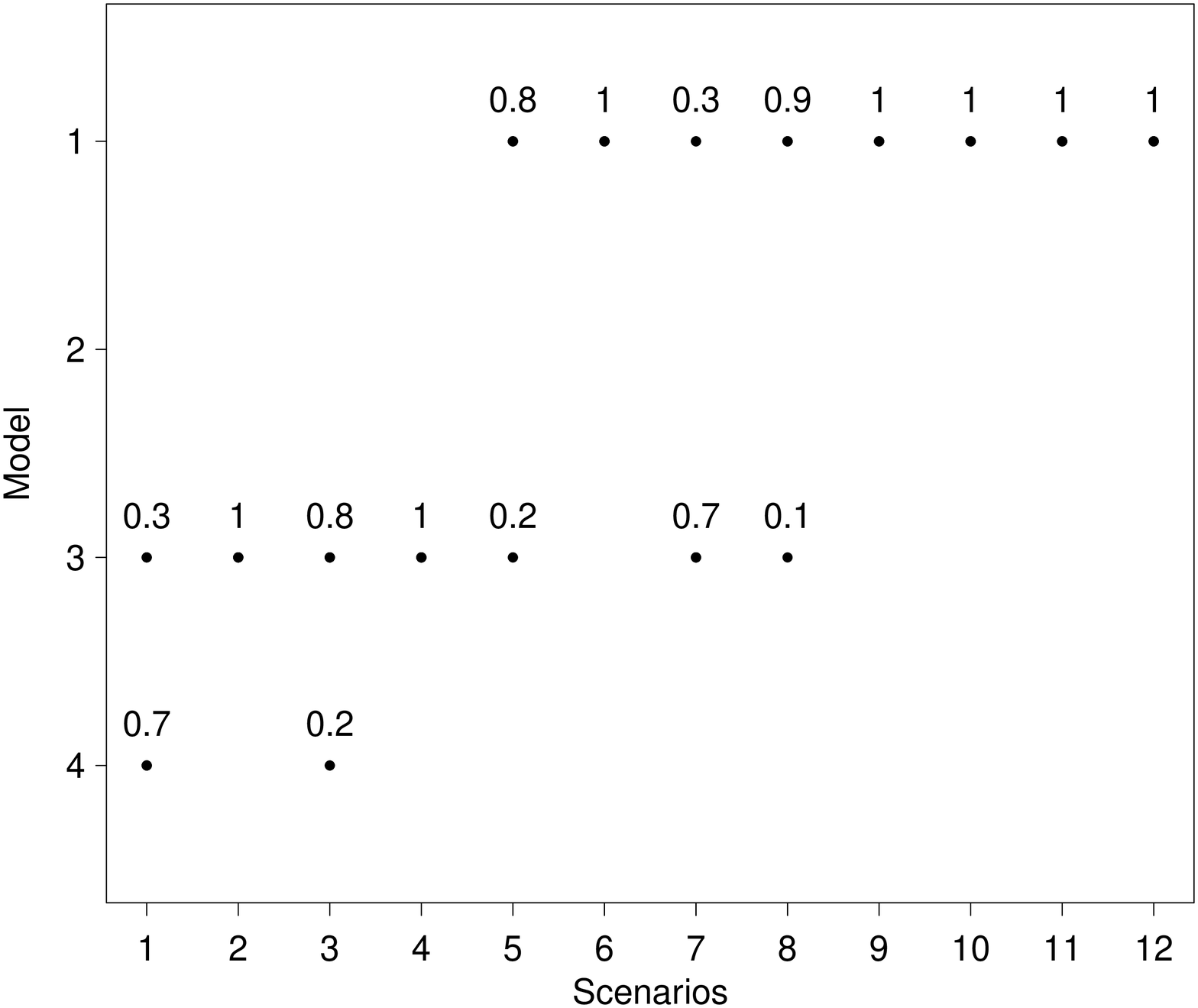}\\ [0.6em]
		{\hspace{100pt} (c)} &  \\ [-0.5em]
		\includegraphics[width=220pt,height=200pt]{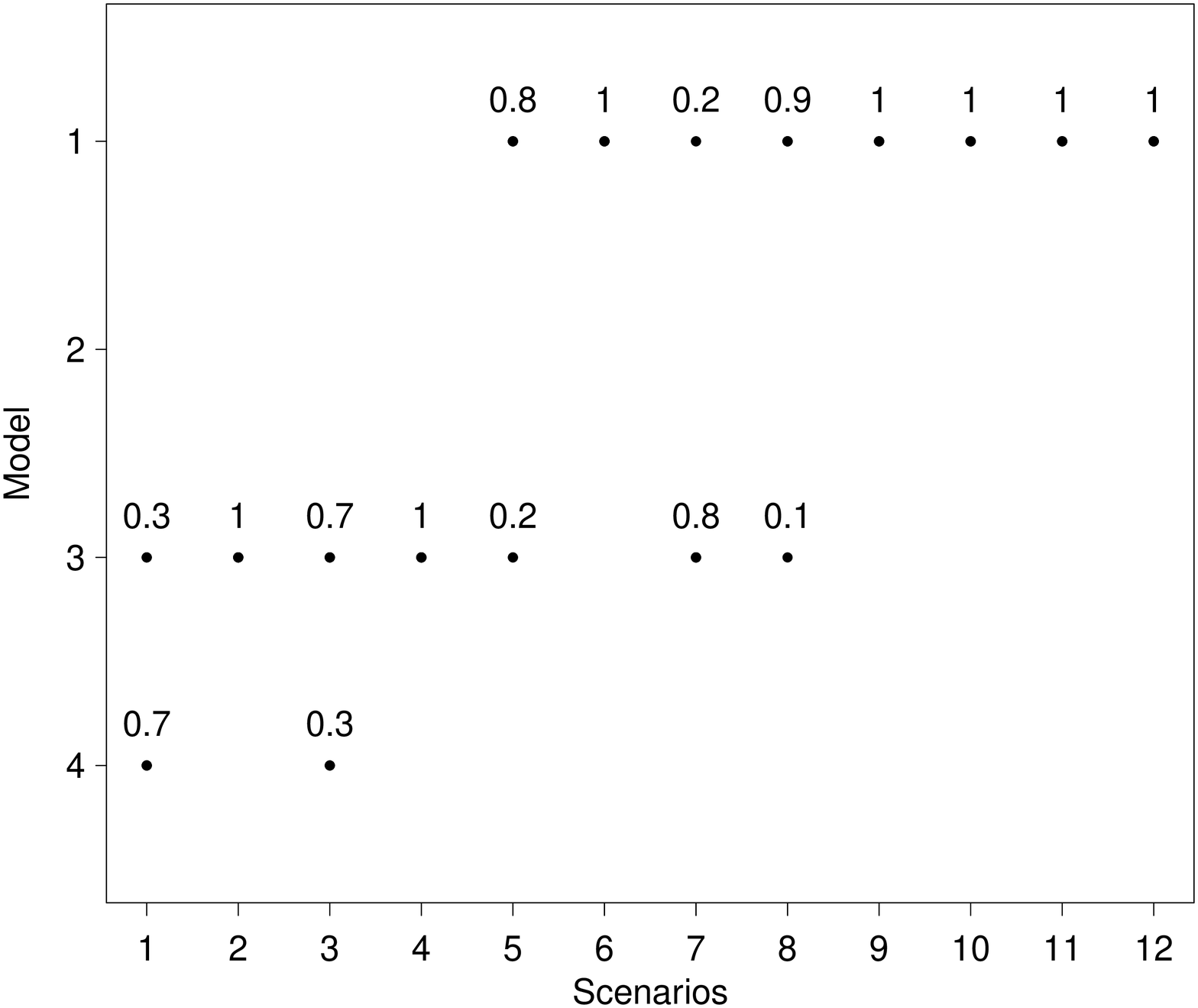} &  \\ [-0.2em]          
	\end{tabular}
	\caption{\small{The proportion of times Bayes factor favours any particular model using the MAP approximation approach. Plots (a) - (c) are obtained by computing Gelfand-Dey estimator with three different choices for $g$, viz.,  truncated normal density with $\alpha = 0.9, 0.95, 0.99$. }}
	\label{fig.modelsel2.BMSE}
\end{figure}

\begin{figure}[H] 
	\centering
	\begin{tabular}{l @{\extracolsep{20pt}} l }
		{\hspace{100pt} (a)} &  {\hspace{100pt} (b)} \\ [-0.5em]
		\includegraphics[width=220pt,height=180pt]{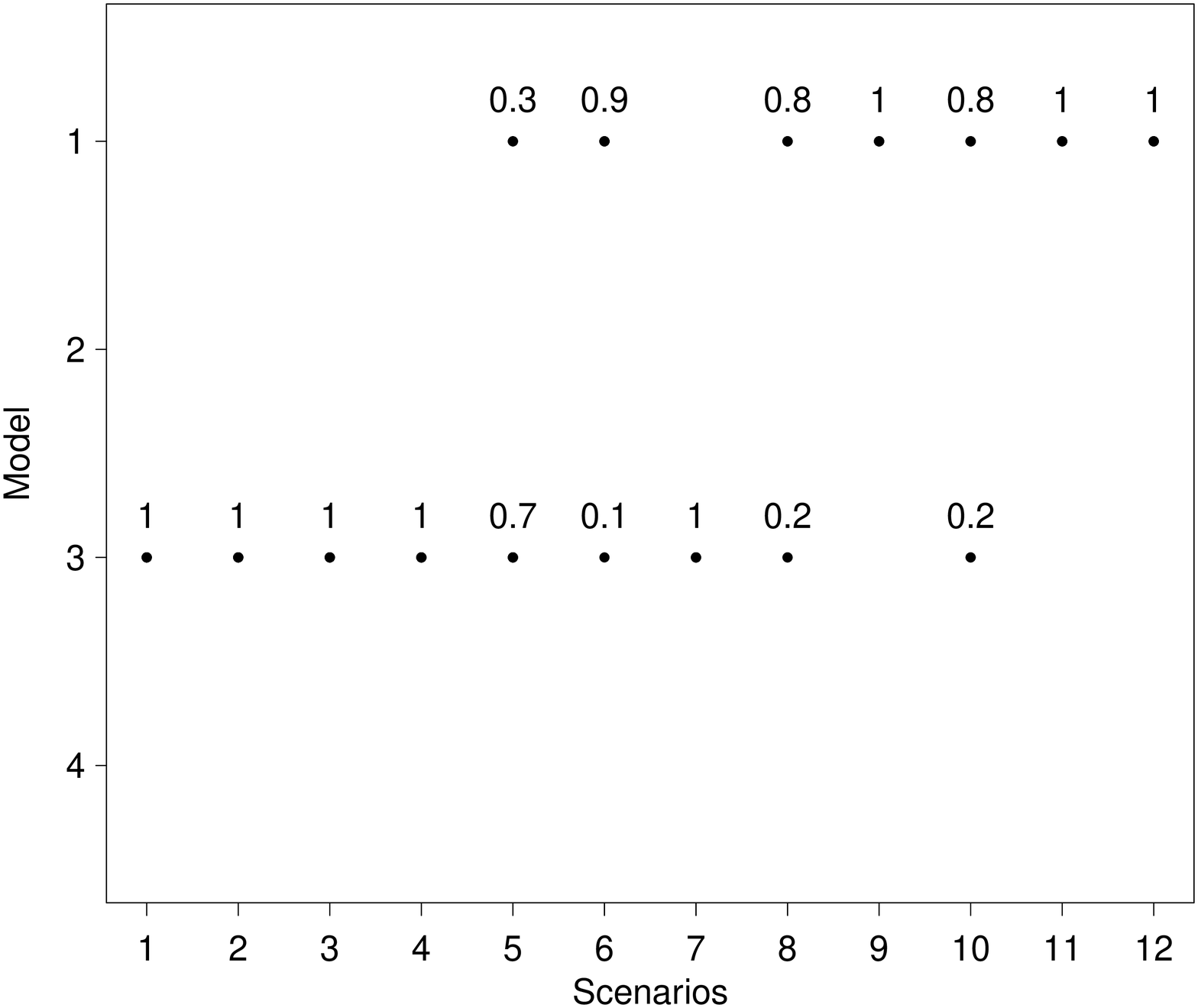} &  \includegraphics[width=220pt,height=180pt]{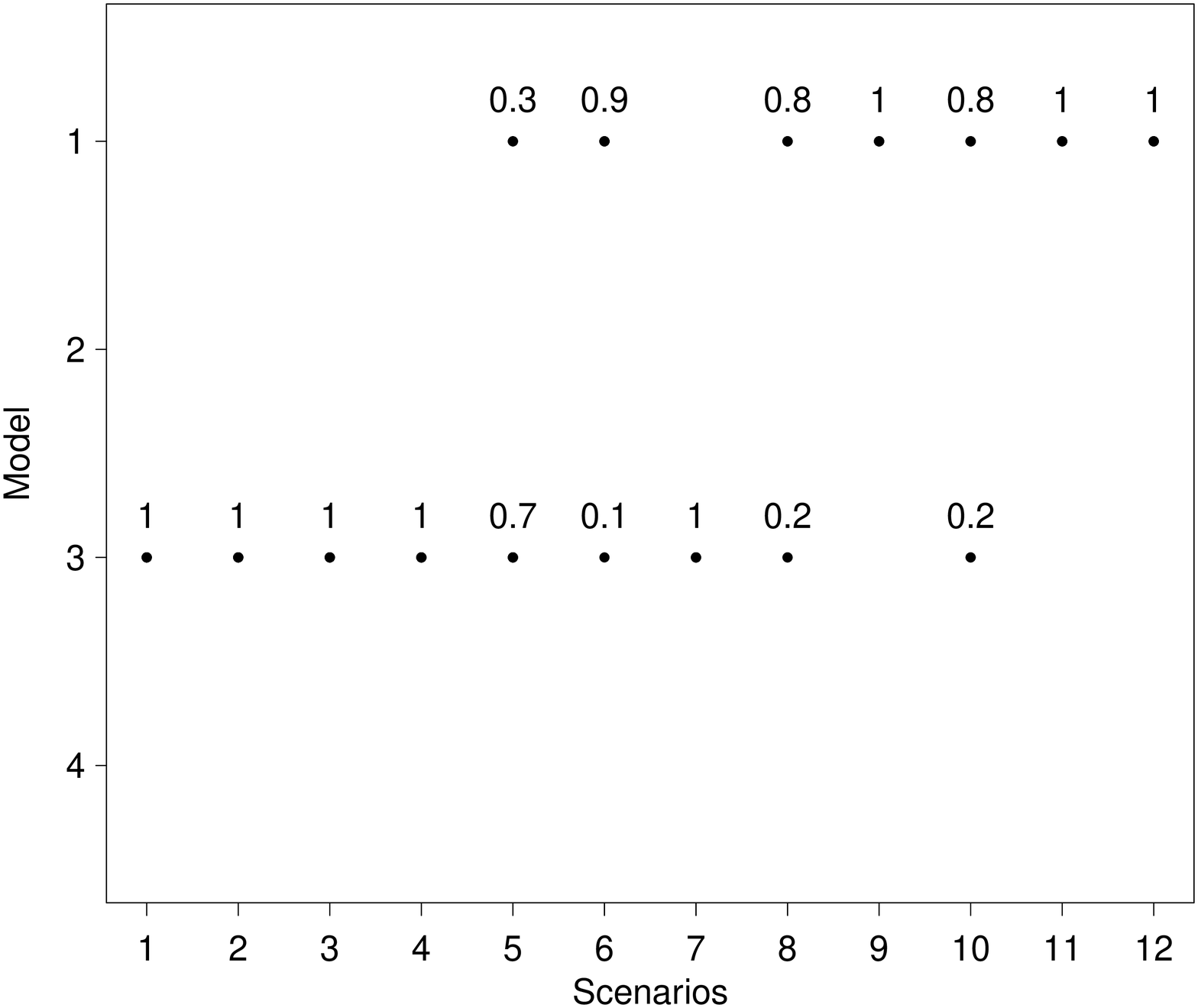} \\ [0.6em]
		{\hspace{100pt} (c)} &  {\hspace{100pt} (d)} \\ [-0.5em]
		\includegraphics[width=220pt,height=180pt]{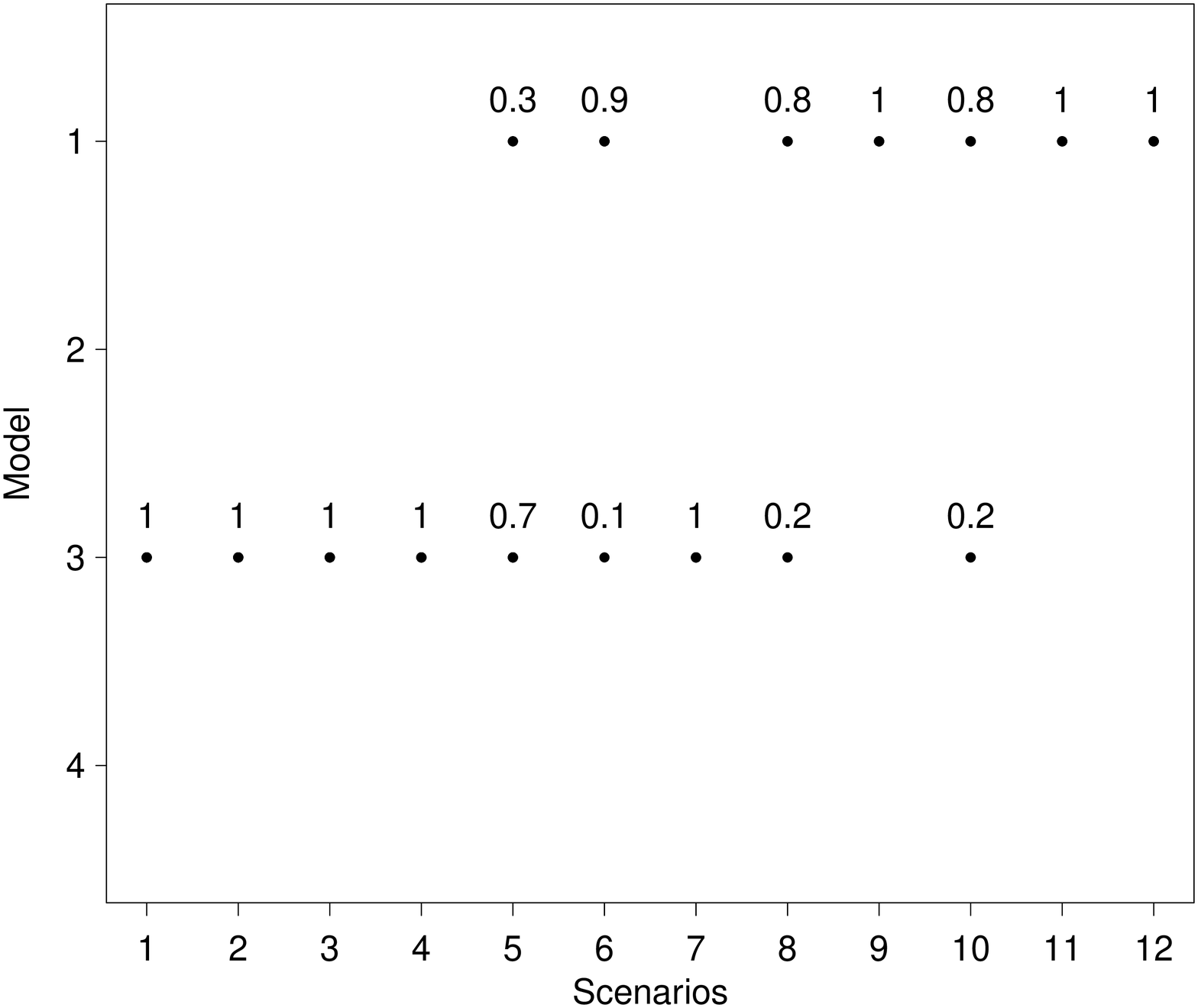} & \includegraphics[width=220pt,height=180pt]{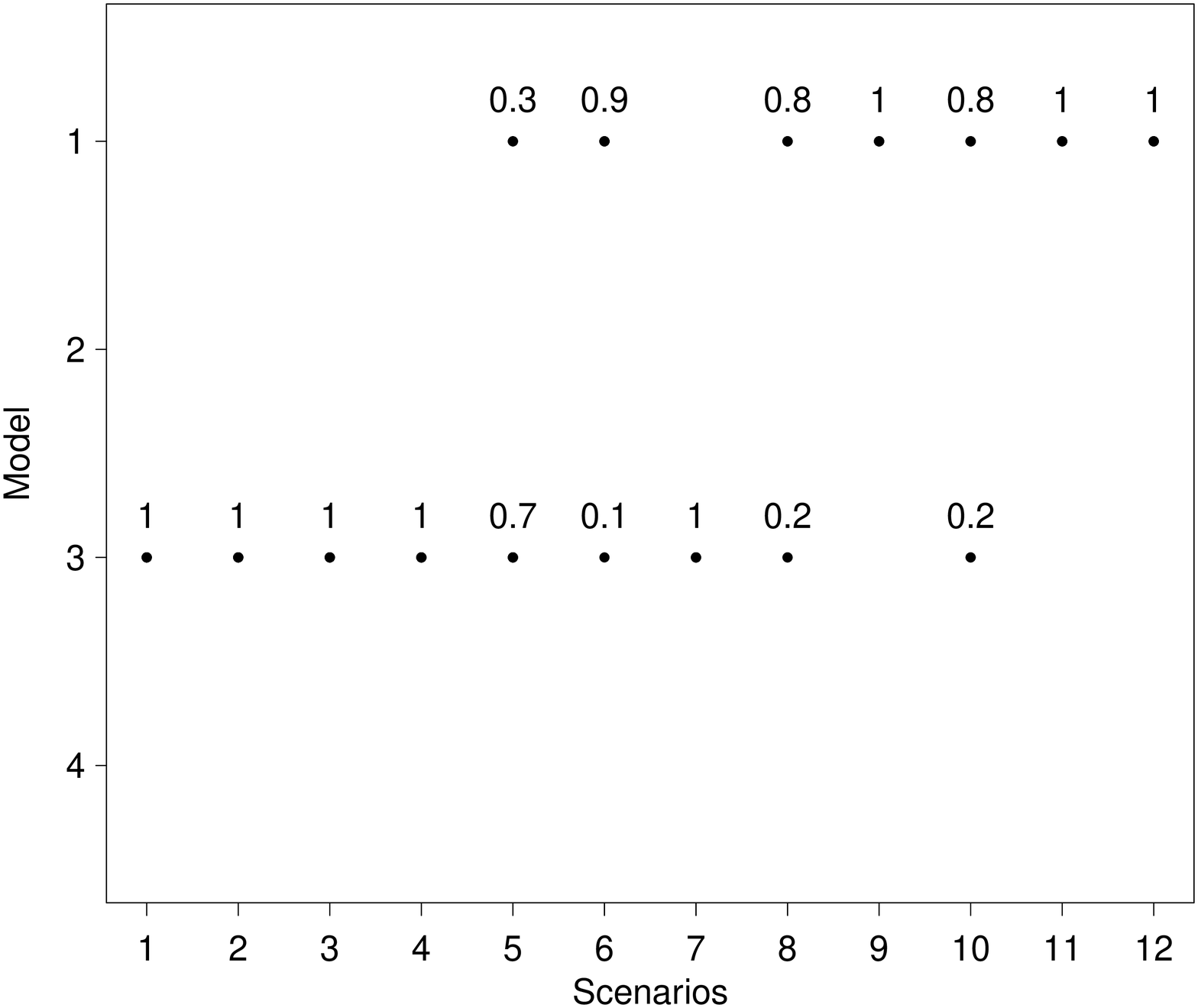}  \\ [0.6em]  
		{\hspace{100pt} (e)} &  {\hspace{100pt} (f)} \\ [-0.5em]
		\includegraphics[width=220pt,height=180pt]{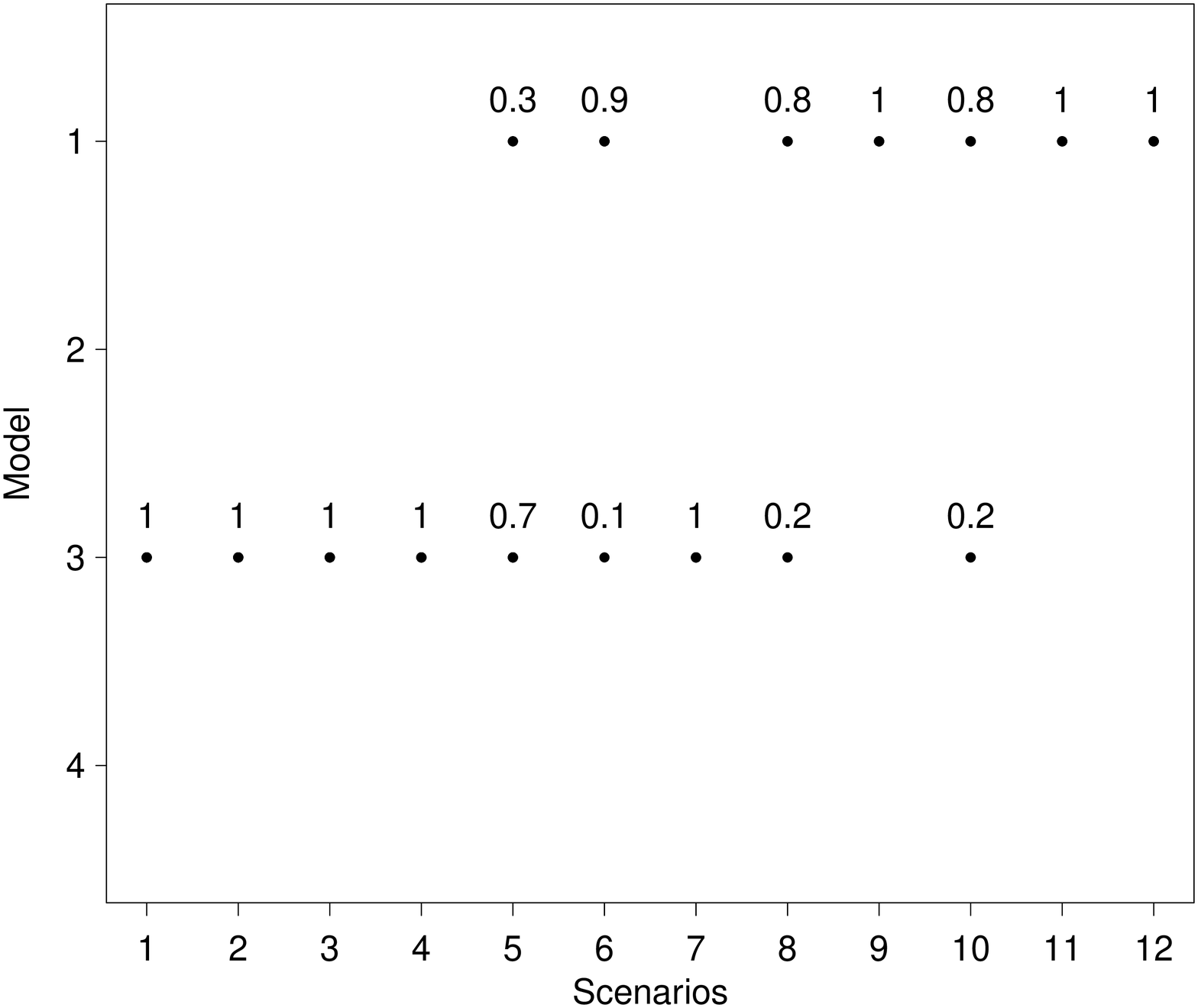} & \includegraphics[width=220pt,height=180pt]{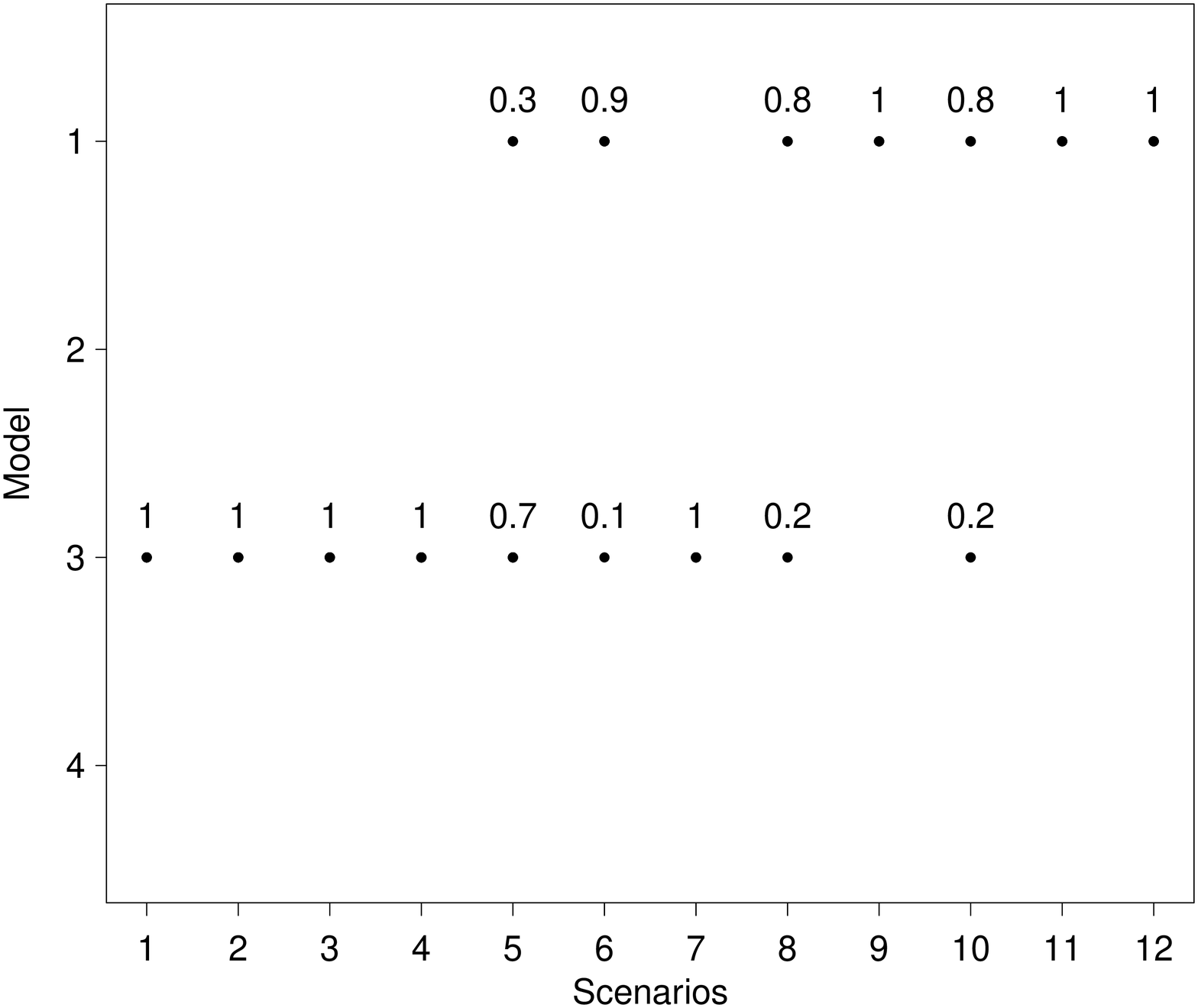}  \\ [0.6em]          
	\end{tabular}
	\caption{\small{The proportion of times Bayes factor favours any particular model using the integrated likelihood approximation approach. Plot (a): Gelfand-Dey estimator with a multivariate normal density for $g$. Plots (b)-(f) : Gelfand-Dey estimator with five different choices for $g$, viz., densities of multivariate-$t$ distribution with degrees of freedom 10, 100, 500, 1000, 10000 respectively.}}
	\label{fig.modelsel3.BMSE}
\end{figure}

\begin{figure}[H] 
	\centering
	\begin{tabular}{l @{\extracolsep{0pt}} l }
		{\hspace{100pt} (a)} &  {\hspace{100pt} (b)} \\ [-0.5em]
		\includegraphics[width=220pt,height=172pt]{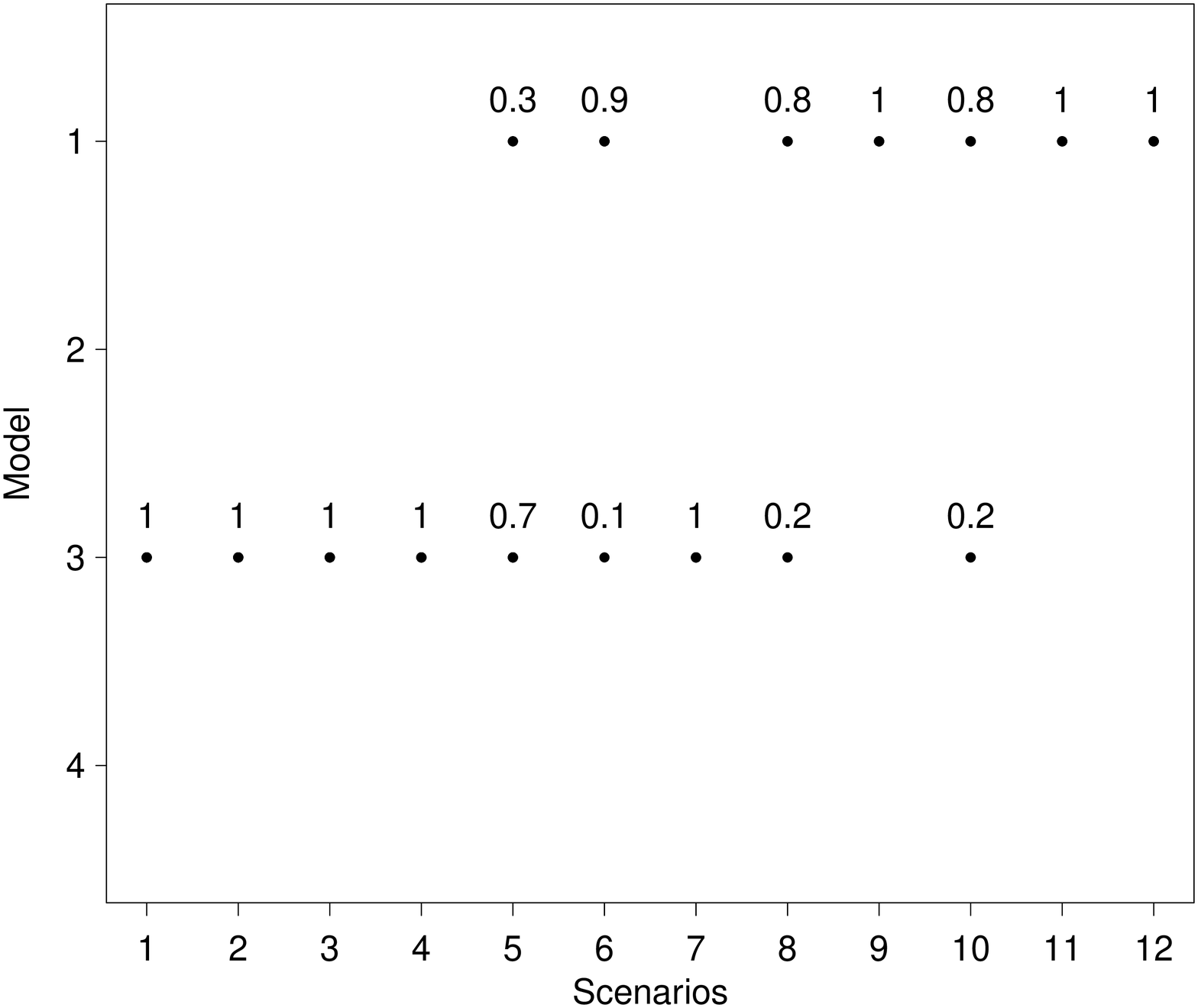} &
		\includegraphics[width=220pt,height=172pt]{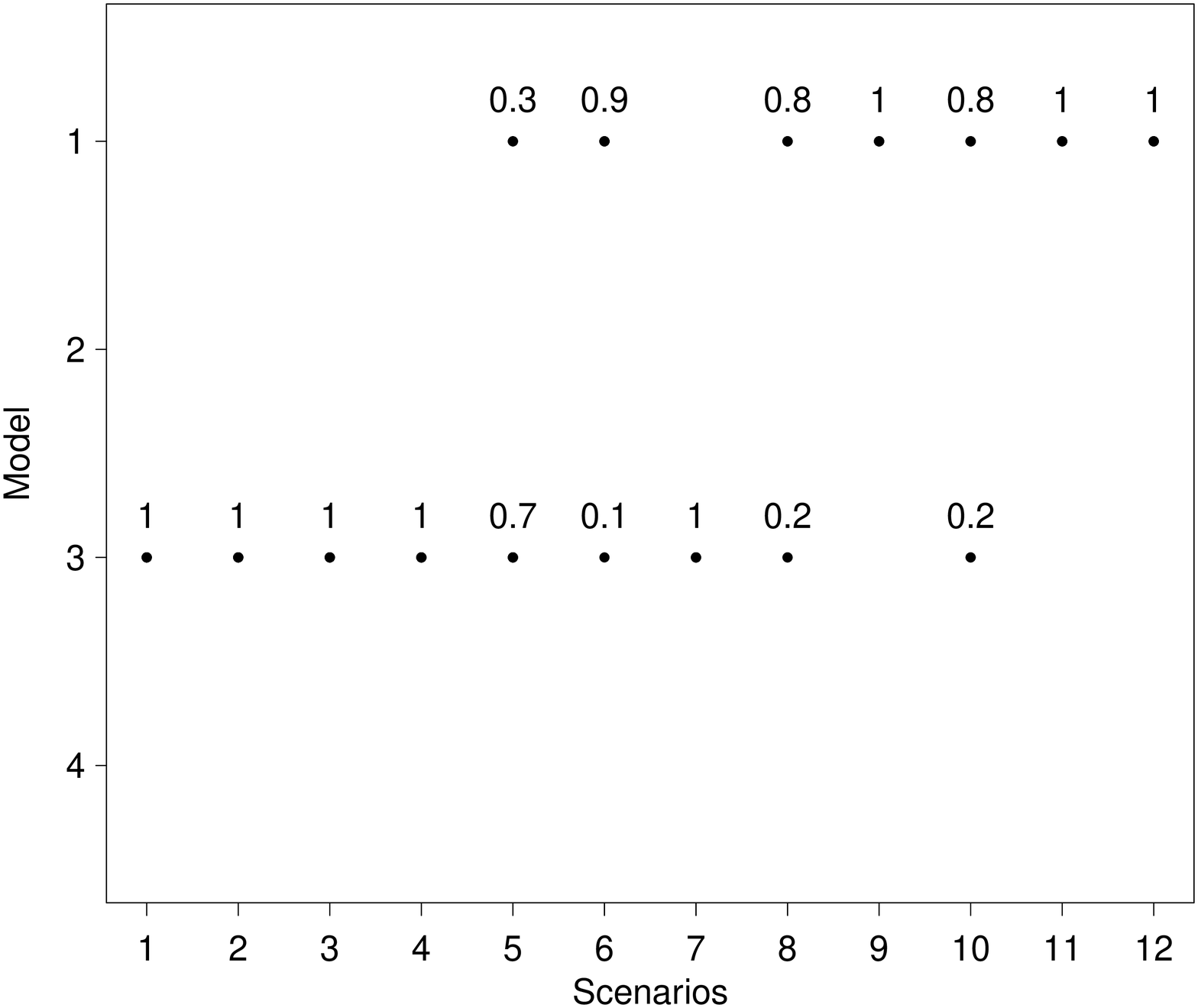} \\ [0.6em]
		{\hspace{100pt} (c)} &   \\ [-0.5em]
		\includegraphics[width=220pt,height=172pt]{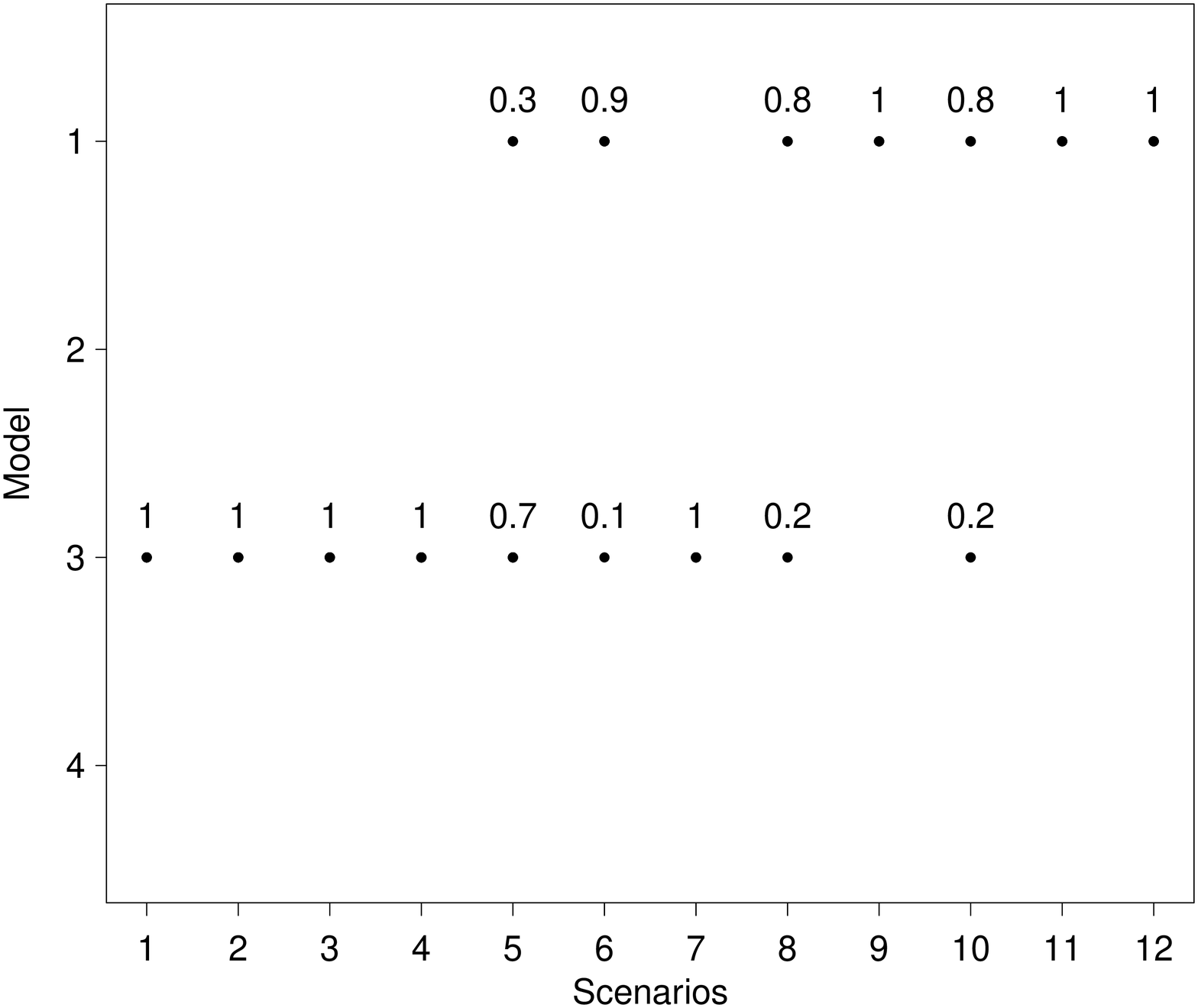} & \\ [-0.2em]          
	\end{tabular}
	\caption{\small{The proportion of times Bayes factor favours any particular model using the integrated likelihood approximation approach. Plots (a) - (c) : Gelfand-Dey estimator with three different choices for $g$, viz.,  truncated normal density with $\alpha = 0.9, 0.95, 0.99$.}}
	\label{fig.modelsel4.BMSE}
\end{figure}  

\begin{figure}[H] 
	\centering
	\includegraphics[width=220pt,height=172pt]{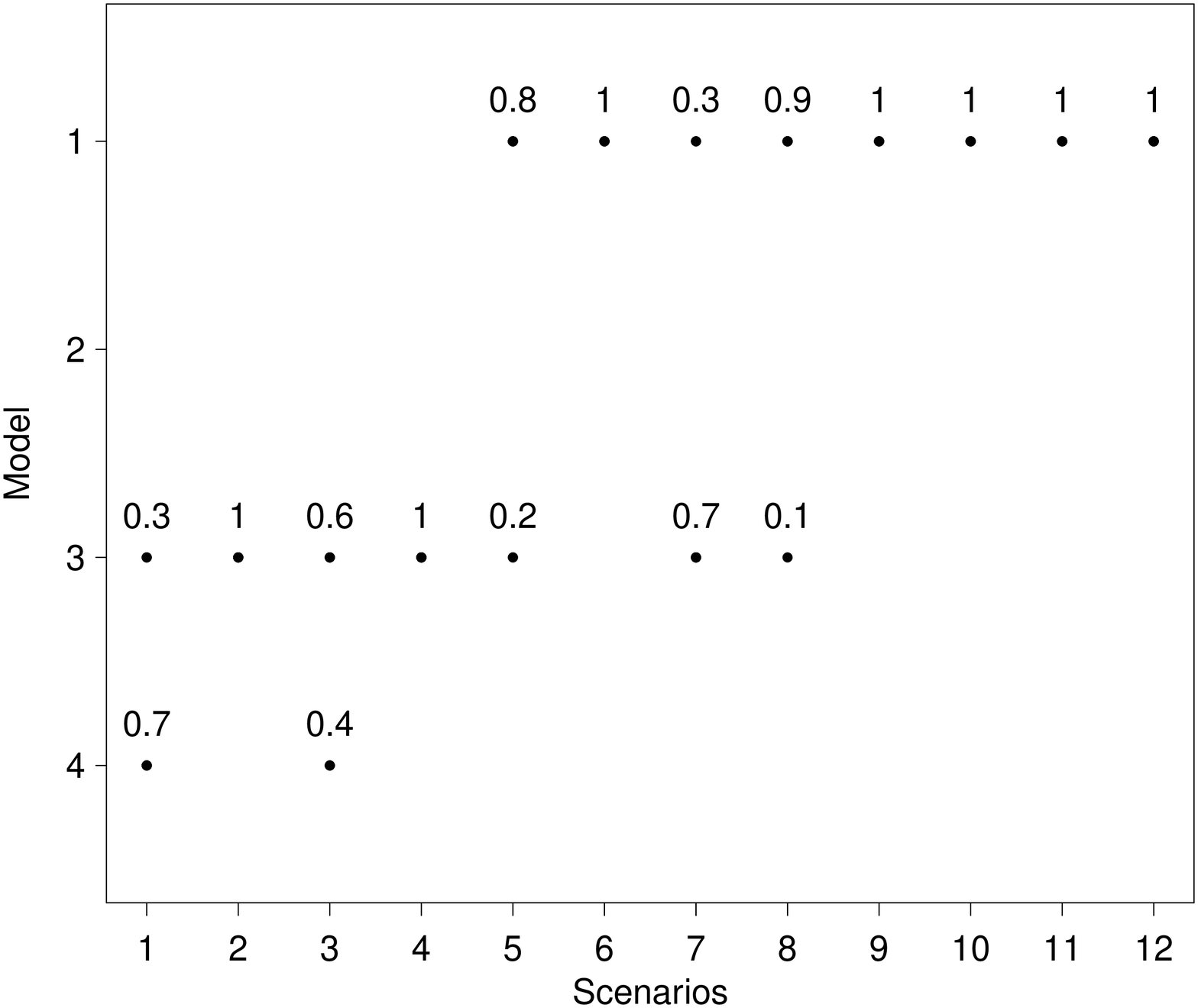}\\ [-0.2em]  
	\caption{\small{The proportion of times Bayes factor (using the harmonic mean estimator) favours any particular model.}}			
	\label{fig.modelsel_hm.BMSE}
\end{figure}  

\begin{figure}[H] 
	\centering
	\begin{tabular}{l @{\extracolsep{0pt}} l }     
		{\hspace{100pt} (a)} &  {\hspace{100pt} (b)} \\ [-0.5em]
		\includegraphics[width=220pt,height=180pt]{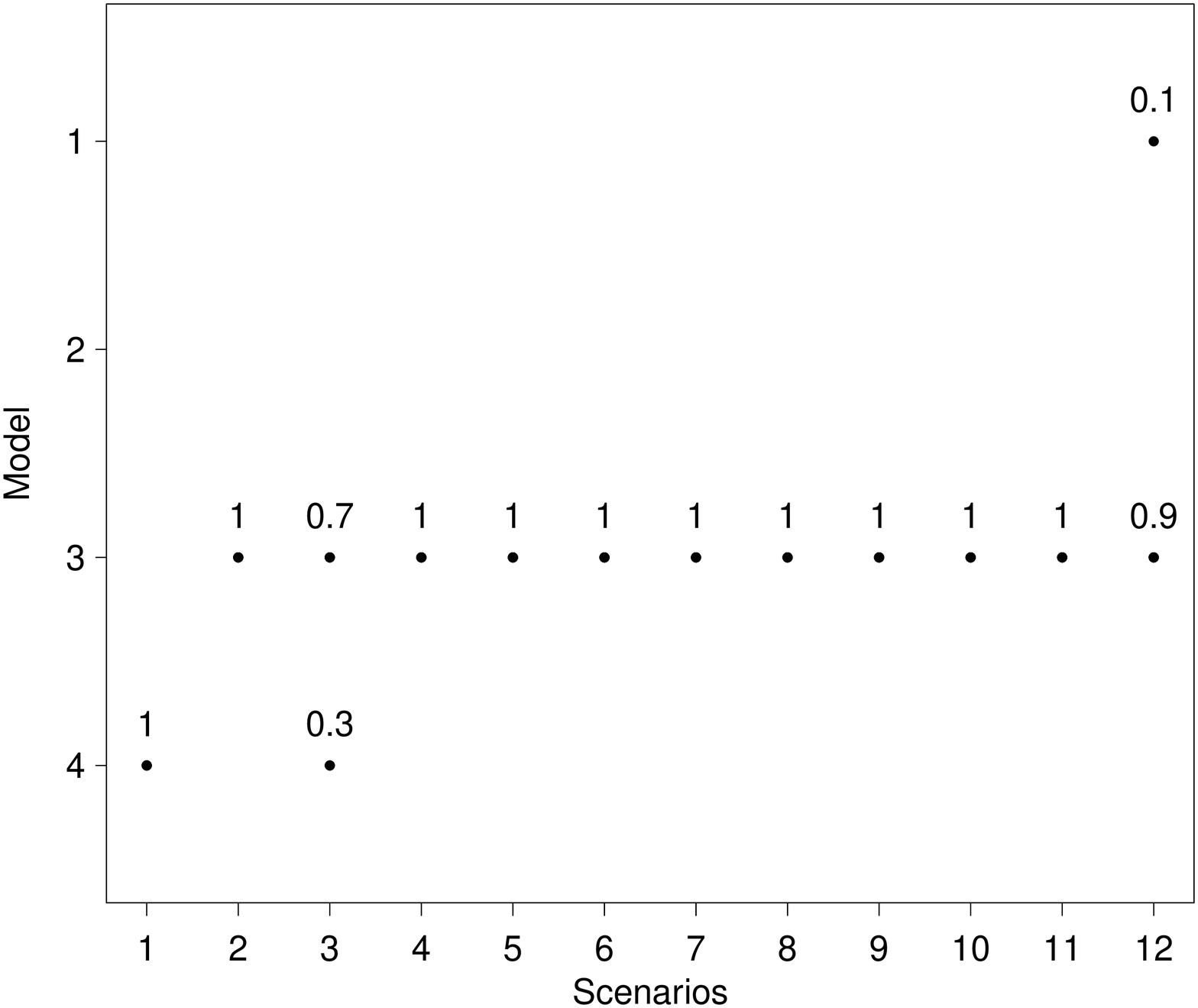} & \includegraphics[width=220pt,height=180pt]{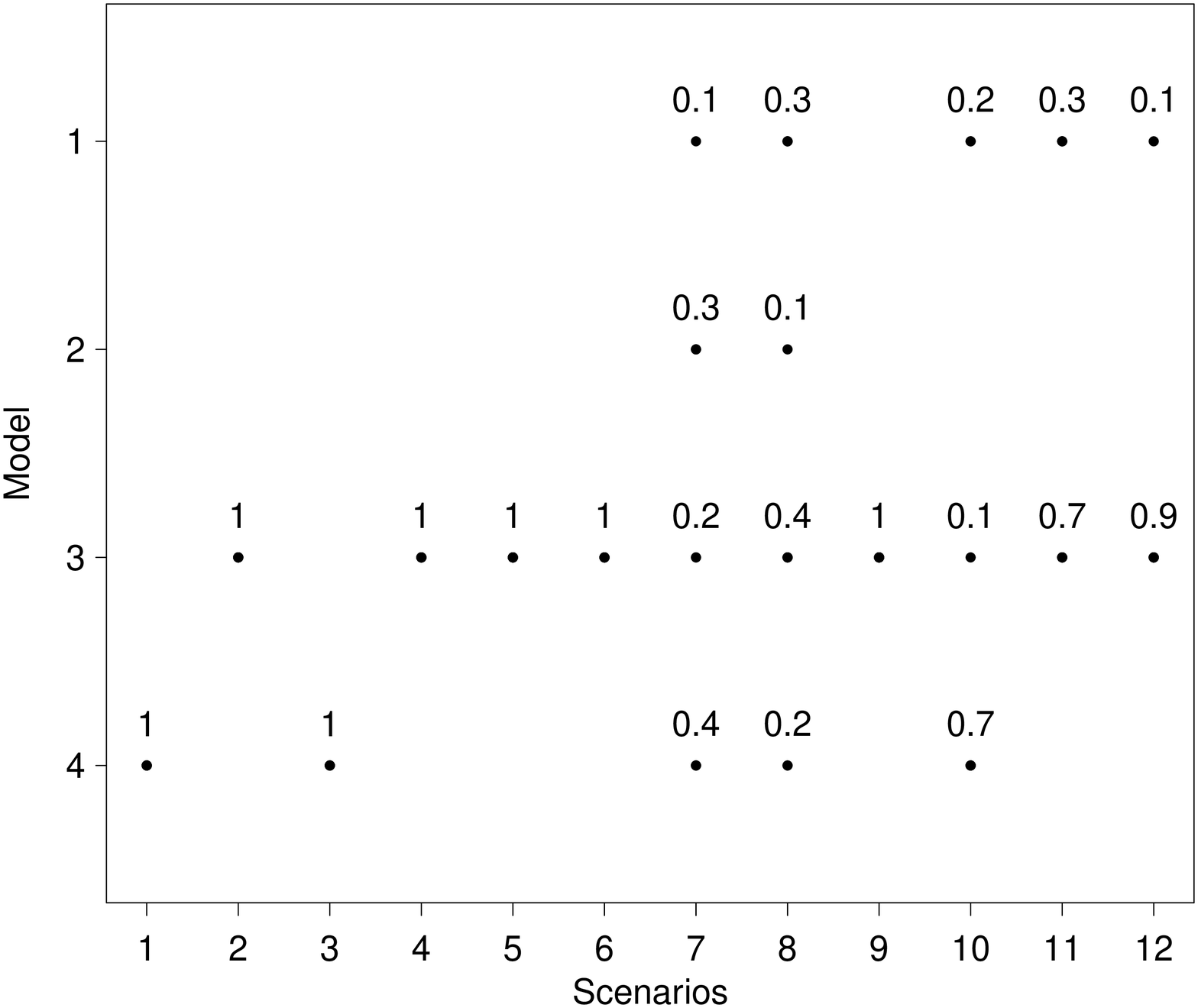}  \\ [0.6em]          
		{\hspace{100pt} (c)} &  {\hspace{100pt} (d)} \\ [-0.5em]
		\includegraphics[width=220pt,height=180pt]{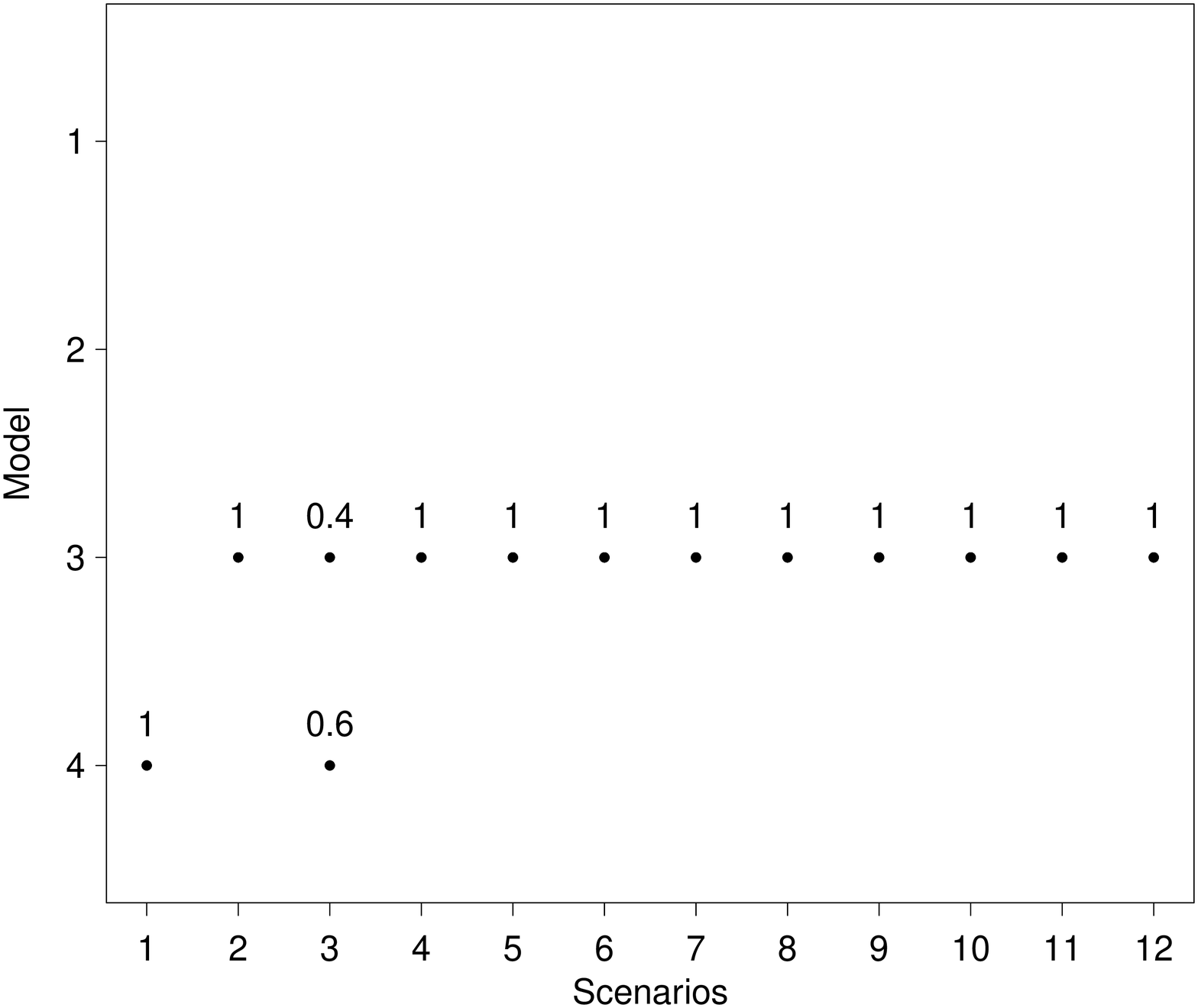} & \includegraphics[width=220pt,height=180pt]{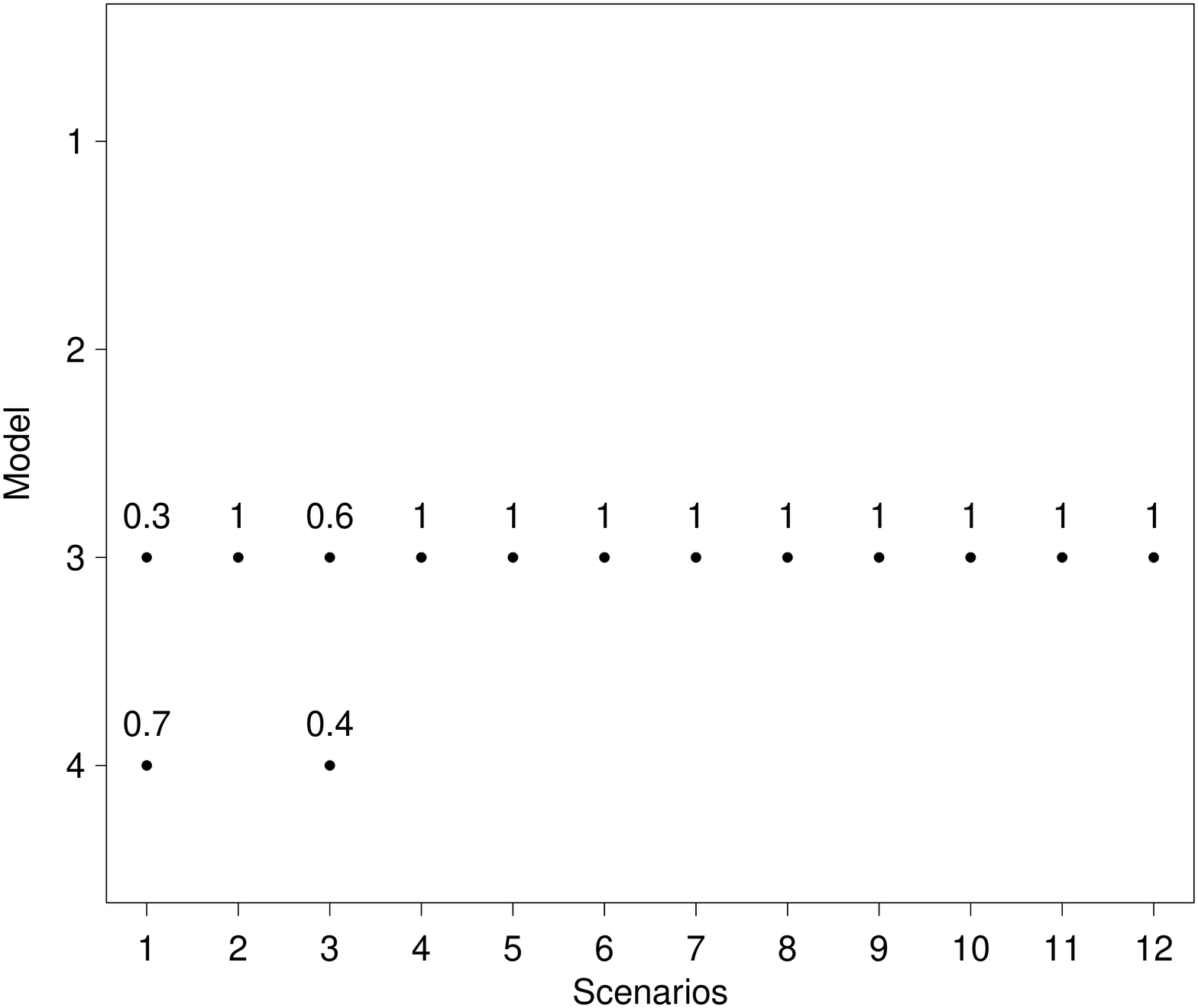}  \\ [0.6em]
		{\hspace{100pt} (e)} &  {\hspace{100pt} (f)} \\ [-0.5em]
		\includegraphics[width=220pt,height=180pt]{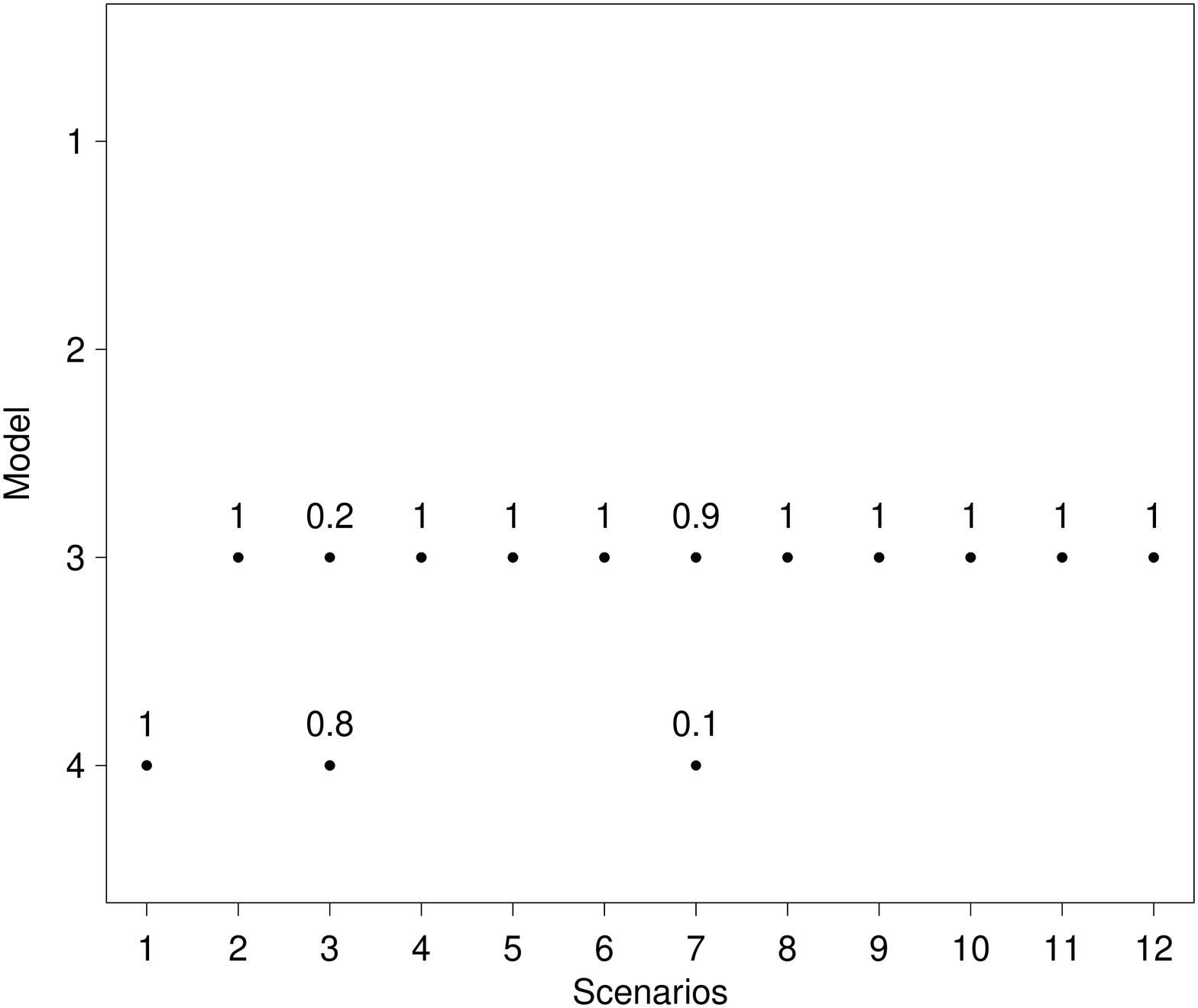} & \includegraphics[width=220pt,height=180pt]{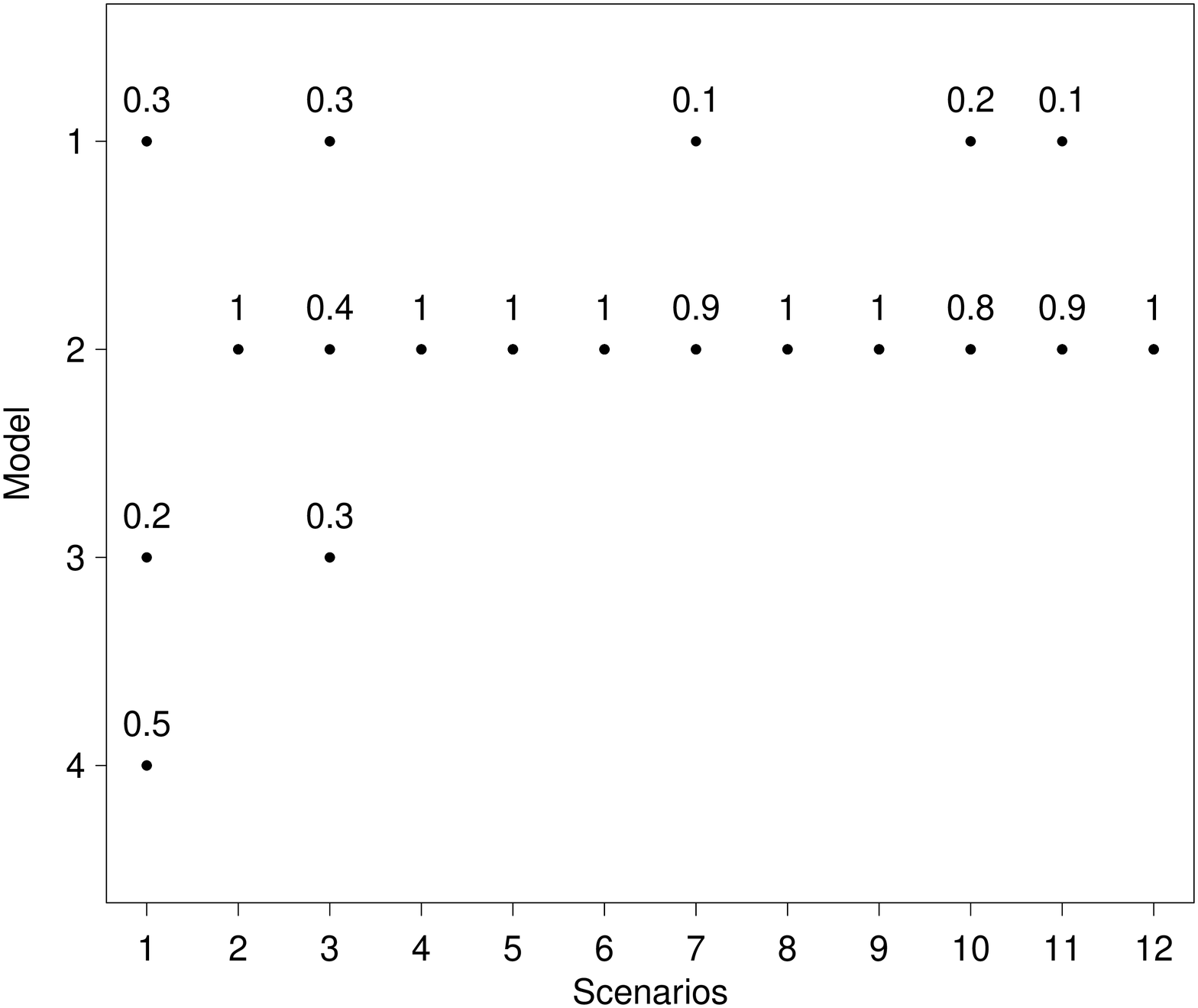} 
	\end{tabular}
	\caption{\small{The proportion of time a model selection method favours any particular model. Plots (a)-(f) correspond to WAIC1, WAIC2, WAIC3, DIC1, DIC2 and posterior predictive loss respectively.}}			
	\label{fig.modelsel5.BMSE}
\end{figure} 

\begin{figure}[H] 
	\centering
	\begin{tabular}{l @{\extracolsep{20pt}} l }
		{\hspace{110pt}$N$} & {\hspace{110pt} $\psi$} \\ [-0.75em]
		\includegraphics[width=210pt,height=210pt]{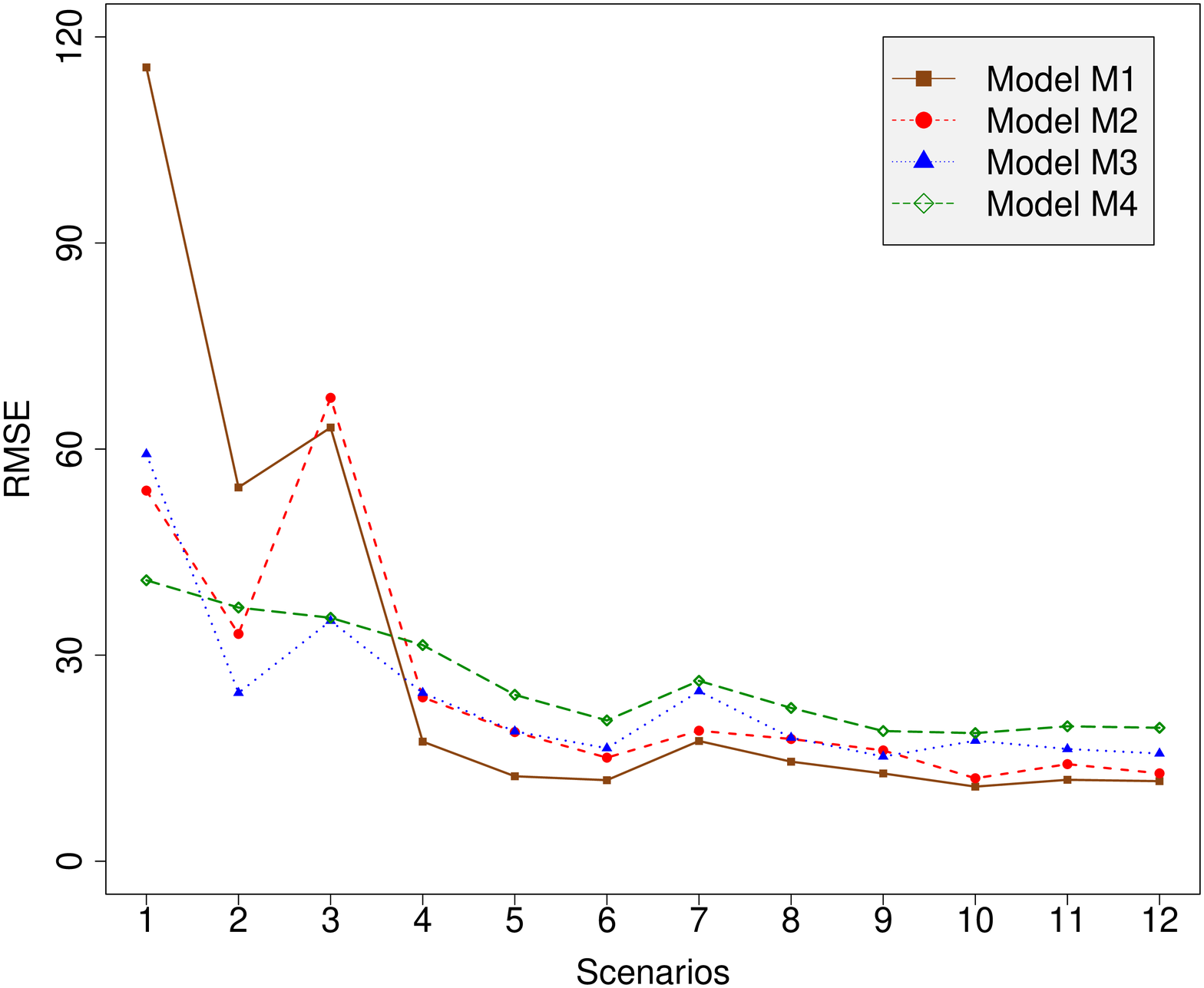} & \includegraphics[width=210pt,height=210pt]{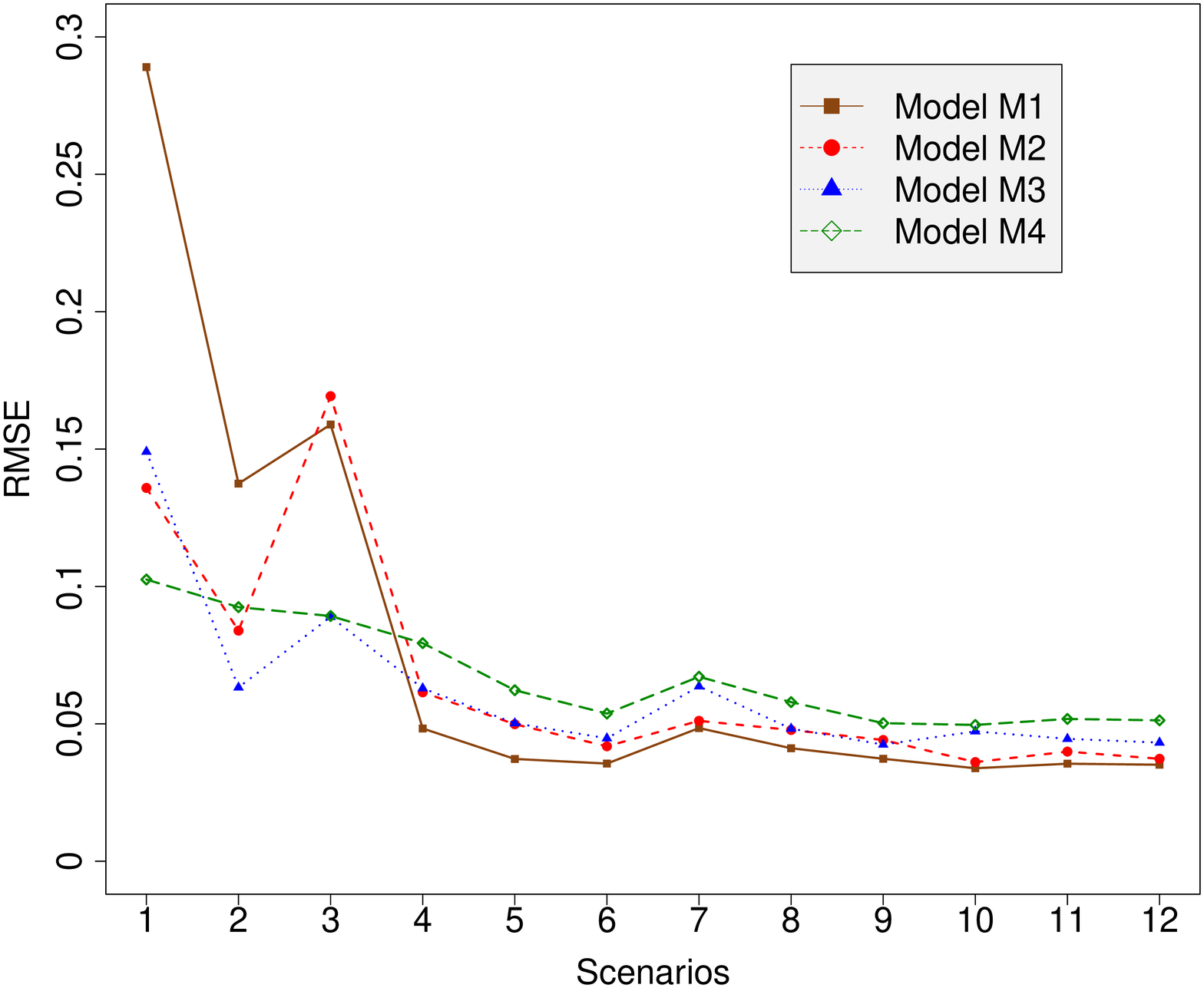}\\
		{\hspace{95pt} $N_{Male}$} & {\hspace{110pt} $\theta$} \\ [-0.75em]
		\includegraphics[width=210pt,height=210pt]{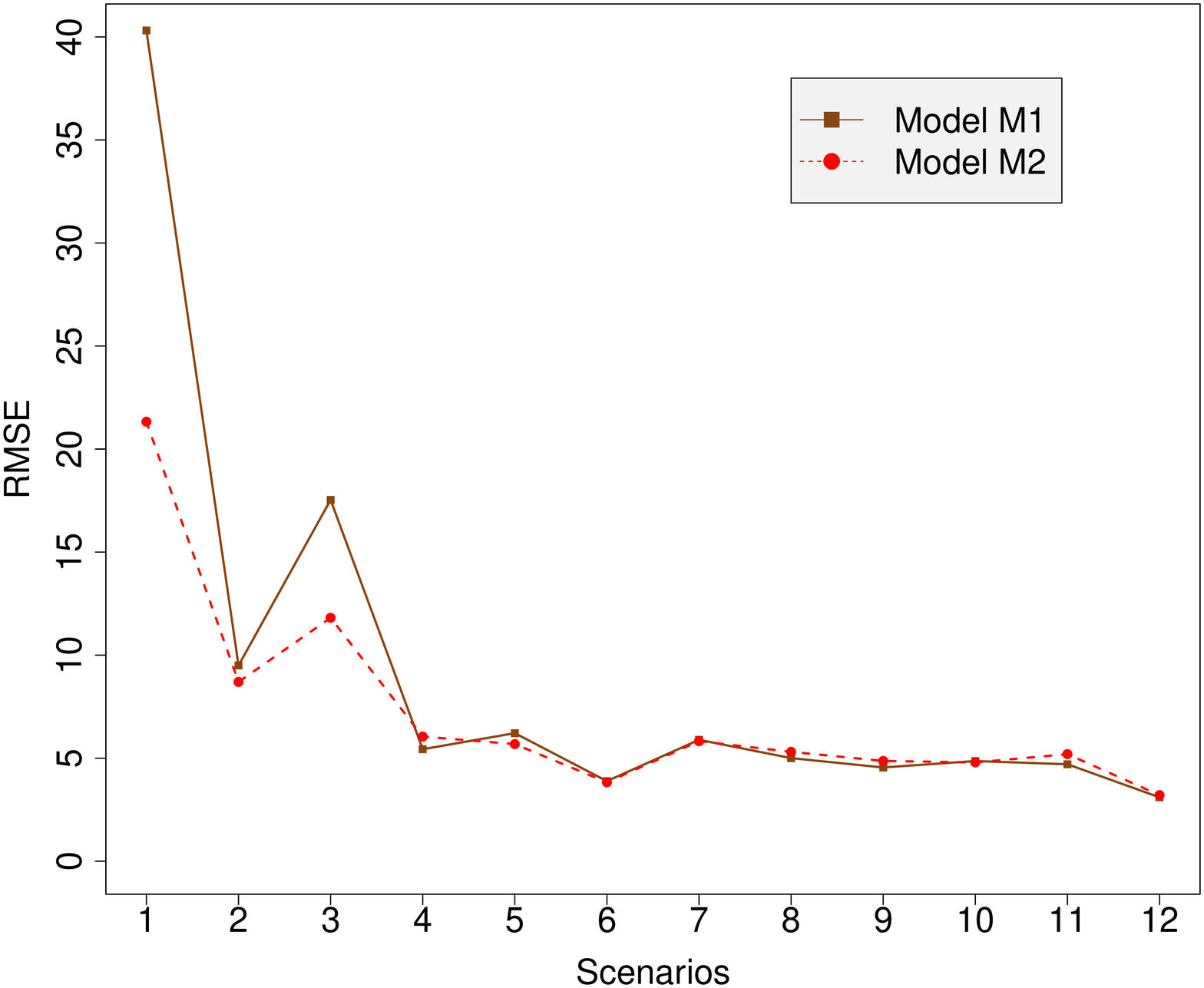} & \includegraphics[width=210pt,height=210pt]{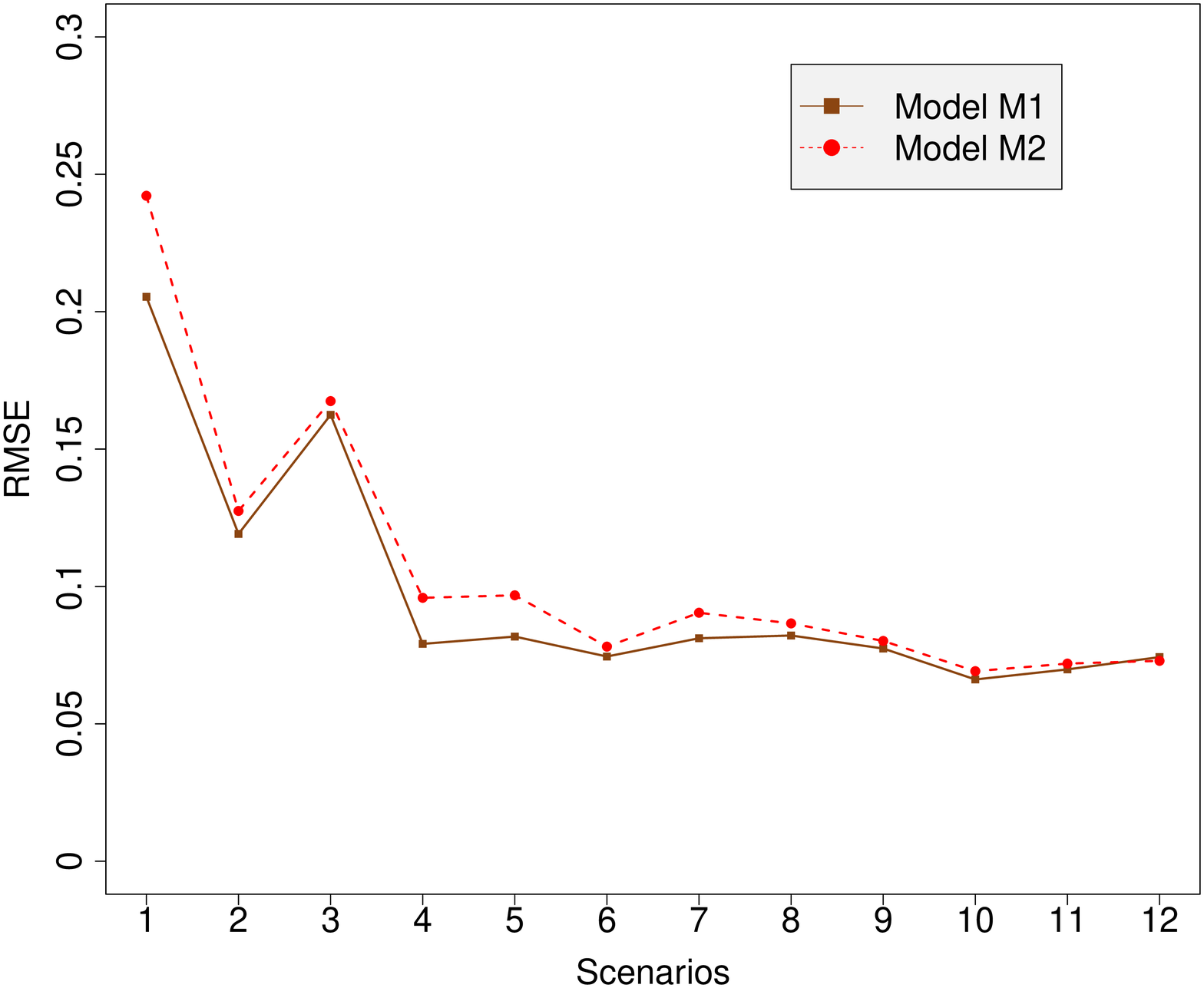}\\
		{\hspace{110pt} $\phi$} &  \\ [-0.75em]
		\includegraphics[width=210pt,height=210pt]{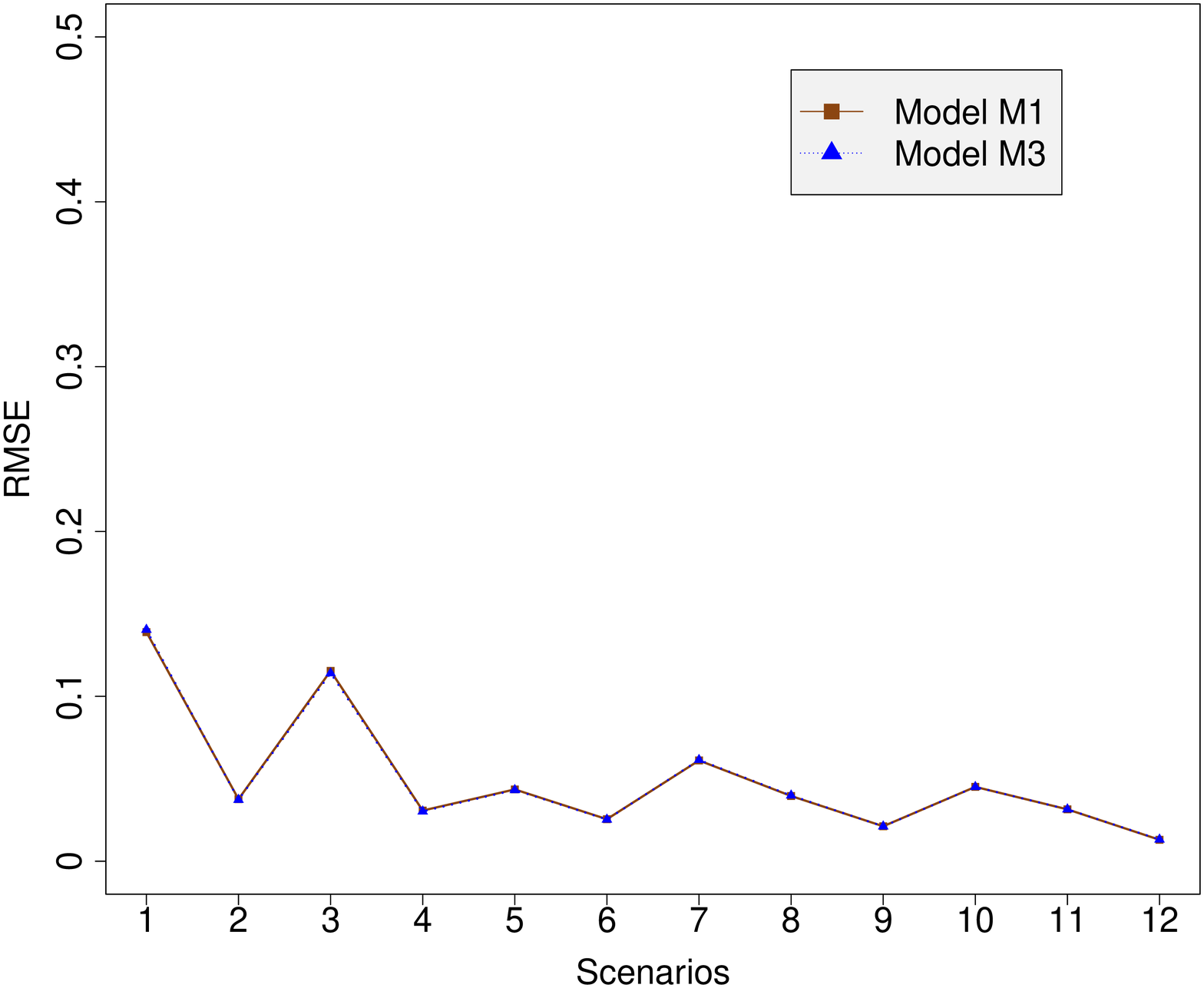} &
	\end{tabular}
	\caption{Plot of average RMSE estimates of $N$, $\psi$, $N_{Male}$, $\theta$ and $\phi$ over different simulation scenarios.}
	\label{fig.rmse1.BMSE}
\end{figure}

\begin{figure}[H] 
	\centering
	\begin{tabular}{l @{\extracolsep{20pt}} l }
		{\hspace{110pt} $\w_0$} & {\hspace{95pt} $p_0$}\\ [-0.75em]
		\includegraphics[width=210pt,height=210pt]{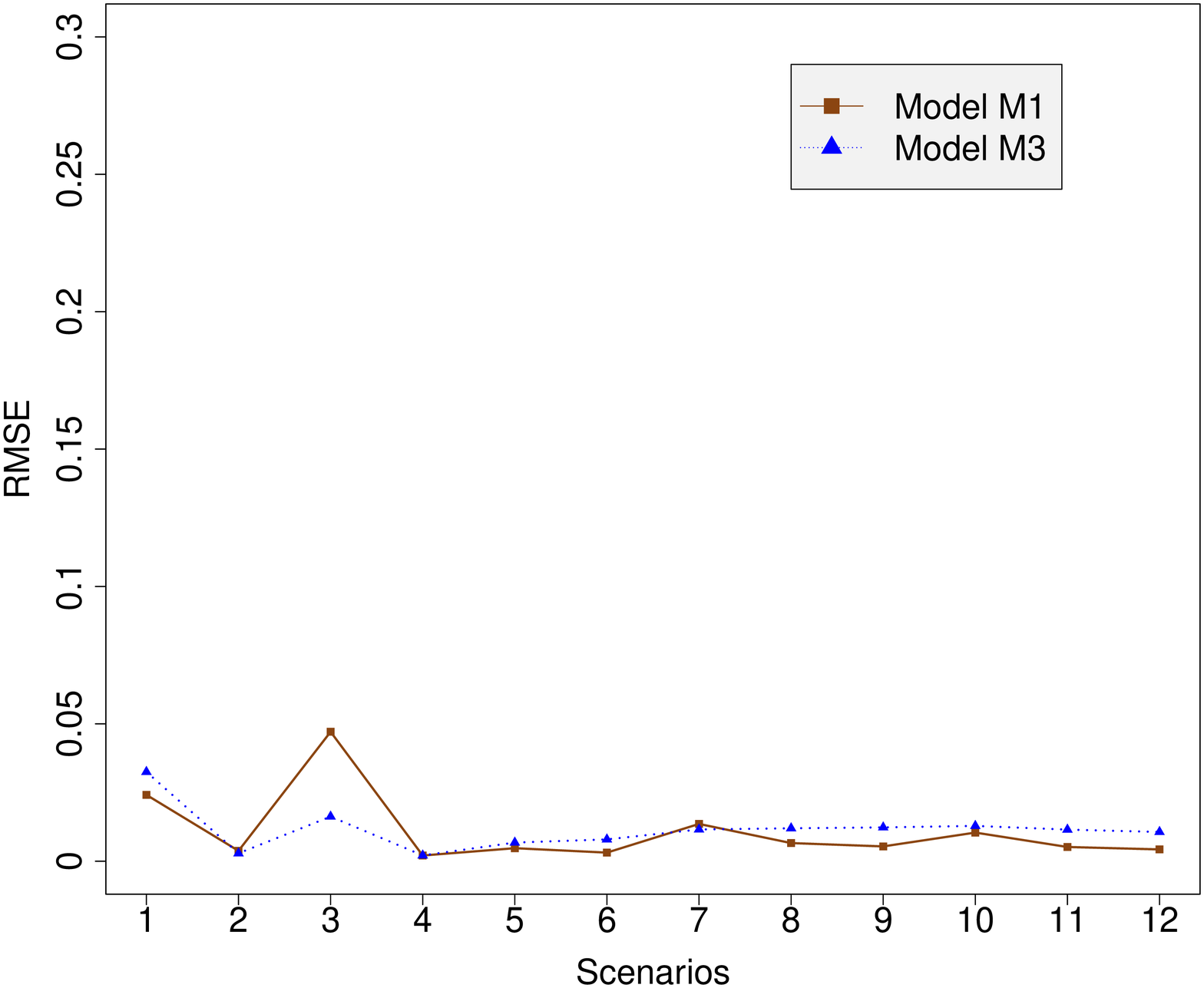} & \includegraphics[width=210pt,height=210pt]{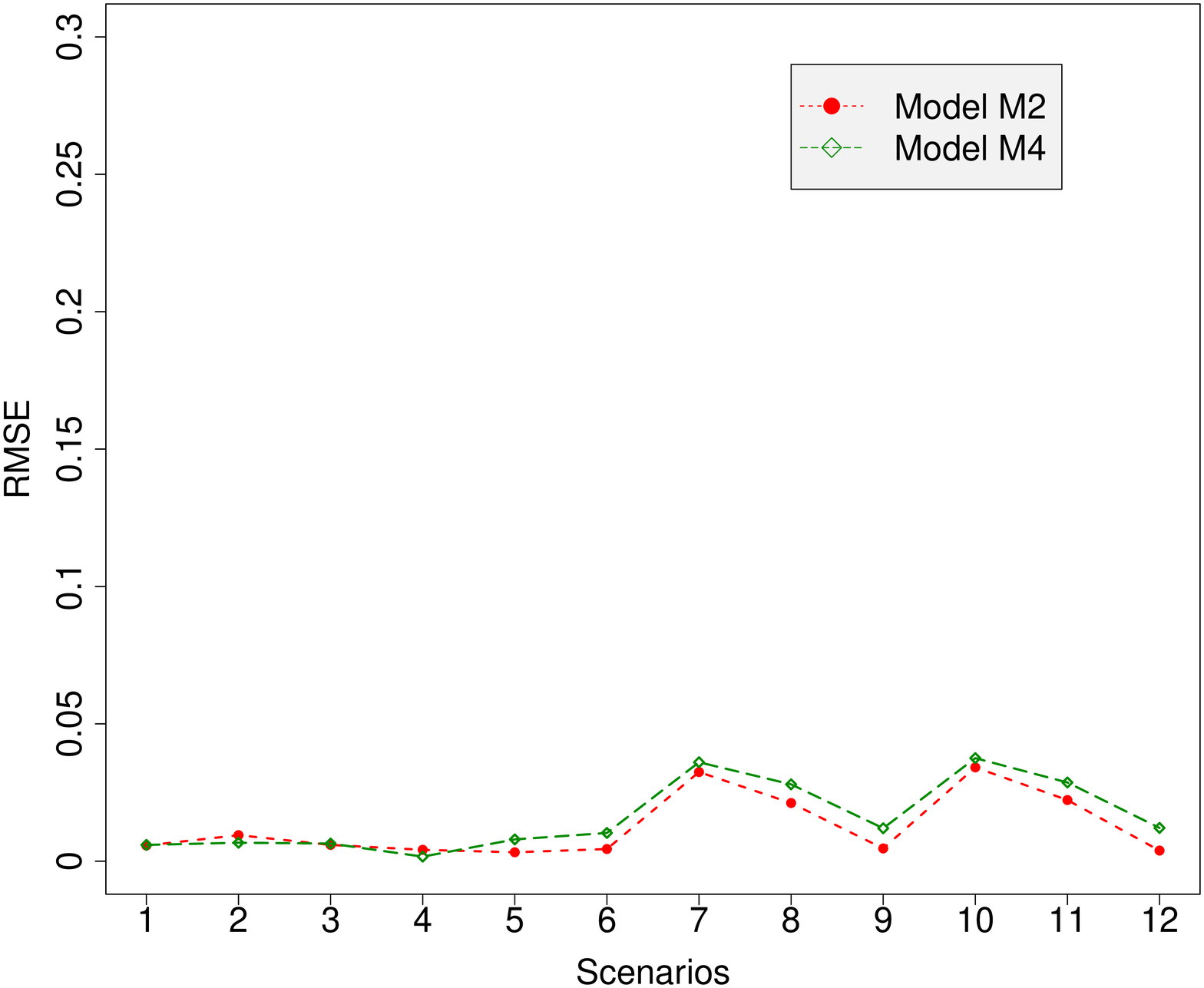}\\
		{\hspace{110pt} $\sigma_m$} & {\hspace{95pt} $\sigma_f$} \\ [-0.75em]
		\includegraphics[width=210pt,height=210pt]{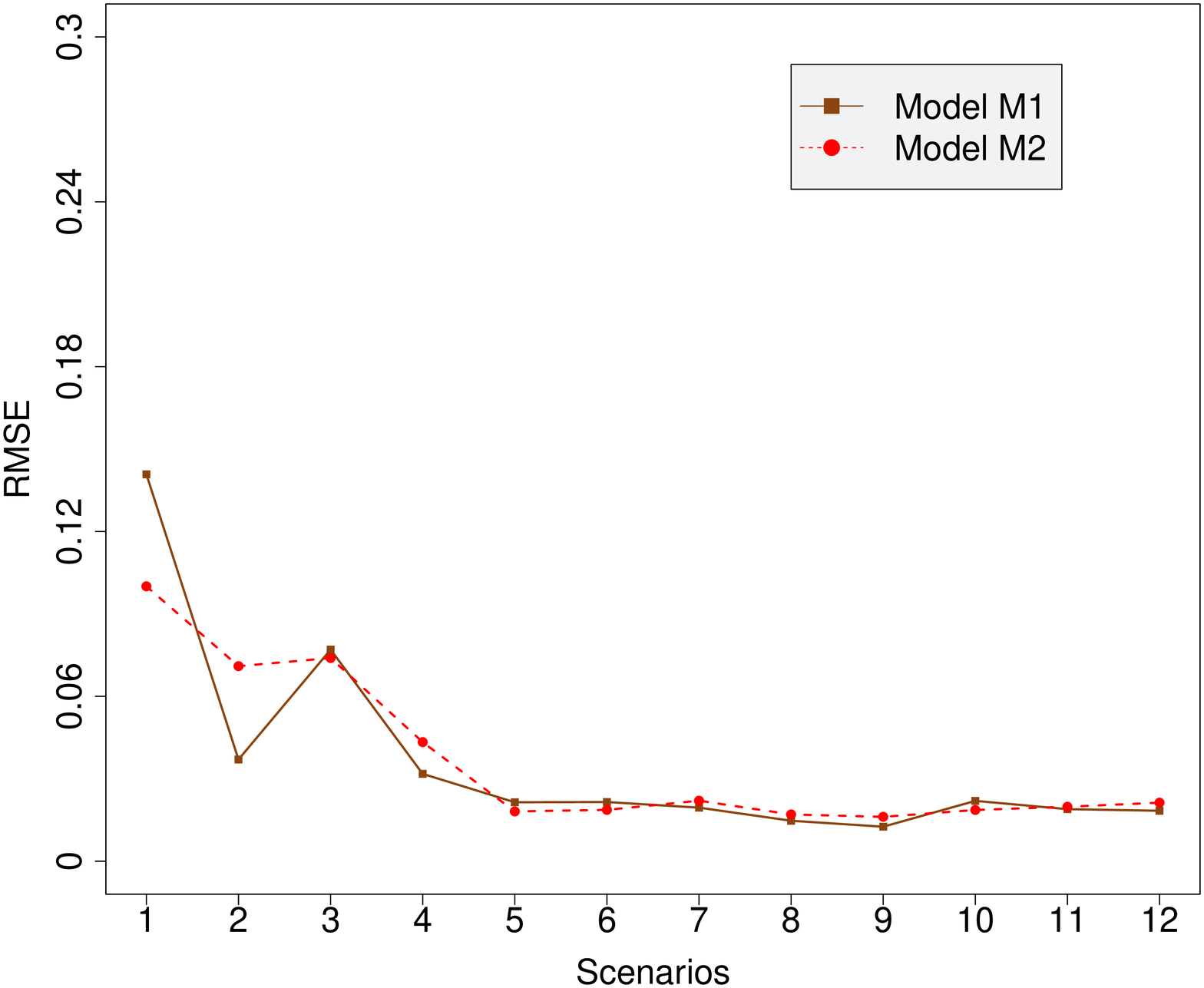} & \includegraphics[width=210pt,height=210pt]{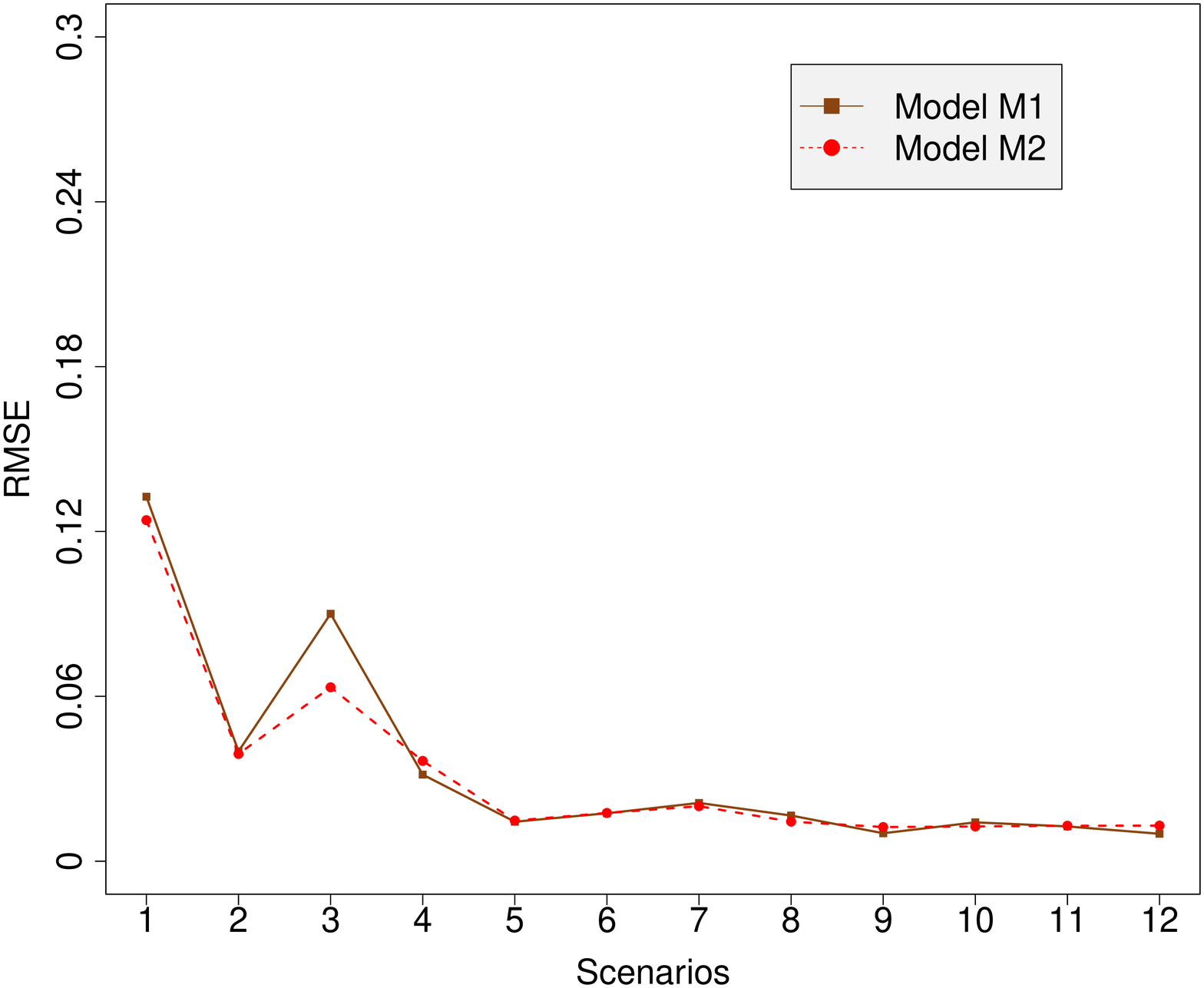}\\
		{\hspace{110pt} $\sigma$} & \\ [-0.75em]
		\includegraphics[width=210pt,height=210pt]{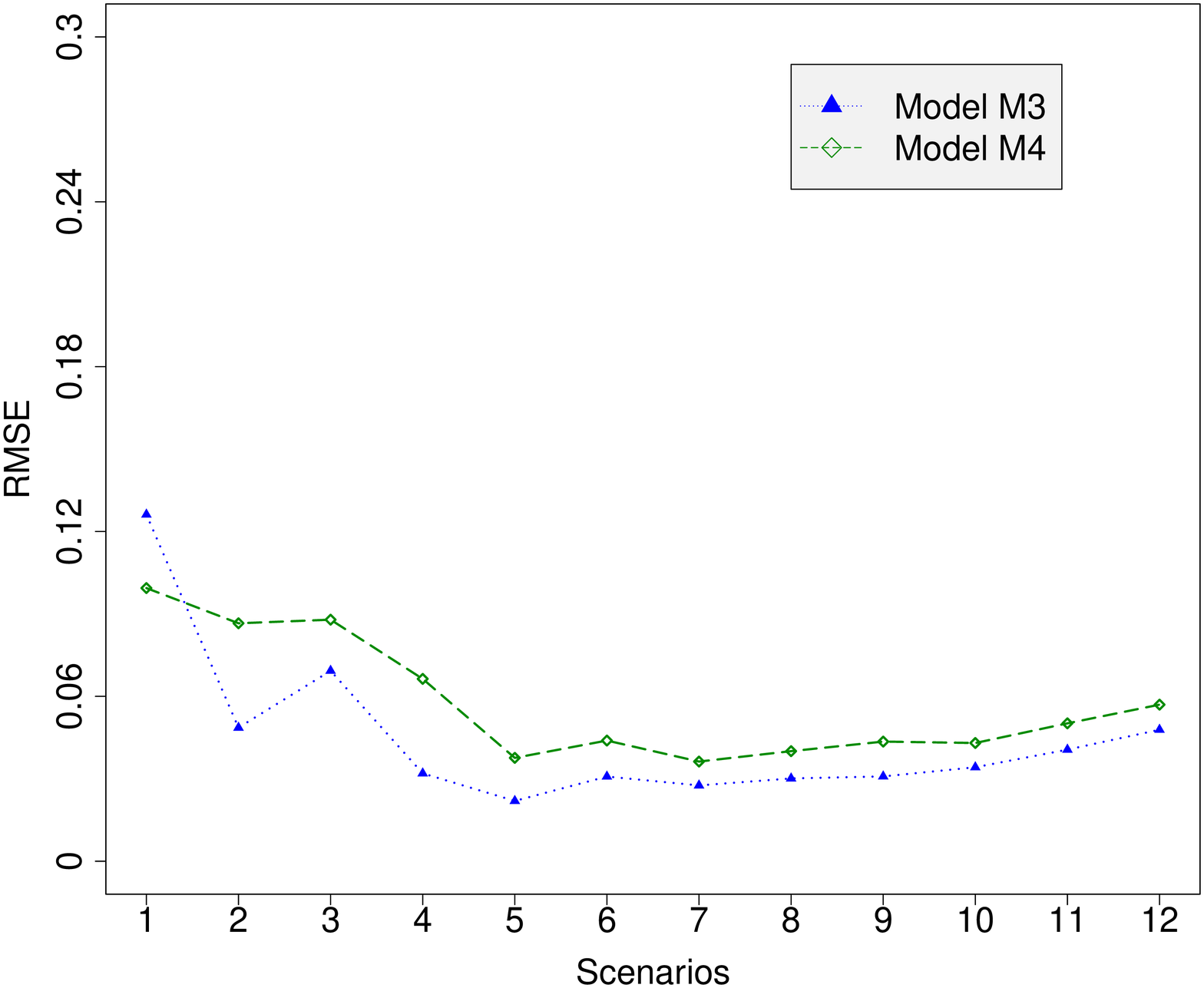}  &     
	\end{tabular}
	\caption{Plot of average RMSE estimates of $\w_0$, $p_0$, $\sigma_m$, $\sigma_f$ and $\sigma$ over different simulation scenarios.}
	\label{fig.rmse2.BMSE}
\end{figure} 
%
%
%

\begin{figure}[H] 
	\centering
	\includegraphics[width=470pt,height=470pt]{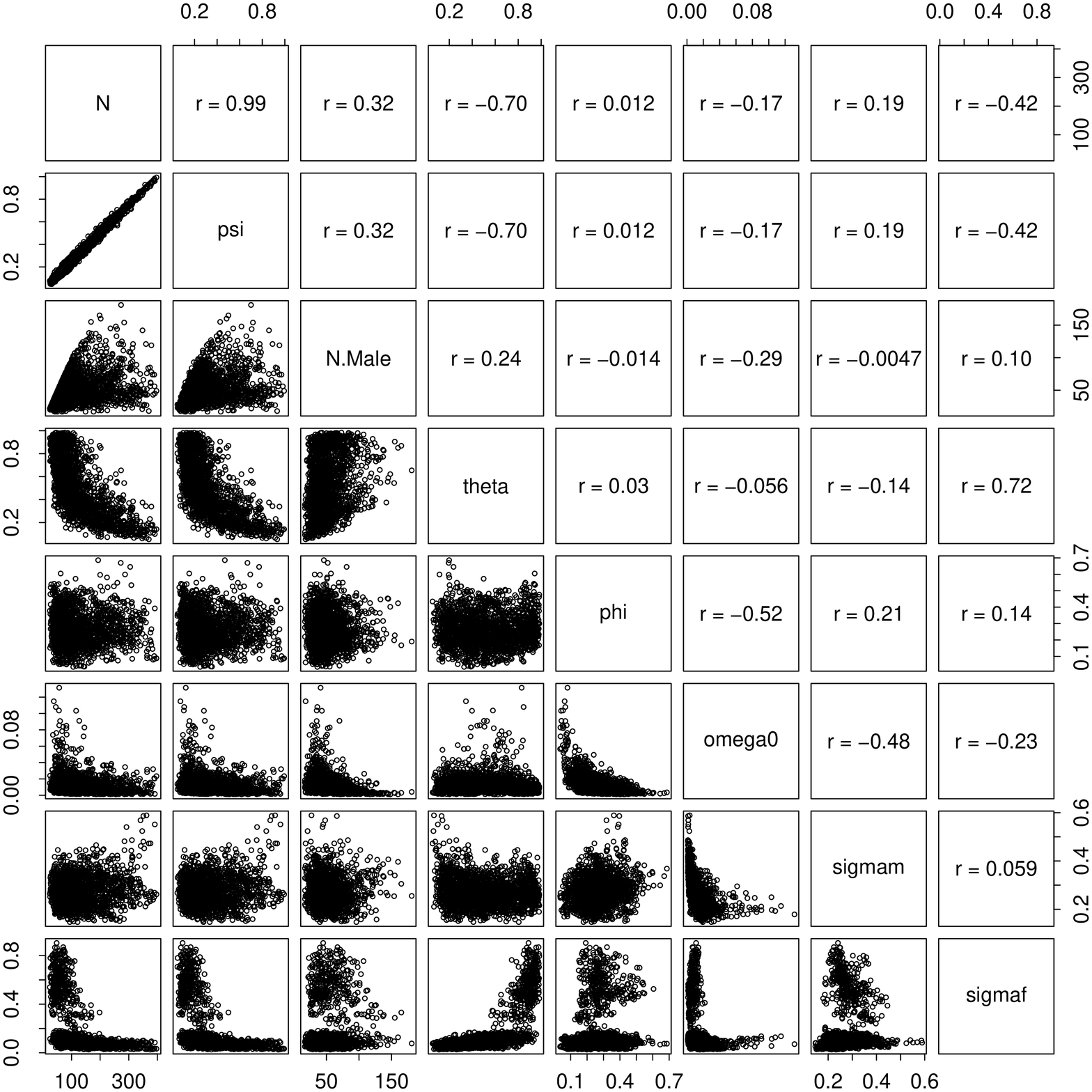}  \caption{Scatter plot of the parameters obtained from MCMC chains of a simulated data analysis of $M_1$ corresponding to scenario 1. Here $r$ denotes the correlation coefficient and is computed by using the MCMC chains of the respective parameters.}
	\label{fig.scatplotsc1m1.BMSE}
\end{figure} 
\begin{figure}[H] 
	\centering
	\includegraphics[width=470pt,height=470pt]{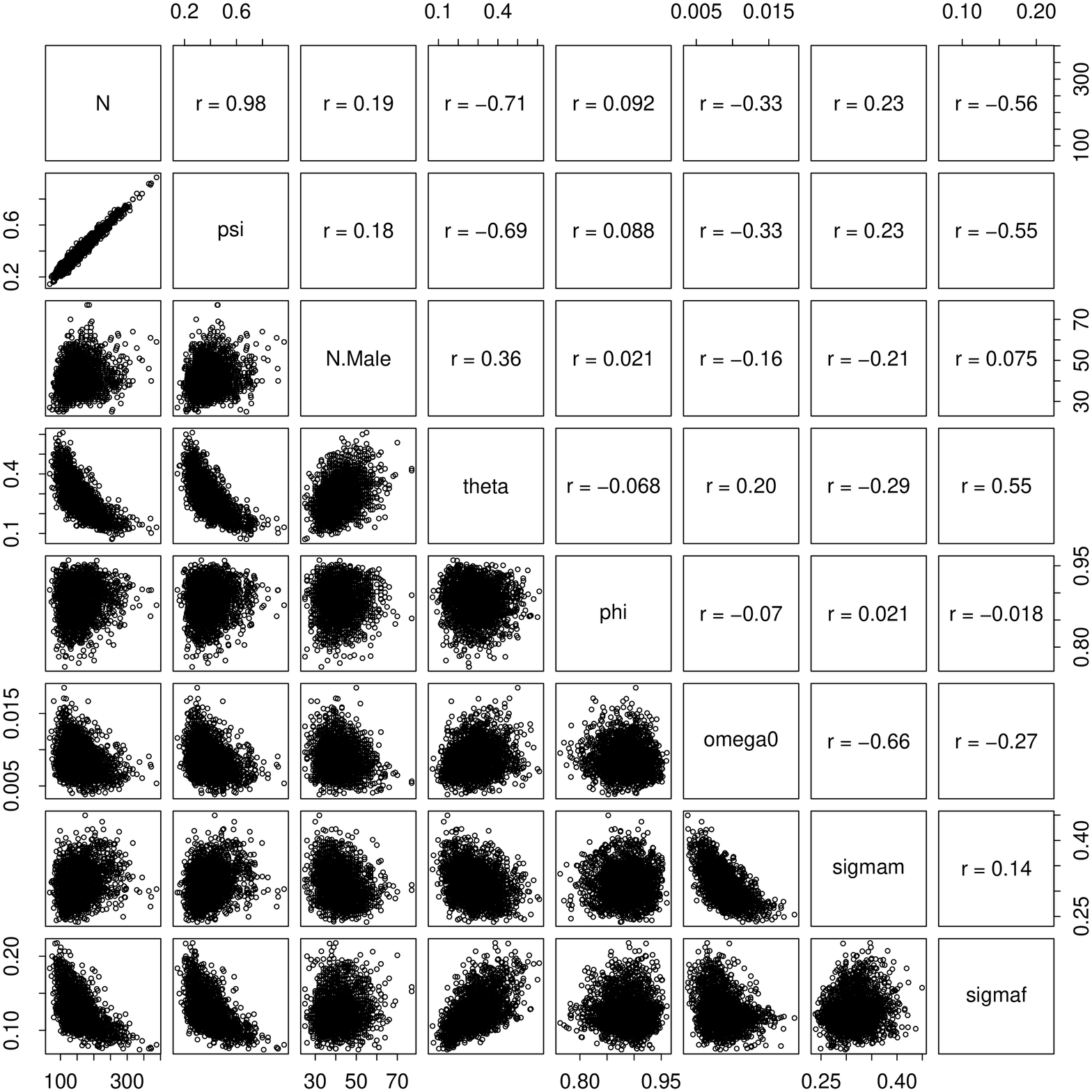}  \caption{Scatter plot of the parameters obtained from MCMC chains of a simulated data analysis of $M_1$ corresponding to scenario 2. Here $r$ denotes the correlation coefficient and is computed by using the MCMC chains of the respective parameters.}
	\label{fig.scatplotsc2m1.BMSE}
\end{figure} 
\begin{figure}[H] 
	\centering
	\includegraphics[width=470pt,height=470pt]{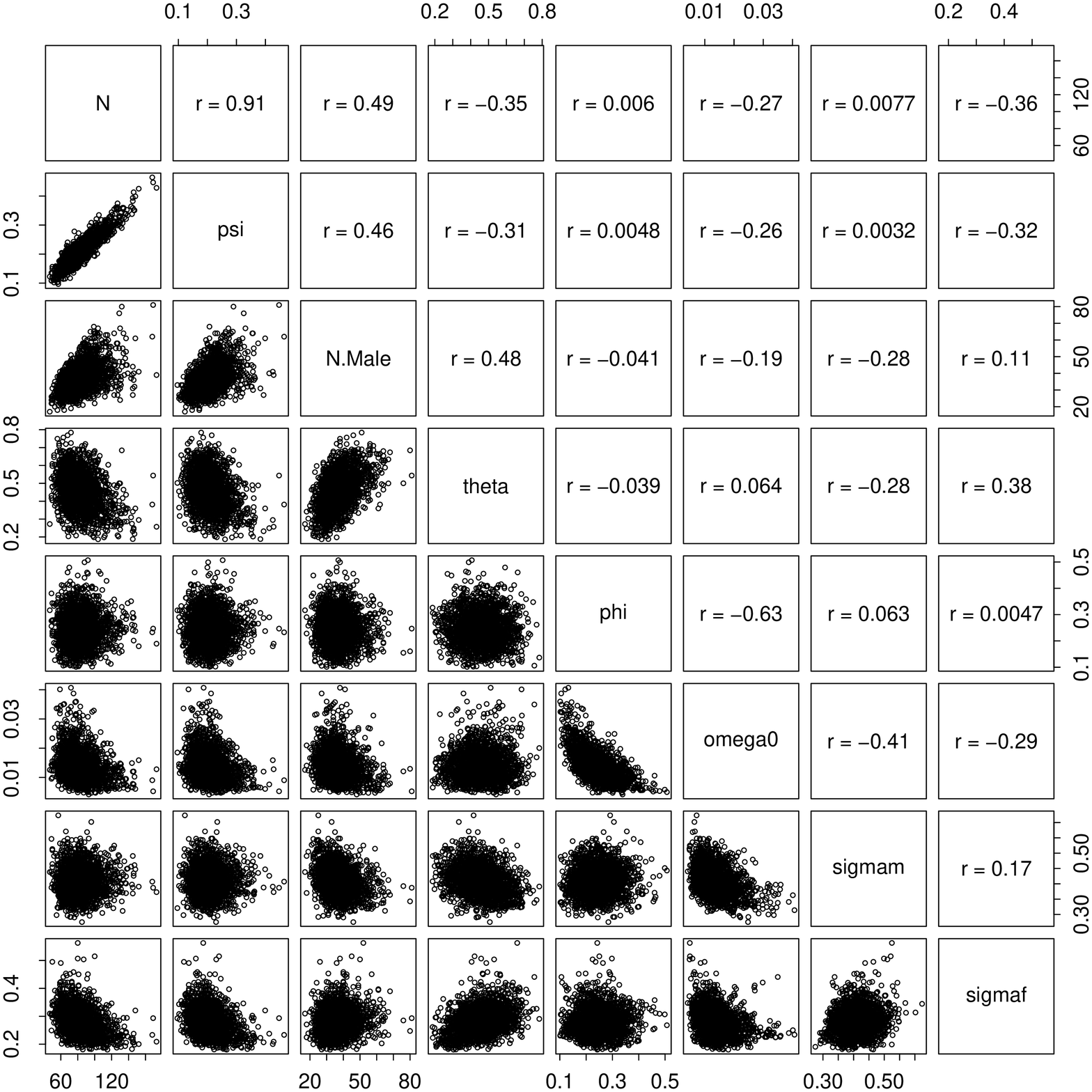}  \caption{Scatter plot of the parameters obtained from MCMC chains of a simulated data analysis of $M_1$ corresponding to scenario 3. Here $r$ denotes the correlation coefficient and is computed by using the MCMC chains of the respective parameters.}
	\label{fig.scatplotsc3m1.BMSE}
\end{figure} 
\begin{figure}[H] 
	\centering
	\includegraphics[width=470pt,height=470pt]{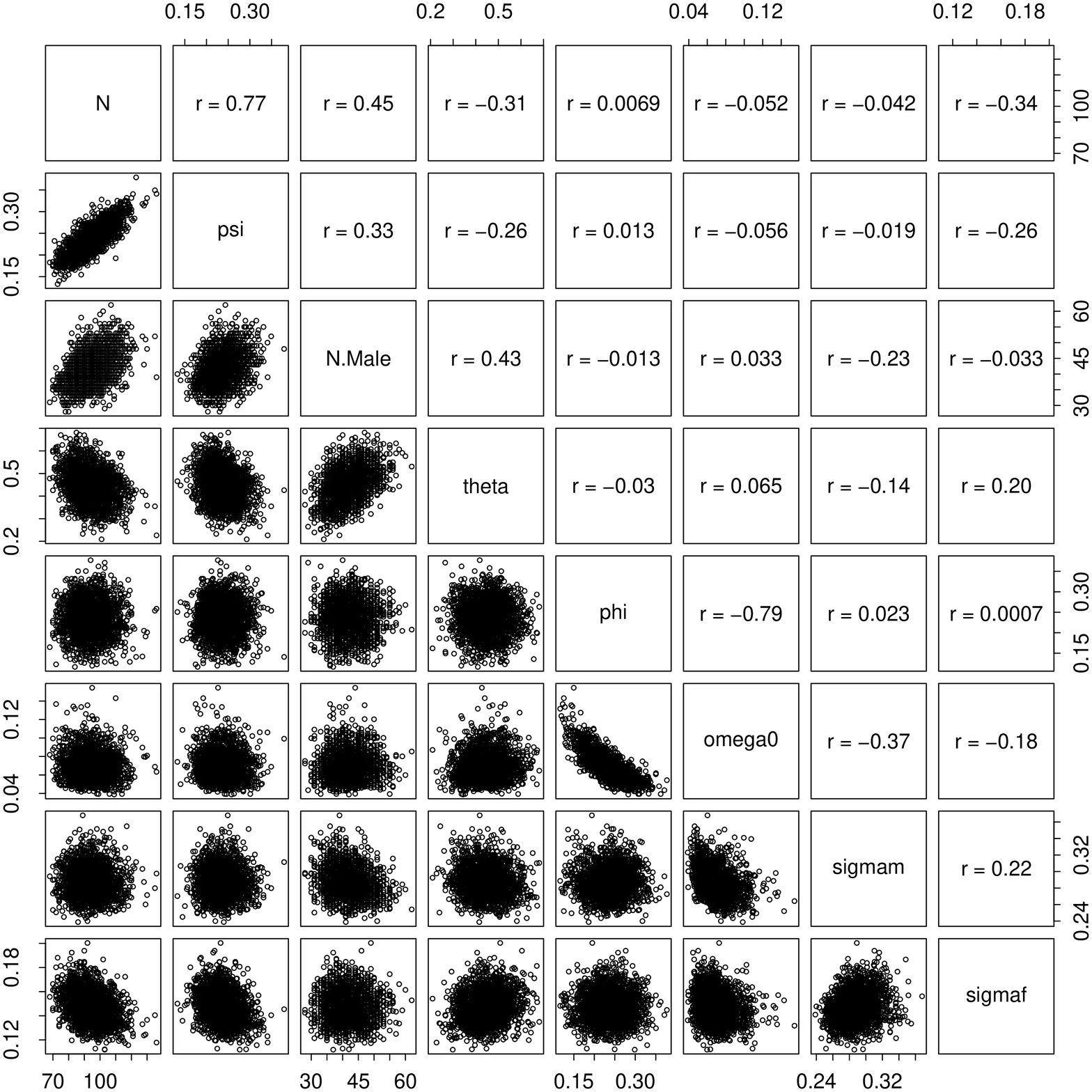}  \caption{Scatter plot of the parameters obtained from MCMC chains of a simulated data analysis of $M_1$ corresponding to scenario 7. Here $r$ denotes the correlation coefficient and is computed by using the MCMC chains of the respective parameters.}
	\label{fig.scatplotsc7m1.BMSE}
\end{figure} 
\begin{figure}[H] 
	\centering
	\includegraphics[width=470pt,height=470pt]{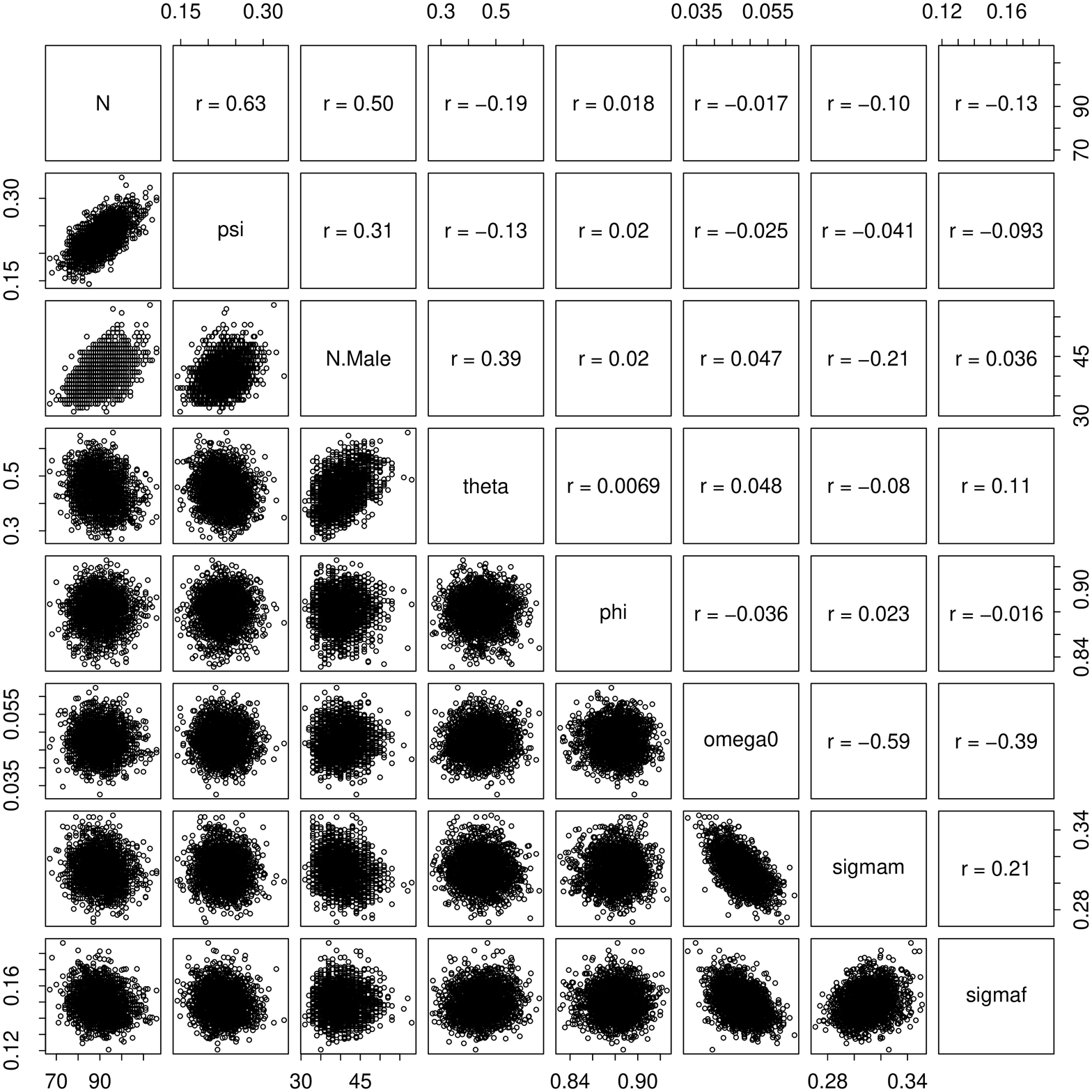}  \caption{Scatter plot of the parameters obtained from MCMC chains of a simulated data analysis of $M_1$ corresponding to scenario 9. Here $r$ denotes the correlation coefficient and is computed by using the MCMC chains of the respective parameters.}
	\label{fig.scatplotsc9m1.BMSE}
\end{figure} 
\begin{figure}[H] 
	\centering
	\includegraphics[width=470pt,height=470pt]{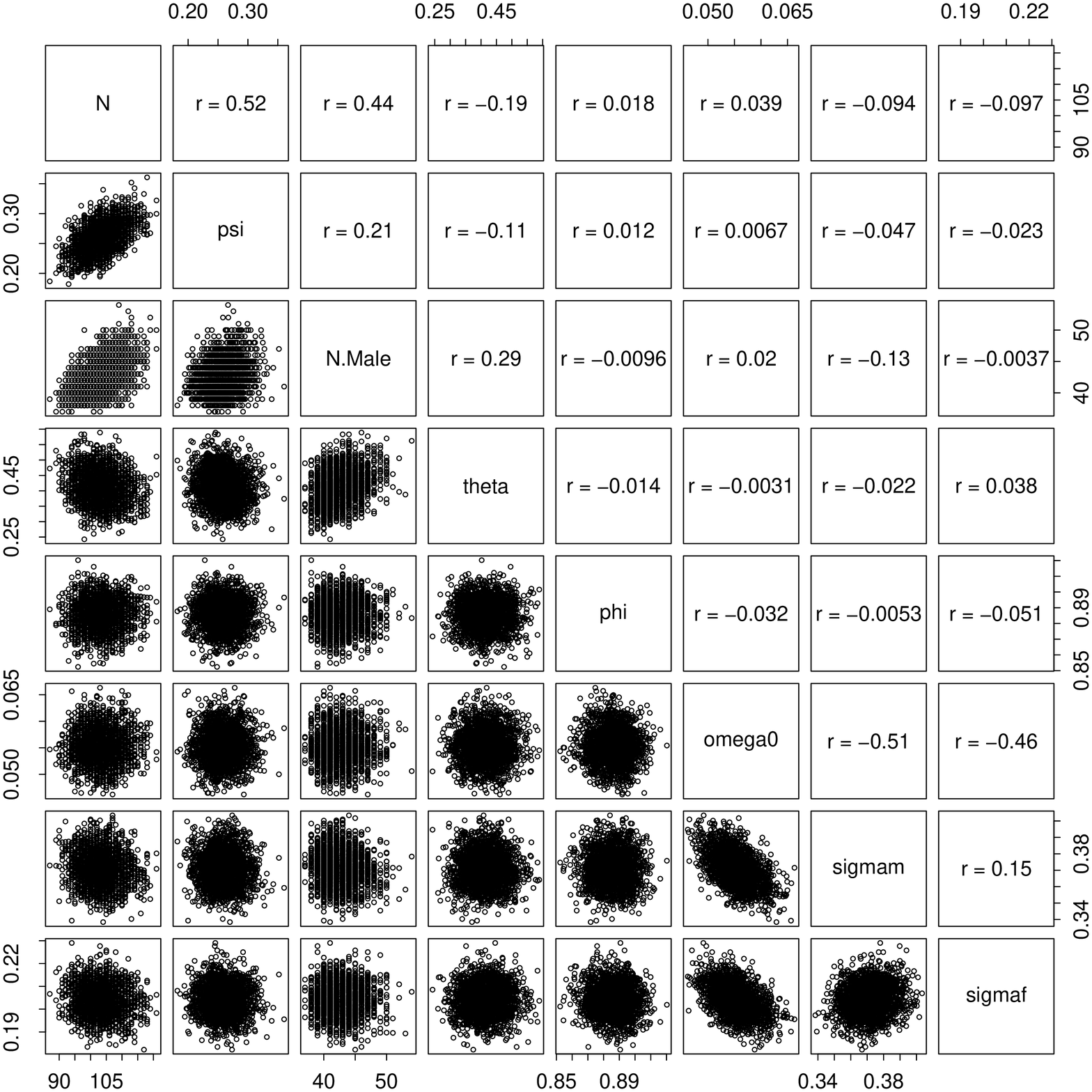}  \caption{Scatter plot of the parameters obtained from MCMC chains of a simulated data analysis of $M_1$ corresponding to scenario 12. Here $r$ denotes the correlation coefficient and is computed by using the MCMC chains of the respective parameters.}
	\label{fig.scatplotsc12m1.BMSE}
\end{figure} 

\pagebreak

 \bibliographystyle{apalike}
 \bibliography{thesis-articles-books} 


 \end{document}